\pdfoutput=1
\documentclass[12pt,a4paper]{iopart} 

\usepackage{graphicx}
\usepackage{color}
\usepackage{amssymb}
\usepackage{iopams}
\usepackage{bm}
\usepackage{units}
\usepackage{fancyhdr}
\usepackage{multicol}
\usepackage{float}
\usepackage{url}
\usepackage{gensymb}
\usepackage{tabularx}
\usepackage{longtable}

\usepackage[margin=1.5cm, bottom=1.5cm]{geometry}

\usepackage{harvard}

\definecolor{darkgreen}{rgb}{0.1,0.7,0.1}
\renewcommand{\vec}{\bm}

\renewcommand{\mat}[1]{\bm{\mathrm{#1}}}

\newcommand{\fourier}{\tilde}

\newcommand{\survey}[1]{{#1}}

\newcommand{\instrument}[1]{{#1}}

\newcommand{\citet}[2][]{#1\citeasnoun{#2}}

\newcommand{\citep}[2][]{(#1\citename{#2} \citeyear*{#2})}

\newcommand{\begtwocol}{\doiftwocol{\begin{multicols}{2}}}
\newcommand{\stoptwocol}{\doiftwocol{\end{multicols}}}

\begin{document}
%%%%%%%%%%%%%%%%%%%%%%%%%%%%%%%%%%%%%%%%%%%%%%%%%%%%%%%%%%%%

\citationmode{abbr} 

% from aa.cls
\def\aj{AJ}%
          % Astronomical Journal
\def\araa{ARA\&A}%
          % Annual Review of Astron and Astrophys
\def\apj{ApJ}%
          % Astrophysical Journal
\def\apjl{ApJ}%
          % Astrophysical Journal, Letters
\def\apjs{ApJS}%
          % Astrophysical Journal, Supplement
\def\ao{Appl.~Opt.}%
          % Applied Optics
\def\apss{Ap\&SS}%
          % Astrophysics and Space Science
\def\aap{A\&A}%
          % Astronomy and Astrophysics
\def\aapr{A\&A~Rev.}%
          % Astronomy and Astrophysics Reviews
\def\aaps{A\&AS}%
          % Astronomy and Astrophysics, Supplement
\def\azh{AZh}%
          % Astronomicheskii Zhurnal
\def\baas{BAAS}%
          % Bulletin of the AAS
\def\jcap{JCAP}%
	  % Journal of Cosmology and Astroparticle Physics
\def\jrasc{JRASC}%
          % Journal of the RAS of Canada
\def\memras{MmRAS}%
          % Memoirs of the RAS
\def\mnras{MNRAS}%
          % Monthly Notices of the RAS
\def\na{New Astron.}%
	  % New Astronomy
\def\nar{New Astron.~Rev.}%
	  % New Astronomy Reviews
\def\pasa{PASA}
	  % Publications of the Astronomical Society of Australia
\def\pra{Phys.~Rev.~A}%
          % Physical Review A: General Physics
\def\prb{Phys.~Rev.~B}%
          % Physical Review B: Solid State
\def\prc{Phys.~Rev.~C}%
          % Physical Review C
\def\prd{Phys.~Rev.~D}%
          % Physical Review D
\def\pre{Phys.~Rev.~E}%
          % Physical Review E
\def\prl{Phys.~Rev.~Lett.}%
          % Physical Review Letters
\def\pasp{PASP}%
          % Publications of the ASP
\def\pasj{PASJ}%
          % Publications of the ASJ
\def\qjras{QJRAS}%
          % Quarterly Journal of the RAS
\def\skytel{S\&T}%
          % Sky and Telescope
\def\solphys{Sol.~Phys.}%
\def\sovast{Soviet~Ast.}%
          % Soviet Astronomy
\def\ssr{Space~Sci.~Rev.}%
          % Space Science Reviews
\def\zap{ZAp}%
          % Zeitschrift fuer Astrophysik
\def\nat{Nature}%
          % Nature
\def\iaucirc{IAU~Circ.}%
          % IAU Cirulars
\def\aplett{Astrophys.~Lett.}%
          % Astrophysics Letters
\def\apspr{Astrophys.~Space~Phys.~Res.}%
          % Astrophysics Space Physics Research
\def\bain{Bull.~Astron.~Inst.~Netherlands}%
          % Bulletin Astronomical Institute of the Netherlands
\def\fcp{Fund.~Cosmic~Phys.}%
          % Fundamental Cosmic Physics
\def\gca{Geochim.~Cosmochim.~Acta}%
          % Geochimica Cosmochimica Acta
\def\grl{Geophys.~Res.~Lett.}%
          % Geophysics Research Letters
\def\jcp{J.~Chem.~Phys.}%
          % Journal of Chemical Physics
\def\jgr{J.~Geophys.~Res.}%
          % Journal of Geophysics Research
\def\jqsrt{J.~Quant.~Spec.~Radiat.~Transf.}%
          % Journal of Quantitiative Spectroscopy and Radiative Trasfer
\def\memsai{Mem.~Soc.~Astron.~Italiana}%
          % Mem. Societa Astronomica Italiana
\def\nphysa{Nucl.~Phys.~A}%
          % Nuclear Physics A
\def\physrep{Phys.~Rep.}%
          % Physics Reports
\def\physscr{Phys.~Scr}%
          % Physica Scripta
\def\planss{Planet.~Space~Sci.}%
          % Planetary Space Science
\def\procspie{Proc.~SPIE}%
          % Proceedings of the SPIE
\let\astap=\aap
\let\apjlett=\apjl
\let\apjsupp=\apjs
\let\applopt=\ao

\pagestyle{fancy}
\lhead{\leftmark}
\rhead{\thepage}
\cfoot{}

%\fancyhead[LE,RO]{\thepage}
%\fancyhead[RE]{\leftmark}
%\fancyhead[LO]{\title}

% One column format
\newcommand{\doifonecol}[1]{#1}
\newcommand{\doiftwocol}[1]{}
% Two column format
%\renewcommand{\doifonecol}[1]{}
%\renewcommand{\doiftwocol}[1]{#1}

%%%%%%%%%%%%%%%%%%%%%%%%%%%%%%%%%%%%%%%%%%%%%%%%%%%%%%%%%%%%
% Title page
%%%%%%%%%%%%%%%%%%%%%%%%%%%%%%%%%%%%%%%%%%%%%%%%%%%%%%%%%%%%

\thispagestyle{empty}
\vspace{-2cm}

\voffset-10pt

\begin{center}
\noindent
\includegraphics[height=3cm]{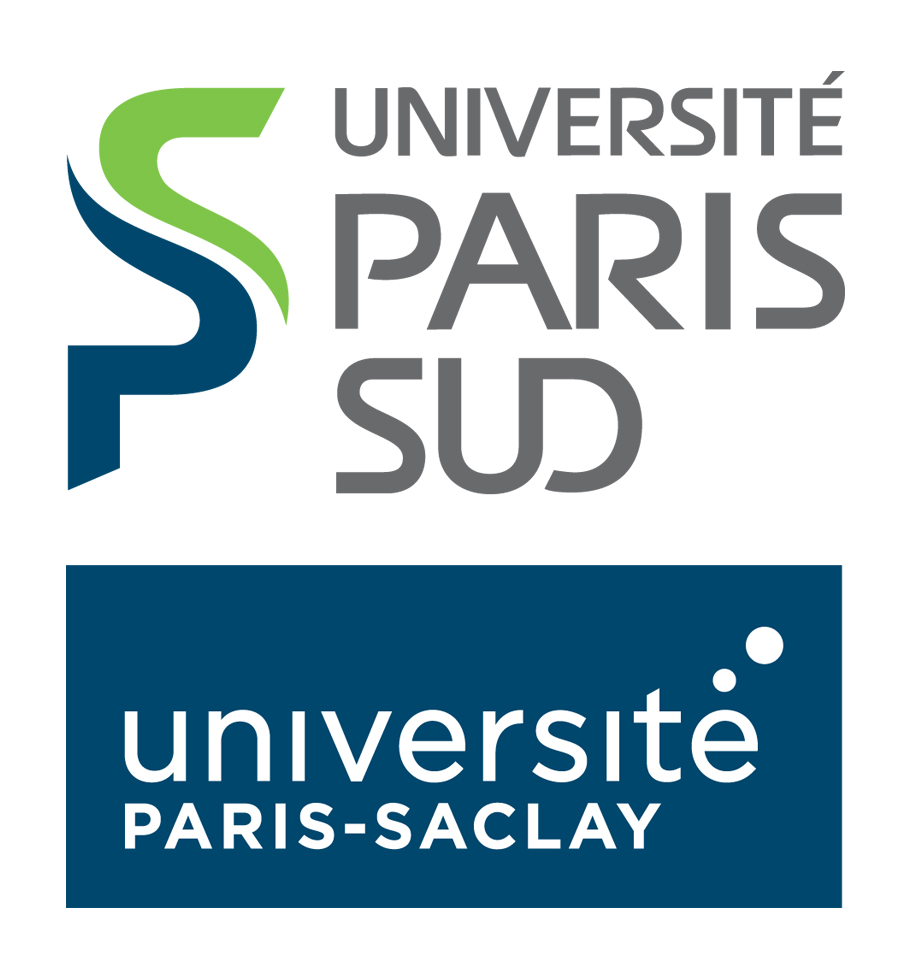}
\hfill
\raisebox{0.5cm}{
\includegraphics[height=2cm]{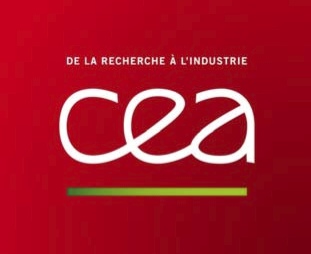}
\includegraphics[height=2.cm]{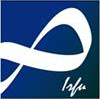}
\includegraphics[height=2.cm]{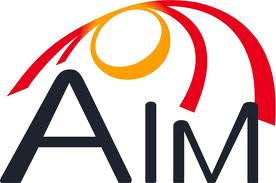}
}
\end{center}

\vspace{7mm}

\begin{center}
 {\Large\sc Universit\'e Paris-Sud}\\
  \vspace*{0.8cm}
% Facult\'e des sciences d'Orsay\\ 
%  \vspace*{0.2cm}
\'Ecole doctorale Astronomie et Astrophysique d'\^Ile-de-France (ED 127) \\
 \vspace*{0.2cm} 
%%%
%% remplacer le cas Ã©chÃ©ant par votre laboratoire d'accueil
%%%
CEA Saclay, Irfu, DAp-AIM (UMR 7158) 
\end{center}
\vspace{5mm}

\begin{center}
M\'emoire pr\'esent\'e pour l'obtention du\\\mbox{}\\
{\Large\bf Dipl\^ome d'habilitation \`a diriger les recherches}
\end{center}

\begin{center}
{\large Discipline : Astrophysique}
\end{center}

\vspace{5mm}

\begin{center}
\textit{par}
\end{center}

\begin{center}
{\large\bf Martin KILBINGER}
\end{center}

\vspace{5mm}
 
\begin{center}
\fbox{\Large Cosmological parameters from weak cosmological lensing}
\end{center}

\vspace{10mm}

%Rapporteurs~: 
%%%
% par ordre alphabÃ©tique
%%%
%\hspace{30mm}
%\begin{tabular}{l}
%{\sc   Alain BLANCHARD}\vspace{1mm}\\
%{\sc   Martin Kunz}\vspace{1mm}\\
%{\sc   Sophia MAUROGORDATO}\vspace{1mm}
%\end{tabular}

\vspace{10mm}

Date de soutenance~: 4 avril 2018

\vspace{8mm}

Lieu: CEA Saclay, Dap

\vspace{8mm}

Composition du jury~:
%%%
% par ordre alphabÃ©tique des membres
%%%
\hspace{15mm}
%%%%
%% version avant soutenance Ã  adapter (un jury peut contenir 
%% tout ou partie des rapporteurs (voir ci-dessous le cas 
%% Ã©chÃ©ant), voire des invitÃ©s) Le prÃ©sident du jury est 
%% choisi par le jury en son sein le jour de la soutenance,
%% donc ne peut apparaitre que dans la version aprÃ¨s soutenance
%%%%
\begin{tabular}{ll}
{\sc   Alain BLANCHARD}&(Rapporteur)\vspace{1mm}\\
{\sc   Martin KUNZ}&(Rapporteur)\vspace{1mm}\\
{\sc   Christophe PICHON}&(Rapporteur)\vspace{1mm}\\
{\sc   St\'ephane PLASZCZYNSKI}&(Pr\'esident \& Examinateur) \vspace{1mm}\\
{\sc   James G.~BARTLETT}&(Examinateur) \vspace{1mm}\\
{\sc   Nicholas KAISER}&(Examinateur)\vspace{1mm}\\
\end{tabular}

%\author{Martin Kilbinger}

%\address{Laboratoire AIM, CEA Saclay - CNRS - Paris 6, Irfu/SAp, F-91191 Gif-sur-Yvette, France}
%\eads{\mailto{martin.kilbinger@cea.fr}}

\begtwocol

\newpage

\newpage

%\input acknowledgements
%\addcontentsline{toc}{section}{Acknowledgements}
%\newpage

%\addcontentsline{toc}{section}{Table of contents}
\tableofcontents

\newpage

% Chapter 1: Introduction
% History
% introduction.tex

%%%%%%%%%%%%%%%%%%%%%%%%%%%%%%%%%%%%%%%%%%%%%%%%%%%%%%%%%%%%
\section{Introduction}
%%%%%%%%%%%%%%%%%%%%%%%%%%%%%%%%%%%%%%%%%%%%%%%%%%%%%%%%%%%%

%%%%%%%%%%%%%%%%%%%%%%%%%%%%%%%%%%%%%%%%%%%%%%%%%%%%%%%%%%%%
\subsection{One hundred years of gravitational lensing}
%%%%%%%%%%%%%%%%%%%%%%%%%%%%%%%%%%%%%%%%%%%%%%%%%%%%%%%%%%%%

On May, 29, 1919, during a solar eclipse, the deflection of light rays of stars
due to the Sun's gravitational field was measured \cite{1920RSPTA.220..291D},
marking the first successful test of the theory of general relativity
\citep[GR; ]{1916AnP...354..769E}. Only much later, in 1979 the first discovery
of extra-galactic gravitational lensing was obtained, with the detection of a
doubly-imaged quasar lensed by a galaxy \cite{1979Natur.279..381W}. Lensing
distortions have been known since 1987 with the observation of giant arcs ---
strongly distorted galaxies behind massive galaxy clusters
\cite{1987A&A...172L..14S}. Three years later in 1990, weak gravitational
lensing was detected for the first time as statistical tangential alignments of
galaxies behind massive clusters \cite{1990ApJ...349L...1T}. It took another 10
years until, in 2000, coherent galaxy distortions were measured in blind
fields, showing the existence of weak gravitational lensing by the large-scale
structure, or cosmic shear
\cite{2000MNRAS.318..625B,kaiser00,2000A&A...358...30V,2000Natur.405..143W}.
And so, nearly 100 years after its first measurement, the technique of
gravitational lensing has evolved into a powerful tool for challenging GR on
cosmological scales.

All observed light from distant galaxies is subject to gravitational lensing.
This is because light rays propagate through a universe that is inhomogeneous
due to the ubiquitous density fluctuations at large scales. These fluctuations
create a tidal gravitational field that causes light bundles to be deflected
differentially. As a result, images of light-emitting galaxies that we observe
are distorted. The direction and amount of distortion is directly related to
the size and shape of the matter distribution projected along the line of
sight. The deformation of high-redshift galaxy images in random lines of sight
therefore provides a measure of the large-scale structure (LSS) properties,
which consists of a network of voids, filaments, and halos. The larger the
amplitude of the inhomogeneity of this cosmic web is, the larger the
deformations are. This technique of \emph{cosmic shear}, or \emph{weak
cosmological lensing} is the topic of this review.

The typical distortions of high-redshift galaxies by the cosmic web are on the
order of a few percent, much smaller than the width of the intrinsic shape and
size distribution. Thus, for an individual galaxy, the lensing effect is not
detectable, placing cosmic shear into the regime of \emph{weak gravitational
lensing}. The presence of a tidal field acting as a gravitational lens
results in a coherent alignment of galaxy image orientations. This alignment
can be measured statistically as a correlation between galaxy shapes.

Cosmic shear is a very versatile probe of the LSS. It measures the clustering
of the LSS from the highly non-linear, non-Gaussian sub-megaparsec
(Mpc) regime, out to very large, linear scales of more than a hundred Mpc. By
measuring galaxy shape correlations between different redshifts, the evolution
of the LSS can be traced, enabling us to detect the effect of dark energy on
the growth of structure. Together with the ability to measure the geometry of
the Universe, cosmic shear can potentially distinguish between dark energy and
modified gravity theories \cite{1999ApJ...522L..21H}. Since gravitational
lensing is not sensitive to the dynamical state of the intervening masses, it
yields a direct measure of the total matter, dark plus luminous. By adding
information about the distribution of galaxies, cosmic shear can shed light on
the complex relationship between galaxies and dark matter.

Since the first detection over a few square degrees of sky area a decade and a
half ago, cosmic shear has matured into an important tool for cosmology.
Current surveys span hundreds of square degrees, and thousands of square
degrees more to be observed in the near future. Cosmic shear is a major science
driver of large imaging surveys from both ground and space.

This document follows in parts my recent review ``Cosmological parameters from weak cosmological
lensing'' \cite{K15}. Various other review articles on weak gravitational lensing have covered this and
related topics, see \citet[e.g.~]{BS01}, \citet{SaasFee}, \citet{2008ARNPS..58...99H},
\citet{2008PhR...462...67M}, \citet{2010CQGra..27w3001B}, \citet{2015IJMPD..2430011F}, and
\citet{2017arXiv171003235M}.

% Cosmo background
% cosmo_background.tex

%%%%%%%%%%%%%%%%%%%%%%%%%%%%%%%%%%%%%%%%%%%%%%%%%%%%%%%%%%%%
\subsection{Cosmological background}
\label{sec:cosmo_back}
%%%%%%%%%%%%%%%%%%%%%%%%%%%%%%%%%%%%%%%%%%%%%%%%%%%%%%%%%%%%

This section provides a very brief overview of the cosmological
concepts and equations relevant for weak gravitational lensing. Detailed derivations
of the following equations can be found in standard cosmology
textbooks, \citet[e.g.~]{pee80}, \citet{CL:96}, \citet{2003moco.book.....D}. 

%%%%%%%%%%%%%%%%%%%%%%%%%%%%%%%%%%%%%%%%%%%%%%%%%%%%%%%%%%%%
\subsubsection{Standard cosmological model}
\label{sec:standard_model}
%%%%%%%%%%%%%%%%%%%%%%%%%%%%%%%%%%%%%%%%%%%%%%%%%%%%%%%%%%%%

In the standard cosmological model, the field equations of General Relativity
(GR) describe the relationship between space-time geometry and the
matter-energy content of the Universe governed by gravity. A solution to these
non-linear differential equations exists representing a homogeneous and
isotropic universe.

To quantify gravitational lensing, however, we need to consider light
propagation in an inhomogeneous universe. For a general metric that describes
an expanding universe including first-order perturbations, the line element
${\rm d} s$ is given as
\begin{equation}
\doiftwocol{\fl}
  {\rm d} s^2 = \left(1 + \frac {2 \Psi}{c^2} \right) c^2 {\rm d} t^2 -
  a^2(t) \left(1 - \frac {2 \Phi}{c^2} \right)
  {\rm d} l^2,
  \label{eq:metric_gen}
\end{equation}
where the scale factor $a$ is a function of cosmic time $t$ (we set
$a$ to unity at present time $t = t_0$), and $c$ is the speed of light.
The spatial part of the metric is given by the comoving coordinate $l$, which remains constant as
the Universe expands.
The two Bardeen gravitational potentials $\Psi$
and $\Phi$ are considered to describe weak fields,
$\Psi, \Phi \ll c^2$.
The potential of a lens with mass $M$ and radius $R$ can be approximated by
$G M / R = (c^2 / 2) (R_{\rm S} / R)$, where $G$ is Newton's gravitational constant
and $R_{\rm S}$ is the Schwarzschild radius. The weak-field condition is fulfilled for
most mass distributions, excluding only those very compact objects whose extent $R$
is comparable to their Schwarzschild radius.

In GR, and in the absence of anisotropic stress which is the case on
large scales, the two potentials are equal, $\Psi = \Phi$. If the
perturbations vanish, (\ref{eq:metric_gen}) reduces to the 
Friedmann-Lema\^itre-Robertson-Walker (FLRW) metric.

The spatial line element ${\rm d} l^2$ can be separated into a radial and angular
part, ${\rm d} l^2 = {\rm d} \chi^2 + f^2_K(\chi) {\rm d} \omega$. Here, $\chi$ is the comoving
coordinate and $f_K$ is the comoving angular distance, the functional form of which
is given for the three distinct cases of three-dimensional space with curvature $K$ as
\doiftwocol{%
\end{multicols}
\par\noindent\rule{\dimexpr(0.5\textwidth-0.5\columnsep-0.4pt)}{0.4pt}%
\rule{0.4pt}{6pt}
}
\begin{equation}
  f_K(\chi)
  = \left\{ \begin{array}{llll} 
      K^{-1/2} \sin{\left( K^{1/2} \, \chi \right) } \;\;\;\;\;\;
      & \mbox{for} & K>0  & \mbox{(spherical)} \\
      \chi & \mbox{for} & K=0 & \mbox{(flat)} \\
      (-K)^{-1/2} \sinh{\left[ (-K)^{1/2} \, \chi \right]} & \mbox{for}
      & K<0 & \mbox{(hyperbolic)} \; .
    \end{array}
  \right.
\end{equation}
\doiftwocol{%
\vspace{\belowdisplayskip}\hfill\rule[-6pt]{0.4pt}{6.4pt}%
\rule{\dimexpr(0.5\textwidth-0.5\columnsep-1pt)}{0.4pt}
\begin{multicols}{2}
}
that are characterised by their corresponding equation-of-state relation
between pressure $p$ and density $\rho$, given by the parameter $w$ as
\begin{equation}
  p = w \, c^2 \rho.
  \label{eq:eos}
\end{equation}
The present-day density of each species is further scaled by the present-day
critical density of the Universe $\rho_{{\rm c}, 0} = 3 H_0^2 / (8 \pi G)$, for
which the Universe has a flat geometry. The Hubble constant $H_0 = H(a=1) =
(\dot a / a)_{t=t_0} = 100 \, h \, \mbox{km} \, \mbox{s}^{-1} \mbox{Mpc}^{-1}$
denotes the present-day value of the Hubble parameter $H$, and the parameter $h
\sim 0.7$ characterizes the uncertainty in our knowledge of $H_0$. The density
parameter of non-relativistic matter is $\Omega_{\rm m} = \rho_{\rm m, 0} /
\rho_{\rm crit, 0}$, which consists of cold dark matter (CDM), baryonic matter,
and possibly heavy neutrinos as $\Omega_{\rm m} = \Omega_{\rm c} + \Omega_{\rm
b} + \Omega_{\nu}$\footnote{Unless written as function of $a$, density
parameters are interpreted at present time; the subscript '0' is omitted.}.
Finally, the component driving the accelerated expansion (``dark energy'') is
denoted by $\Omega_{\rm de}$. Lacking a well-motivated physical model, the
dark-energy equation-of-state parameter $w$ is often parametrized by the first
or first few coefficients of a Taylor expansion, e.g.~$w(a) = w_0 + w_1 (1 -
a)$ \cite{2001IJMPD..10..213C,2003PhRvL..90i1301L}. In the case of the
cosmological constant, $\Omega_{\rm de} \equiv \Omega_\Lambda$ and $w=-1$.

The sum of all density parameters defines the \emph{curvature density parameter}
$\Omega_K$, with $\Omega_{\rm m} + \Omega_{\rm de} + \Omega_{\rm r} = 1 - \Omega_K$,
where $\Omega_K = - (c/H_0)^2 K$ has opposite sign compared to the curvature $K$.

%%%%%%%%%%%%%%%%%%%%%%%%%%%%%%%%%%%%%%%%%%%%%%%%%%%%%%%%%%%%
\subsubsection{Structure formation}
\label{sec:structure_formation}
%%%%%%%%%%%%%%%%%%%%%%%%%%%%%%%%%%%%%%%%%%%%%%%%%%%%%%%%%%%%

In an expanding universe, density fluctuations evolve with time. Tiny quantum
fluctuations in the primordial inflationary cosmos generate small-amplitude
density fluctuations. Subsequently, these fluctuations grow into
the large structures we see today, in the form of clusters, filaments, and
galaxy halos.

At early enough times or on large enough scales, those density fluctuations
are small, and their evolution can be treated using linear
perturbation theory. Once those fluctuations grow to become non-linear, other
approaches to describe them are necessary --- for example higher-order
perturbation theory, renormalization group mechanisms, analytical models of
gravitational collapse, the so-called halo model, or $N$-body simulations.

Fluctuations of the density $\rho$ around the mean density $\bar \rho$
are parametrized by the density contrast
\begin{equation}
  \delta = \frac{\rho - \bar \rho}{\bar \rho}.
  \label{eq:delta}
\end{equation}
For non-relativistic perturbations in the matter-dominated era on scales smaller than the horizon,
i.e.~the light travel distance since $t=0$,
Newtonian physics suffices to describe the evolution of $\delta$ \cite{pee80}.
The density contrast of an ideal fluid of zero pressure
is related to the gravitational potential via the Poisson
equation,
\begin{equation}
  \vec \nabla^2 \Phi = 4 \pi G a^2 \bar \rho \, \delta .
  \label{eq:Poisson}
\end{equation}
The differential equation describing the evolution of $\delta$ typically has to be solved
numerically, although in special cases analytical solutions exist. The
solution that increases with time is called \emph{growing mode}.
The time-dependent function is the \emph{linear growth
factor} $D_+$, which relates the density contrast at time $a$ to an earlier, initial epoch $a_{\rm i}$, with
$\delta(a) \propto D_+(a) \delta_(a_{\rm i})$.
In a matter-dominated Einstein-de-Sitter Universe,
$D_+$ is proportional to the scale factor $a$. The presence of dark
energy results in a suppressed growth of structures.

%%%%%%%%%%%%%%%%%%%%%%%%%%%%%%%%%%%%%%%%%%%%%%%%%%%%%%%%%%%%
\subsubsection{Modified gravity models}
\label{sec:mod_grav}
%%%%%%%%%%%%%%%%%%%%%%%%%%%%%%%%%%%%%%%%%%%%%%%%%%%%%%%%%%%%

A very general, phenomenological characterisation of deviations from
GR is to add parameters to the Poisson equation, and to
treat the two Bardeen potentials as two independent quantities. This
leads to two modified, distinct Poisson equations, which, expressed in
Fourier space, are \cite{2006astro.ph..5313U,2008JCAP...04..013A}
\begin{eqnarray}
k^2 \tilde \Psi(k, a) & = & 4 \pi G a^2 \left[ 1 + \mu(k, a) \right] \rho
\, \tilde \delta(k, a);
\label{eq:Poisson_mod_Psi} \\
k^2 \left[ \tilde \Phi(k, a) + \tilde \Psi(k, a) \right] & = & 8  \pi G a^2
\left[ 1 + \Sigma(k, a) \right] \rho \, \tilde \delta(k, a).
\label{eq:Poisson_mod_Phi_plus_Psi}
\end{eqnarray}
The tilde denotes the Fourier transform.
Non-zero values of the free functions $\mu$ and $\Sigma$ represent
deviations from GR. This flexible parametrization can account for a
variety of modified gravity models, for example a change in the
gravitational force from models with extra-dimensions as in DGP
\nocite{2000PhLB..484..112D}
(Dvali, Gabadadze \& Porrati 2000),
massive gravitons \cite{1994PhRvL..73.2950Z},
$f(R)$ extensions of the Einstein-Hilbert action
\cite{2010LRR....13....3D}, or Tensor-Vector-Scalar (TeVeS) theories
\cite{2009CQGra..26n3001S}.
Non-zero anisotropic stress is predicted from a variety of
higher-order gravity theories, but also expected from models of
clustered dark energy \cite{1998ApJ...506..485H,2011PhRvD..83b3011C}.
See \citet{2012PhR...513....1C} and \citet{2012IJMPD..2130002Y} for further models of modified gravity.

The above-introduced parametrization has the advantage of separating the effect
of the metric on non-relativistic particles (which are influenced by density
fluctuations through (\ref{eq:Poisson_mod_Psi})), and light deflection (which
is governed by both geometry and density fluctuations via
(\ref{eq:Poisson_mod_Phi_plus_Psi}), see e.g.~\citet{2001PhRvD..64h3004U},
\citet{2008PhRvD..78f3503J}). Thus, data from galaxy clustering, redshift-space
distortions, and velocity fields (testing the former relation on the one hand)
and weak-lensing observations (testing the latter equation on the other hand)
are complementary in their ability to constrain modified gravity models.

% WL basics
% lensing_formalism.tex

%%%%%%%%%%%%%%%%%%%%%%%%%%%%%%%%%%%%%%%%%%%%%%%%%%%%%%%%%%%%
\subsection{Weak cosmological lensing formalism}
\label{sec:wl_formalism}
%%%%%%%%%%%%%%%%%%%%%%%%%%%%%%%%%%%%%%%%%%%%%%%%%%%%%%%%%%%%

This section introduces the basic concepts of weak cosmological lensing, and
discusses the relevant observables and their relationships to theoretical
models of the large-scale structure. More details about those concepts can be
found in \citet[e.g. ]{BS01}.

%%%%%%%%%%%%%%%%%%%%%%%%%%%%%%%%%%%%%%%%%%%%%%%%%%%%%%%%%%%%
\subsubsection{Light deflection and the lens equation}
\label{sec:light_deflection}
%%%%%%%%%%%%%%%%%%%%%%%%%%%%%%%%%%%%%%%%%%%%%%%%%%%%%%%%%%%%

There are multiple ways to derive the equations describing the deflection of
light rays in the presence of massive bodies. An intuitive approach is the use
of Fermat's principle of minimal light travel time
\cite{1992grle.book.....S,1985A&A...143..413S,1986ApJ...310..568B}.

Photons propagate on null geodesics, given by a vanishing line
element ${\rm d} s$. In the case of GR we get the light ray travel time from
the metric (\ref{eq:metric_gen}) as
\begin{equation}
  t = \frac 1 c \int \left(1 - \frac{2 \Phi}{c^2} \right) {\rm d} r,
\end{equation}
where the integral is along the light path in physical or proper coordinates
${\rm d} r$. Analogous to geometrical optics, the potential acts as a medium
with variable refractive index $n = 1 - 2 \Phi / c^2$ (with $\Phi < 0$),
changing the direction of the light path. (This effect is what gives
gravitational \emph{lensing} its name.) We can apply Fermat's principle,
$\delta t = 0$, to get the Euler-Lagrange equations for the refractive index.
Integrating these equations along the light path results in the
\emph{deflection angle} $\vec{\hat \alpha}$ defined as the difference between
the directions of emitted and received light rays,
\begin{equation}
  \vec{\hat \alpha} = - \frac{2}{c^2} \int \vec \nabla_\perp^{\rm p} \Phi \, {\rm d} r.
  \label{eq:alpha_hat}
\end{equation}
The gradient of the potential is taken perpendicular to the light path,
with respect to physical coordinates.
The deflection angle is twice the classical prediction in Newtonian dynamics if
photons were massive particles \cite{Soldner1804}.

%%%%%%%%%%%%%%%%%%%%%%%%%%%%%%%%%%%%%%%%%%%%%%%%%%%%%%%%%%%%
\subsubsection{Light propagation in the universe}
\label{sec:light_propagation}
%%%%%%%%%%%%%%%%%%%%%%%%%%%%%%%%%%%%%%%%%%%%%%%%%%%%%%%%%%%%

In this section we quantify the relation between light deflection and
gravitational potential on cosmological scales. 
To describe differential propagation of rays within an
infinitesimally thin light bundle, we consider the difference
between two neighbouring geodesics, which is given by the
\emph{geodesic deviation equation}. In a homogeneous FLRW Universe,
the transverse comoving separation $\vec x_0$ between two light rays
as a function of comoving distance from the observer $\chi$ is proportional to the 
comoving angular distance
\begin{equation}
  \vec x_0(\chi) = f_K(\chi) \bm \theta,
  \label{eq:prop-hom}
\end{equation}
where the separation vector $\vec x_0$ is seen by the observer under the (small) angle $\btheta$
\cite{1992grle.book.....S,1994CQGra..11.2345S}.

This separation vector is modified 
by density perturbations in the Universe.
We have already seen (\ref{eq:alpha_hat}) that a light ray is deflected by an amount
${\rm d} \vec{\hat \alpha} = -2/c^2 \, \bm \nabla_\perp \Phi(\vec x, \chi^\prime) {\rm d} \chi^\prime$
in the presence of the potential $\Phi$ at distance $\chi^\prime$ from the observer.
Note that this equation is now expressed in a comoving frame, as well as the gradient.
From the vantage point of the deflector the induced change in separation vector at
source comoving distance $\chi$ is ${\rm d} \vec x = f_K(\chi - \chi^\prime) {\rm d} \vec
{\hat \alpha}$ (see Fig.~\ref{fig:propagation} for a sketch). The total separation is
obtained by integrating over the line of sight along $\chi^\prime$. Lensing
deflections modify the path of both light rays, and we denote with the superscript $^{(0)}$
the potential along the second, fiducial ray. The result is
\begin{equation}
\vec{x}(\chi) = f_K(\chi) \vec \theta - \frac{2}{c^2} \int_0^\chi {\rm d}\chi^\prime
   f_K(\chi-\chi^\prime) \left[ \vec \nabla_\perp\Phi(\vec{x}(\chi^\prime), \chi^\prime)
  - \vec \nabla_\perp\Phi^{(0)}(\chi^\prime) \right].
\label{eq:prop-int}
\end{equation}
In the absence of lensing the separation vector $\vec x$ would be seen by the
observer under an angle
$\vec \beta = \vec x(\chi) / f_K(\chi)$. The difference between the apparent
angle $\vec \theta$ and $\vec \beta$
is the total, scaled deflection angle $\vec \alpha$, defining the \emph{lens equation}
\begin{equation}
  \vec \beta = \vec \theta - \vec \alpha ,
  \label{eq:lens}
\end{equation}
with
\begin{equation}
  \vec \alpha = \frac{2}{c^2} \int_0^\chi {\rm d}\chi^\prime
  \frac{f_K(\chi-\chi^\prime)}{f_K(\chi)} \left[ \vec \nabla_\perp\Phi(\vec{x}(\chi^\prime), \chi^\prime)
  - \vec \nabla_\perp\Phi^{(0)}(\chi^\prime)
  \right].
  \label{eq:deflection_angle}
\end{equation}
\Eref{eq:lens} is analogous to the standard lens equation in the case
of a single, thin lens, in which case $\vec \beta$ is the source
position.

\stoptwocol
\begin{figure}
  \begin{center}
    \resizebox{0.8\hsize}{!}{
      \includegraphics{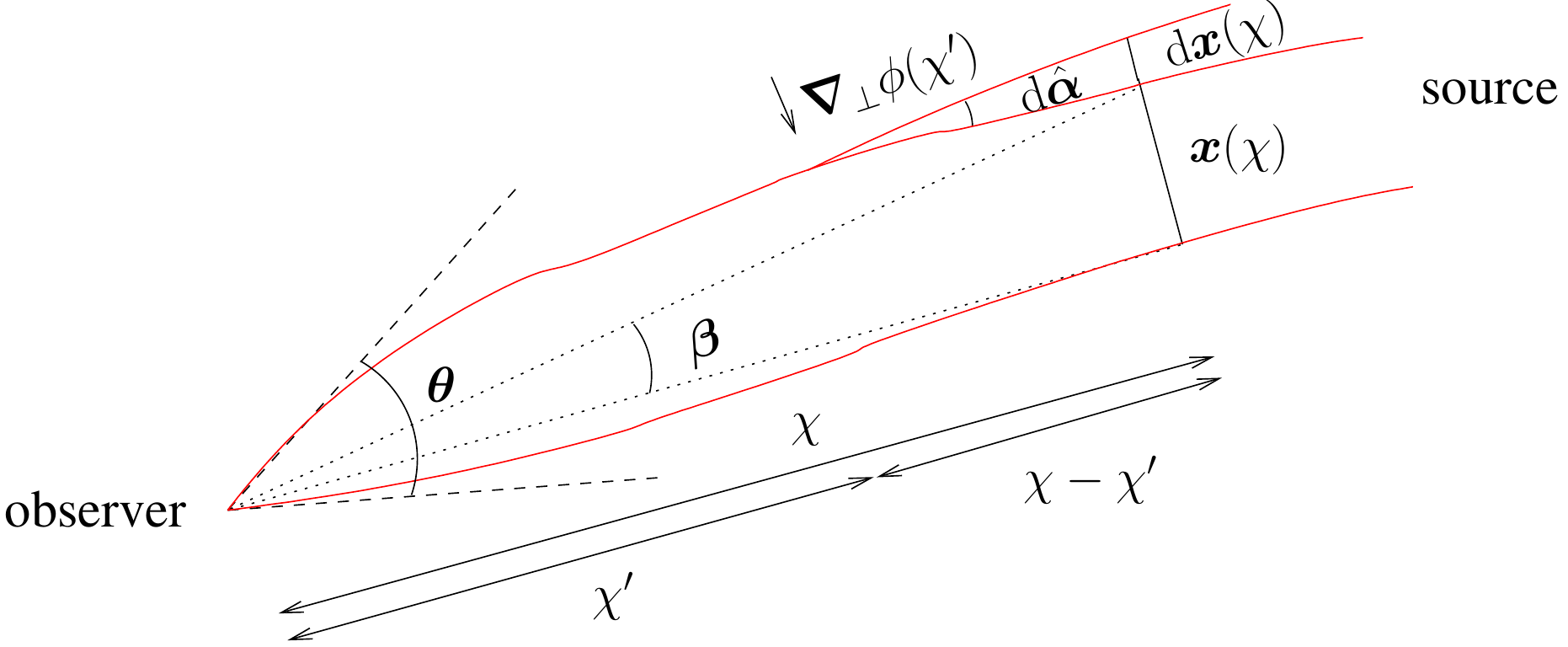}
    }
  \end{center}

  \caption{Propagation of two light rays (red solid lines), converging
    on the observer on the left. The light rays are separated by the transverse comoving distance 
    $\vec x$, which varies with distance $\chi$ from the observer. An exemplary deflector at distance $\chi^\prime$
    perturbes the geodescics proportional to the transverse gradient $\vec \nabla_\perp \phi$
    of the potential.
    The dashed lines indicate the apparent direction of the light rays,
    converging on the observer under the angle $\vec \theta$. The dotted lines show the unperturbed
    geodesics, defining the angle $\vec \beta$ under which the unperturbed transverse comoving separation $\vec x$
    is seen.}

  \label{fig:propagation}

\end{figure}
\begtwocol

%%%%%%%%%%%%%%%%%%%%%%%%%%%%%%%%%%%%%%%%%%%%%%%%%%%%%%%%%%%%
\subsubsection{Linearized lensing quantities}
\label{sec:linear_lensing}
%%%%%%%%%%%%%%%%%%%%%%%%%%%%%%%%%%%%%%%%%%%%%%%%%%%%%%%%%%%%

The integral equation (\ref{eq:prop-int}) can be approximated by substituting
the separation vector
$\vec x$ in the integral by the
$0^{\rm th}$-order solution $\vec x_0(\chi) = f_K(\chi) \vec \theta$ (\ref{eq:prop-hom}).
This corresponds to integrating the potential gradient along the
unperturbed ray, which is called the \emph{Born approximation} (see Sect.~\ref{sec:corrections}
for higher-order corrections). 
Further, we linearise the lens equation (\ref{eq:lens}) and define the (inverse) amplification matrix as the Jacobian $\mat
A = \partial \vec \beta / \partial \vec \theta$, which describes a
linear mapping from lensed (image) coordinates $\vec \theta$ to
unlensed (source) coordinates $\vec \beta$,
\begin{eqnarray}
  A_{ij} & = & \frac{\partial \beta_i}{\partial \theta_j} = \delta_{ij} -
  \frac{\partial \alpha_i}{\partial \theta_j}
  \nonumber \\
  & = & \delta_{ij} - \frac{2}{c^2} \int_0^\chi {\rm d}\chi^\prime
  \frac{f_K(\chi-\chi^\prime) f_K(\chi^\prime)}{f_K(\chi)} \frac{\partial^2}{\partial x_i
\partial x_j} \Phi(f_K(\chi^\prime) \vec \theta, \chi^\prime) .
  \label{eq:Jacobi}
\end{eqnarray}
The second term in (\ref{eq:deflection_angle}) drops out since it does not
depend on the
angle $\vec \theta$.

In this approximations the deflection angle can be written as the gradient of a
2D potential, the \emph{lensing potential} $\psi$,
\begin{equation}
  \psi(\vec \theta, \chi) = \frac 2 {c^2} \int_0^\chi {\rm d} \chi^\prime
    \frac{f_K(\chi - \chi^\prime)}{f_K(\chi) f_K(\chi^\prime)}\,
  \Phi(f_K(\chi^\prime) \vec \theta, \chi^\prime) .
  \label{eq:lensing_potential}
\end{equation}
With this definition, the Jacobi matrix can be expressed as
\begin{equation}
  A_{ij} = \delta_{ij} - \partial_i \partial_j \psi, 
  \label{eq:Jacobi_psi}
\end{equation}
where the partial derivatives are understood with respect to $\vec \theta$.
The symmetrical matrix ${\mat A}$ is parametrized in terms of the scalar
\emph{convergence}, $\kappa$, and the two-component spin-two
\emph{shear}, $\gamma = (\gamma_1, \gamma_2)$, as
\begin{equation}
  \mat A = \left( \begin{array}{cc} 1 - \kappa - \gamma_1 & - \gamma_2
    \\ - \gamma_2 & 1 - \kappa + \gamma_1 \end{array} \right).
\label{eq:jacobi}
\end{equation}
This defines the convergence and shear as second derivatives of the potential,
\begin{equation}
\kappa = \frac 1 2 \left( \partial_1 \partial_1 + \partial_2 \partial_2 \right) \psi
         = \frac 1 2 \nabla^2 \psi; \;\;;
  \gamma_1 = \frac 1 2 \left( \partial_1 \partial_1 - \partial_2 \partial_2 \right) \psi;
  \;\;
  \gamma_2 = \partial_1 \partial_2 \psi.
  \label{eq:kappa_gamma_psi}
\end{equation}
The inverse Jacobian $\mat A^{-1}$ describes the local mapping of the source
light distribution to image coordinates. The convergence, being the diagonal
part of the matrix, is an isotropic increase or decrease of the observed size
of a source image. Shear, the trace-free part, quantifies an anisotropic
stretching, turning a circular into an elliptical light distribution.

It is mathematically convenient to write the shear as complex number, $\gamma =
\gamma_1 + \rm i \gamma_2 = |\gamma| \exp(2 \rm i \varphi)$, with $\varphi$
being the polar angle between the two shear components. Shear transforms as a
spin-two quantity: a rotation about $\pi$ is the identity transformation of an
ellipse (see Fig.~\ref{fig:wheel} for an illustration).

\begin{figure}

  \begin{minipage}{0.45\textwidth}
  \begin{center}
   \resizebox{\hsize}{!}{
	  \includegraphics{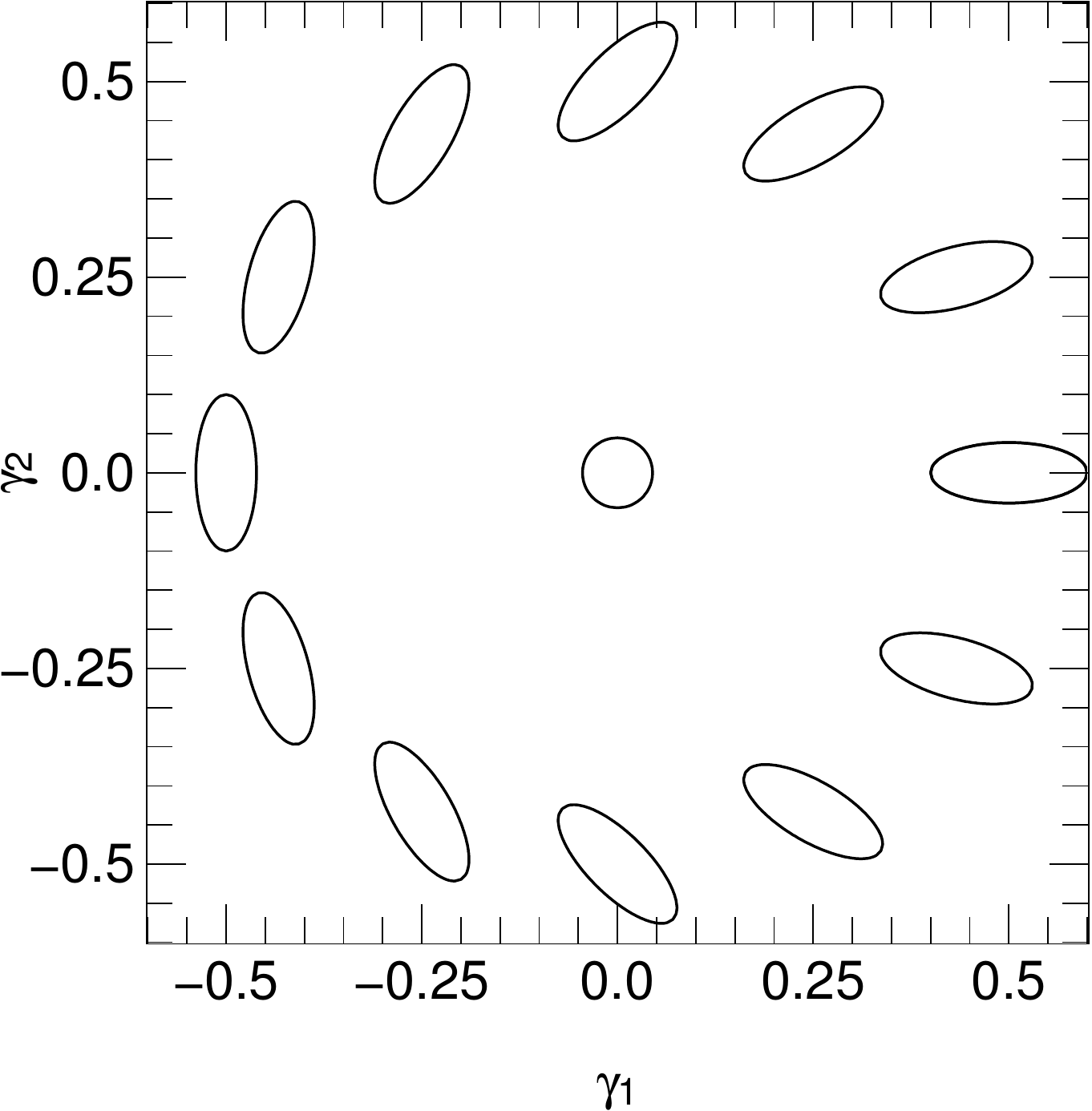}
   }
  \end{center}
  \end{minipage}%
  \hspace*{-0.1\textwidth}%
  \begin{minipage}{0.65\textwidth}

   \caption{The orientation of the ellipses given by the Cartesian
      coordinates ${\gamma_1}$ and ${\gamma_2}$ of the
      shear. While the polar angle $\varphi$ passes through the range
      $[0;\, 2\pi]$, the shear ellipse rotates around
      $\pi$.}
   \label{fig:wheel}

    \end{minipage}%

\end{figure}

In the context of cosmological lensing by large-scale structures,
images are very weakly lensed, and the values of $\kappa$ and $\gamma$
are on the order of a few percent or less. Each source is mapped uniquely onto one image,
there are no multiple images, and the matrix $\mat A$ is indeed invertible.

We can factor out $(1-\kappa)$ from $\mat A$ (\ref{eq:jacobi}), since this multiplier
only affects the size but not the shape of the source. Cosmic
shear is based on the measurement of galaxy shapes (see Sect.~\ref{sec:shapes}), and
therefore the observable in question is not the shear $\gamma$ but the
\emph{reduced shear},
\begin{equation}
  g = \frac{\gamma}{1 - \kappa},
  \label{eq:reduced_shear}
\end{equation}
which has the same spin-two transformation properties as shear.
Weak lensing is the regime where the effect of gravitational lensing is very small,
with both the convergence and the shear much smaller than unity.
%$\kappa \ll 1$
Therefore, shear is a good approximation of
reduced shear to linear order (see Sect.~\ref{sec:corrections} for
its validity).

%%%%%%%%%%%%%%%%%%%%%%%%%%%%%%%%%%%%%%%%%%%%%%%%%%%%%%%%%%%%
\subsubsection{Projected overdensity}
\label{sec:projected_overdensity}
%%%%%%%%%%%%%%%%%%%%%%%%%%%%%%%%%%%%%%%%%%%%%%%%%%%%%%%%%%%%

Since the convergence $\kappa$ is related to the lensing potential $\psi$
(\ref{eq:lensing_potential}) via a
2D Poisson equation (\ref{eq:kappa_gamma_psi}), it can be interpreted
as a (projected) surface density. To introduce the 3D density contrast $\delta$,
we apply the 2D Laplacian
of the lensing potential (\ref{eq:lensing_potential}) to the 3D potential $\Phi$
and add the second-order
deriviate along the comoving coordinate, $\partial^2 / \partial \chi^2$.
This additional term vanishes, since positive and negative contributions cancel out
to a good approximation when integrating along the line of sight.
Next, we replace the
3D Laplacian of $\Phi$ with the over-density $\delta$ using the Poisson
equation (\ref{eq:Poisson}), and $\bar \rho \propto a^{-3}$. Writing the mean matter
density in terms of 
the critical density, we get
\begin{equation}
  \kappa(\vec \theta, \chi) = \frac{3 H_0^2 \Omega_{\rm m}}{2 c^2}
  \int\limits_0^\chi \frac{{\rm d} \chi^\prime}{a(\chi^\prime)} 
  \frac{f_K(\chi - \chi^\prime)}{f_K(\chi)} f_K(\chi^\prime) \,
\delta(f_K(\chi^\prime) \vec \theta, \chi^\prime).
  \label{eq:kappa_chi}
\end{equation}
This expression is a projection of the density along comoving coordinates,
weighted by geometrical factors involving
the distances between source, deflector, and observer. In the case of a flat universe,
the geometrical weight $(\chi - \chi^\prime) \chi^\prime$ is a parabola with maximum at
$\chi^\prime = \chi/2$.
Thus, structures at around half the distance to the source are most efficient to generate
lensing distortions.

The mean convergence from a population of source galaxies is obtained by weighting the above
expression with the galaxy probability distribution in comoving distance, $n(\chi) {\rm d} \chi$,
\begin{equation}
  \kappa(\vec \theta) = \int\limits_0^{\chi_{\rm lim}} {\rm d} \chi \, n(\chi) \,
  \kappa(\vec \theta, \chi).
  \label{eq:kappa}
\end{equation}
The integral extends out to the limiting comoving distance $\chi_{\rm lim}$ of
the galaxy sample. Inserting (\ref{eq:kappa_chi}) into (\ref{eq:kappa}) and
interchanging the integral order results
in the following expression, 
\begin{equation}
  \kappa(\vec \theta) = \frac{3 H_0^2 \Omega_{\rm m}}{2 c^2} \int\limits_0^{\chi_{\rm lim}}
  \frac{{\rm d} \chi}{a(\chi)}
  q(\chi) f_K(\chi) \, \delta(f_K(\chi) \vec \theta, \chi).
  \label{eq:kappa_final}
\end{equation}
The lens efficiency $q$ is defined as
\begin{equation}
  q(\chi) = \int\limits_\chi^{\chi_{\rm lim}} {\rm d} \chi^\prime \, n(\chi^\prime)
  \frac{f_K(\chi^\prime - \chi)}{f_K(\chi^\prime)},
  \label{eq:lens_efficiency}
\end{equation}
and indicates the lensing strength at a distance $\chi$ of the combined
background galaxy distribution. Thus, the convergence is a linear measure of
the total matter density, projected along the line of sight with dependences on
the geometry of the universe via the distance ratios, and the source galaxy
distribution $n(\chi) {\rm d} \chi = n(z) {\rm d} z$. The latter is usually
obtained using photometric redshifts (Sect.~\ref{sec:photo-z}). We will see in
Sect.~\ref{sec:tomography} how to recover information in the redshift
direction.

By construction, the expectation value of shear and convergence are zero, since
$\langle \delta \rangle = 0$. The first non-trivial statistical measure of the
distribution of $\kappa$ and $\gamma$ are second moments. Practical estimators
of weak-lensing second-order statistics in real and Fourier-space are discussed
in Sects.~\ref{sec:power_spectrum} and \ref{sec:real_space_2nd}.

%%%%%%%%%%%%%%%%%%%%%%%%%%%%%%%%%%%%%%%%%%%%%%%%%%%%%%%%%%%%
\subsubsection{Estimating shear from galaxies}
\label{sec:estim_shear}
%%%%%%%%%%%%%%%%%%%%%%%%%%%%%%%%%%%%%%%%%%%%%%%%%%%%%%%%%%%%

In the case of cosmic shear, not the convergence but the shear is measured from
the observed galaxy shapes, as discussed in this section. Theoretical
predictions of the convergence (\ref{eq:kappa_final}) can be related to the
observed shear using the relations (\ref{eq:kappa_gamma_psi}). Further, a
convergence field can be estimated by reconstruction from the observed galaxy
shapes. 

We can attribute an intrinsic, complex \emph{source ellipticity}
$\varepsilon^{\rm s}$
to a galaxy. Cosmic shear modifies this ellipticity as a function of the complex reduced shear,
which depends on the definition of $\varepsilon^{\rm s}$. If we define this quantity for an image
with elliptical isophotes, minor-to-major axis ratio $b/a$, and position angle $\phi$, as
$\varepsilon = (a - b)/(a + b) \times \exp(2 {\rm i} \phi)$,
the observed ellipticity $\varepsilon$ (for $|g| \le 1)$ is given as \cite{1997A&A...318..687S}
\begin{equation}
  \varepsilon = \frac{\varepsilon^{\rm s} + g}{1 + g^\ast\varepsilon^{\rm s}}.
  \label{eq:eps_g}
\end{equation}
The asterisk ``$^\ast$'' denotes complex conjugation.
In the weak-lensing regime, this relation is approximated to
\begin{equation}
 \varepsilon \approx \varepsilon^{\rm s} + \gamma .
  \label{eq:eps_eps_s_gamma}
\end{equation}

If the intrinsic ellipticity of galaxies has no preferred orientation, the
expectation value of $\varepsilon^{\rm s}$ vanishes, $\langle \varepsilon^{\rm
s} \rangle = 0$, and the observed ellipticity is an unbiased estimator of the
reduced shear,
\begin{equation}
\left\langle \varepsilon \right\rangle = g .
\label{eq:estimator_shear}
\end{equation}
This relation breaks down in the presence of intrinsic galaxy alignments (Sect.~\ref{sec:ia}).

Another commonly used ellipticity estimator has been proposed by
\cite{1995A&A...294..411S}. This estimator has a slightly simpler dependence on
second moments of galaxy images, which have been widely used for shape estimation,
see Sect.~\ref{sec:shapes}. However, it it does not provide an unbiased
estimator of $g$, but explicitly depends on the intrinsic ellipticity
distribution.

In the weak-lensing regime, the shear cannot be detected from an individual
galaxy. With distortions induced by the LSS of the order $\gamma \sim 0.03$, and the
typical intrinsic ellipticity rms of $\sigma_\varepsilon = \langle |
\varepsilon |^2 \rangle^{1/2} \sim 0.3$, one needs to average over a number of
galaxies $N$ of at least a few hundred to obtain a signal-to-noise ratio $S/N =
\gamma \times N^{1/2} / \sigma_\varepsilon$ of above unity. 

%%%%%%%%%%%%%%%%%%%%%%%%%%%%%%%%%%%%%%%%%%%%%%%%%%%%%%%%%%%%
\subsubsection{E- and B-modes}
\label{sec:E_B_modes}
%%%%%%%%%%%%%%%%%%%%%%%%%%%%%%%%%%%%%%%%%%%%%%%%%%%%%%%%%%%%

The Born approximation introduced in Sect.~\ref{sec:linear_lensing} results in the
definition of the convergence and shear to be functions of a single scalar
potential (\ref{eq:lensing_potential}). The two shear components
defined in that way (\ref{eq:kappa_gamma_psi}) are not independent, and the
shear field cannot have an arbitrary form. We can define a vector field $\vec u$
as the gradient of the ``potential'' $\kappa$, $\vec u = \vec \nabla \kappa$.
By definition, the curl of this gradient vanishes, $\vec \nabla \times \vec u =
\partial_1 u_2 - \partial_2 u_1 = 0$.
Inserting the relations between $\kappa, \gamma$ and $\psi$ (\ref{eq:kappa_gamma_psi})
into this equality results in second-derivative constraints for $\gamma$.
A shear field fulfilling those relations is called an \emph{E-mode} field,
analogous to the electric field. In real life however, $\vec u$ obtained from observed data
is in general not a pure gradient field but has a non-vanishing curl component.
The corresponding convergence field can be decomposed into its E-mode
component, $\kappa^{\rm E}$, and B-mode, $\kappa^{\rm B}$, given by
$\nabla^2 \kappa^{\rm E} = \vec \nabla u$ and $\nabla^2 \kappa^{\rm B} = \vec \nabla \times u$..
The \emph{B-mode} component can have various origins:
\begin{enumerate}
\item Higher-order terms in the light-propagation equation (\ref{eq:prop-int}),
      e.g.~lens-lens coupling and integration
      along the perturbed light path
      (\ref{eq:jacobi}) \cite{2010A&A...523A..28K}.
\item Other higher-order terms beyond usual approximations of relations such as between shear and reduced shear,
      or between shear and certain ellipticity estimators (see Sect.~\ref{sec:shapes})
       \cite{2010A&A...523A..28K}.
\item Lens galaxy selection biases, such as size and magnitude bias \cite{2003ApJ...583...58W,2009ApJ...702..593S},
      or clustering of lensing galaxies \cite{1998A&A...338..375B,2002A&A...389..729S}.
\item Correlations of the intrinsic shapes of galaxies with each other, and with the structures
    that induce weak-lensing
    distortions (intrinsic alignment, Sect.~\ref{sec:ia}) \cite{2002ApJ...568...20C}.
\item Image and data analysis errors such as PSF correction residuals, systematics in the astrometry.
\end{enumerate}

The astrophysical effects (i) - (iv) cause a B-mode at the percent-level
compared to the E-mode. The intrinsic alignment B-mode amplitude is the least
well-known since the model uncertainty is large \cite{2013MNRAS.435..194C}. Up
to now, cosmic shear surveys do not have the statistical power to reliably
detect those B-modes. Until recently, the amplitude of a B-mode detection has
exclusively been used to assess the quality of the data analysis, assuming that
(v) is the only measurable B-mode contributor. While this is a valid approach,
it only captures those systematics that create a B-mode. A B-mode non-detection
might render an observer over-confident to believe that also the E-mode is
uncontaminated by systematics. Further, the ratio of B- to E-mode should not be
used to judge the data quality, since this ratio is not cosmology-independent
and can bias the cosmological inference of the data.

Some of my past work focused on studying and developing estimators that separate E- from B-mode
in shear data. This will be presented in Sect.~\ref{sec:other_2nd_order}.

%%%%%%%%%%%%%%%%%%%%%%%%%%%%%%%%%%%%%%%%%%%%%%%%%%%%%%%%%%%%
\subsubsection{The lensing power spectrum}
\label{sec:power_spectrum}
%%%%%%%%%%%%%%%%%%%%%%%%%%%%%%%%%%%%%%%%%%%%%%%%%%%%%%%%%%%%

The basic second-order function of the convergence (\ref{eq:kappa_final}) is
the two-point correlation function (2PCF) $\langle \kappa(\vec \vartheta)
\kappa(\vec \vartheta + \vec \theta) \rangle$. The brackets denote ensemble
average, which can be replaced by a spatial average over angular positions
$\vec \vartheta$. With the assumption that the density field $\delta$ on large
scales is statistically homogeneous and isotropic, which follows from the
cosmological principle, the same holds for the convergence. The 2PCF is then
invariant under translation and rotation, and therefore a function of only the
modulus of the separation vector between the two lines of sight $\theta$.
Expressed in Fourier space, the two-point correlation function defines the
flat-sky convergence power spectrum $P_\kappa$ with
\begin{equation}
  \left \langle \fourier \kappa(\vec \ell) \fourier \kappa^\ast(\vec \ell^\prime) \right \rangle
  = (2\pi)^2 \delta_{\rm D}(\vec \ell - \vec \ell^\prime) P_\kappa(\ell).
  \label{eq:p_kappa_def}
\end{equation}

Here, $\delta_{\rm D}$ is the Dirac delta function. The
complex Fourier transform $\fourier \kappa$ of the convergence is a function
of the 2D wave vector $\vec \ell$, the Fourier-conjugate of $\vec \theta$.
Again due to statistical homogeneity and isotropy, the power spectrum only
depends on the modulus $\ell$. For simplicity, we ignore the curvature of the
sky in this expression. For lensing on very large scales, and for 3D lensing
(Sect.~\ref{sec:tomography}), the curvature has to be accounted for by more
accurate expressions \cite{2008PhRvD..78l3506L}, or by applying spherical
harmonics instead of Fourier transforms.

If the convergence field is decomposed into an E-mode $\kappa^{\rm E}$ and B-mode component
$\kappa^{\rm B}$, two expressions analogous to (\ref{eq:p_kappa_def}) define the E- and B-mode power spectra,
$P_\kappa^{\rm E}$ and $P_\kappa^{\rm B}$.

Taking the square of (\ref{eq:kappa_final}) in Fourier space, we get the power spectrum
of the density contrast, $P_\delta$, on the right-hand side of the equation. Inserting the 
result into (\ref{eq:p_kappa_def}) we obtain
the flat-sky convergence power spectrum in terms of the density power spectrum as
\begin{equation}
  P_\kappa(\ell) = \frac 9 4 \, \Omega_{\rm m}^2 \left( \frac{H_0}{c} \right)^4
  \int_0^{\chi_{\rm lim}} {\rm d} \chi \,
  \frac{q^2(\chi)}{a^2(\chi)} P_\delta\left(k = \frac{\ell}{f_K(\chi)}, \chi \right).
  \label{eq:p_kappa_limber}
\end{equation}
This simple result can be derived using a few approximations:
the Limber projection is applied, which only collects modes that lie
in the plane of the sky, thereby neglecting correlations along the line of
sight
\cite{1953ApJ...117..134L,1992ApJ...388..272K,2007A&A...473..711S,2012MNRAS.422.2854G}.
In addition, the small-angle approximation (expanding to first order
trigonometric functions of the angle) and the flat-sky limit (replacing
spherical harmonics by Fourier transforms) are used. A further assumption 
is the absence of galaxy clustering, therefore ignoring
source-source \cite{2002A&A...389..729S}, and source-lens
\cite{1998A&A...338..375B,H02} clustering. Theoretical predictions for the
power spectrum are shown in Fig.~\ref{fig:Pkappa_dPkappa_dp}, using linear
theory, and the non-linear fitting formulae of \citet{2012ApJ...761..152T}. See
Sect.~\ref{sec:tomography} for the definition of the tomographic redshift bins.

The projection (\ref{eq:p_kappa_limber}) mixes different 3D $k$-modes into 2D
$\ell$ wavemodes along the line-of-sight integration, thereby washing out many
features present in the 3D density power spectrum. For example, baryonic
acoustic oscillations are smeared out and are not seen in the lensing spectrum
\cite{2006ApJ...647L..91S,2009NewA...14..507Z}. This reduces the sensitivity of
$P_\kappa$ with respect to cosmological parameters, for example compared to the
CMB anisotropy power spectrum. Examples for some parameters are shown in
Fig.~\ref{fig:Pkappa_dPkappa_dp}. Then two main response modes of $P_\kappa$
for changing parameters are an amplitude change, caused by $\sigma_8$,
$\Omega_{\rm m}$, and $w_0$, and a tilt, generated by $n_{\rm s}$, and $h$
(and, consequently, shifts are seen when varying the physical density
parameters $\omega_{\rm m}$ and $\omega_{\rm b}$). The parameter combination
that $P_\kappa$ is most sensitive to is $\sigma_8 \Omega_{\rm m}^\alpha$, with
$\alpha \approx 0.75$ in the linear regime \cite{1997A&A...322....1B}.

\stoptwocol
\begin{figure}

  \begin{center}
  \hspace*{1em}
    \resizebox{\hsize}{!}{
      \includegraphics[bb=40 10 500 232]{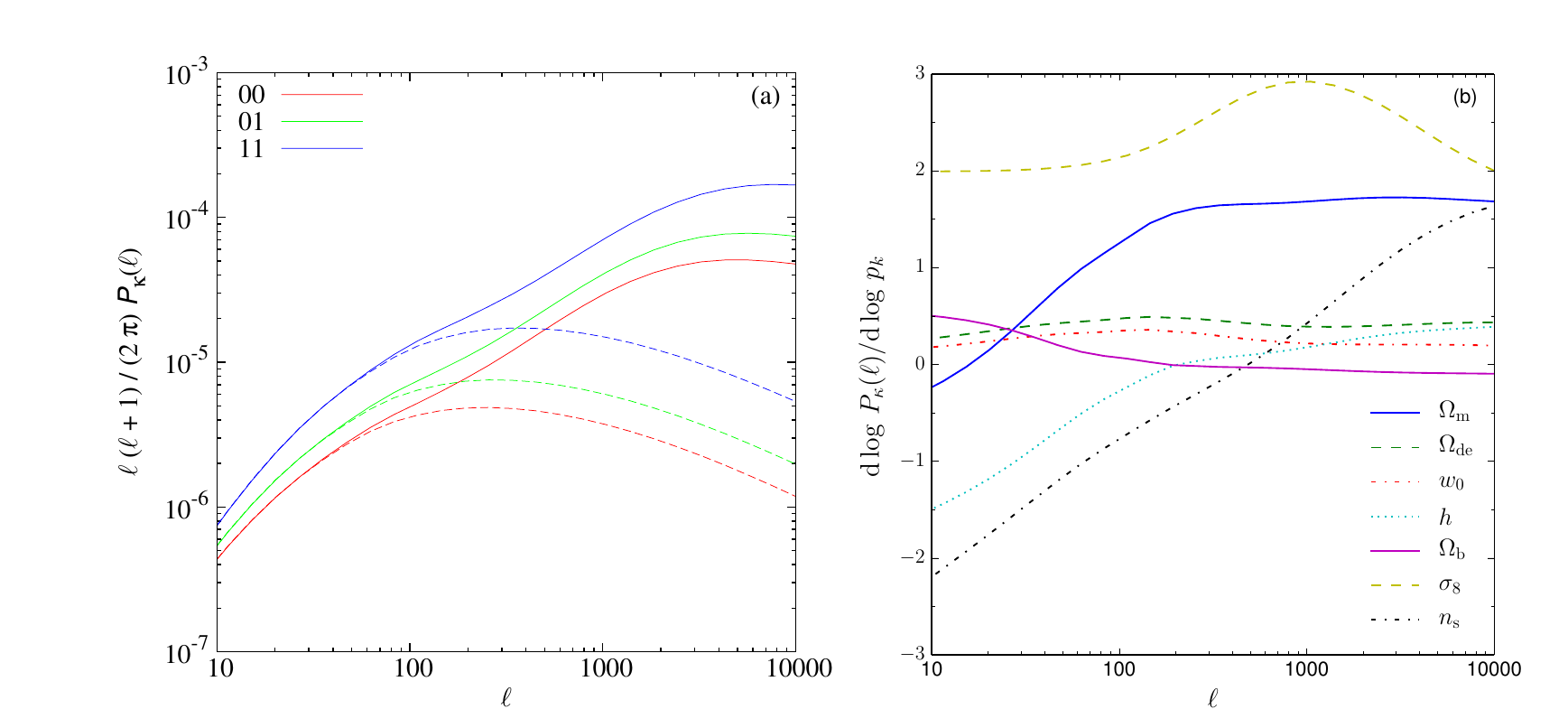}
    }%
  \end{center}

  \caption{(a) The scaled tomographic convergence auto- and cross-power
spectrum $\ell (\ell+1)/(2\pi) P_{\kappa, ij}(\ell)$ for two redshift bins $(i, j) = (0, 1)$
  with redshift ranges $z_0 = [0.5; 0.7]$, and $z_1 = [0.9; 1.1]$ for a Euclid-like source redshift distribution.
  Solid (dashed)
  lines correspond to the non-linear (linear) model.
  (b) Derivatives ${\rm d} \log P_\kappa / {\rm d} \log p_k$ of the convergence
power spectrum with respect to various cosmological parameters $p_k$, as
indicated in the figure. The corresponding redshift
 bin is $[0.9; 1.1]$.
          }

  \label{fig:Pkappa_dPkappa_dp}

\end{figure}
\begtwocol

Writing the relations between $\kappa$, $\gamma$ and the lensing potential $\psi$
(\ref{eq:kappa_gamma_psi}) in Fourier space, and using complex notation for the shear,
one finds for $\ell \ne 0$
\begin{equation}
  \fourier \gamma(\vec \ell) = \frac{ \left(\ell_1 + {\rm i} \, \ell_2\right)^2}{\ell^2}
  \fourier \kappa(\vec \ell)
  = {\rm e}^{2{\rm i}\beta} \fourier \kappa(\vec \ell),
  \label{eq:gamma_kappa_Fourier}
\end{equation}
with $\beta$ being the polar angle of the wave-vector $\vec \ell = (\ell_1,
\ell_2)$, written as complex quantity. % $\vec \ell = \ell_1 + {\rm i} \, \ell_2$.
Therefore, we get the very useful fact that the power spectrum of the
shear equals the one of the convergence, $P_\gamma = P_\kappa$.

The shear power spectrum can in principle be obtained directly from observed
ellipticities \citeaffixed{2001ApJ...554...67H}{e.g.}, or via pixellised
convergence maps in Fourier space that have been reconstructed from the
observed ellipticities, e.g. \citet{1998ApJ...506...64S}. However, the simplest
and most robust way to estimate second-order shear correlations are in real
space, which we will discuss in the following section.

%%%%%%%%%%%%%%%%%%%%%%%%%%%%%%%%%%%%%%%%%%%%%%%%%%%%%%%%%%%%
\subsubsection{Shear tomography}
\label{sec:tomography}
%%%%%%%%%%%%%%%%%%%%%%%%%%%%%%%%%%%%%%%%%%%%%%%%%%%%%%%%%%%%

The redshift distribution of source galaxies determines the redshift range over
which the density contrast is projected onto the 2D convergence and shear. By
separating source galaxies according to their redshift, we obtain lensing
fields with different redshift weighting via the lens efficiency
(\ref{eq:lens_efficiency}), thus probing different epochs in the history of the
Universe with different weights. Despite the two-dimensional aspect of
gravitational lensing, this allows us to recover a 3D \emph{tomographic} view
of the large-scale structure In particular, it helps us to measure subtle
effects that are projected out in 2D lensing, such as the growth of structures,
or a time-varying dark-energy state parameter $w(z)$.

If we denote the redshift distribution in each of $N_z$ bin with $p_i, i = 1
\ldots N_z$, we obtain a new lensing efficiency $q_i$
(\ref{eq:lens_efficiency}) for each case, and a resulting projected overdensity
$\kappa_i$. This leads to $N_z (N_z - 1) / 2$ convergence power spectra
$P_{\kappa, ij}, 1 \le i \le j \le N_z$, including not only the auto-spectra
($i=j$) but also the cross-spectra ($i \ne j$). In (\ref{eq:p_kappa_limber}),
$q^2$ is replaced by the product $q_i q_j$
\cite{1998ApJ...506...64S,1999ApJ...522L..21H}.

%%%%%%%%%%%%%%%%%%%%%%%%%%%%%%%%%%%%%%%%%%%%%%%%%%%%%%%%%%%%
\subsubsection{Intrinsic alignment}
\label{sec:ia}
%%%%%%%%%%%%%%%%%%%%%%%%%%%%%%%%%%%%%%%%%%%%%%%%%%%%%%%%%%%%

Shapes of galaxies can be correlated in the absence of gravitational lensing,
due to gravitational interactions between galaxies and the surrounding tidal
fields. The \emph{intrinsic alignment} (IA) of galaxy shapes adds an excess
correlation to the cosmic shear signal that, if not taken into account
properly, can bias cosmological inferences by tens of per cent. IA is difficult
to account for, since it cannot simply be removed by a sophisticated galaxy
selection, nor can it be easily predicted theoretically since it depends on
details of galaxy formation.

Due to IA, the intrinsic ellipticity of galaxies $\varepsilon^{\rm s}$ no
longer has a random orientation, or phase. This directly contributes to the
measured two-point shear correlation function (\ref{eq:estim_xi_pm}), as
follows. The first term in (\ref{eq:eps_eps_four_terms}) describes the
correlation of intrinsic ellipticities of two galaxies $i$ and $j$. This term
($II$, or shape-shape correlation) is non-zero only for physically close
galaxies. Its contribution to cosmic shear ($GG$, or shear-shear correlation),
the last term in (\ref{eq:eps_eps_four_terms}), can be suppressed by
down-weighting or omitting entirely galaxy pairs at the same redshift
\cite{HH03,KS02,KS03}.

The second and third term in (\ref{eq:eps_eps_four_terms}) correspond to the
correlation between the intrinsic ellipcitiy of one galaxy with the shear of
another galaxy. For either of these terms ($GI$, or shape-shear correlation) to
be non-zero, the foreground galaxy ellipticity has to be correlated via IA to
structures that shear a background galaxy. A lensing mass distribution causes
background galaxies to be aligned tangentially. Foreground galaxies at the same
redshift as the mass distribution are strechted radially towards the mass by
tidal forces. Therefore the ellipticities of background and foreground galaxies
tend to be orthogonal, corresponding to a negative $GI$ correlation. For
typical cosmic shear surveys with not too small redshift bins, $GI$ dominates
over $II$. Overall, the intrinsic alignment of galaxy orientations contribute
to the lensing power spectrum typically to up to 10\%.

%%%%%%%%%%%%%%%%%%%%%%%%%%%%%%%%%%%%%%%%%%%%%%%%%%%%%%%%%%%%
\subsubsection{Higher-order corrections}
\label{sec:corrections}
%%%%%%%%%%%%%%%%%%%%%%%%%%%%%%%%%%%%%%%%%%%%%%%%%%%%%%%%%%%%

The approximations made in Sects.~\ref{sec:linear_lensing} and
\ref{sec:power_spectrum}, resulting in the convergence power spectrum, have to
be tested for their validity. Corrections to the linearised propagation
equation (\ref{eq:jacobi}) include couplings between lens structures at
different redshift (lens-lens coupling), and integration along the perturbed
ray (additional terms to the Born approximation). Further, higher-order
correlations of the convergence take account of the reduced shear as
observable. Similar terms arise from the fact that the observed size and
magnitudes of lensing galaxies are correlated with the foreground convergence
field \citeaffixed{2001MNRAS.326..326H,2009PhRvL.103e1301S}{magnification and
size bias; }. Over the relevant scale range ($\ell \le 10^4$) most of those
effects are at least two orders of magnitude smaller than the first-order
E-mode convergence power spectrum, and create a B-mode spectrum of similar low
amplitude. The largest contribution is the reduced-shear correction, which
attains nearly $10\%$ of the shear power spectrum on arc minute scales
\cite{1997A&A...322....1B,1998MNRAS.296..873S,2006PhRvD..73b3009D,2010A&A...523A..28K}.
In \citet{K10} I present simple fitting formulae that provide the reduced-shear
power spectrum to $2\%$ accuracy for $\ell < 2 \times 10^5$ for $\Lambda$CDM
cosmological parameters within the WMAP7 68\% error ellipsoid.

Thanks to the broad lensing kernel, the Limber approximation is very precise
and deviates from the full integration only on very large scales, for $\ell <
20$ \cite{2012MNRAS.422.2854G,2012PhRvD..86b3001B}.
Similary, the flat-sky approximation for the lensing power spectrum
(\ref{eq:p_kappa_limber}) and correlation function (\ref{eq:xi_pm_pkappa})
provides sub-percent level accuracy on all but the very largest scales
\cite{KH17,2016arXiv161104954K,2017JCAP...05..014L}. The full GR treatment of
fluctuations together with dropping the small-angle approximation was also
found to make a difference only on very large scales
\cite{2010PhRvD..81h3002B}. In \citet{KH17} I showed that using the Limber and
flat-sky approximations, current cosmological results are unaffected. I
developed the second-order Limber approximation for cosmic shear, and
demonstrated that this will also sufficient for future surveys, since the
corresponding errors are sub-dominant compared to cosmic variance on all
scales, see Fig.~\ref{fig:Cl_cases}.

\begin{figure}

  \begin{center}
    \resizebox{1.0\hsize}{!}{
      \includegraphics{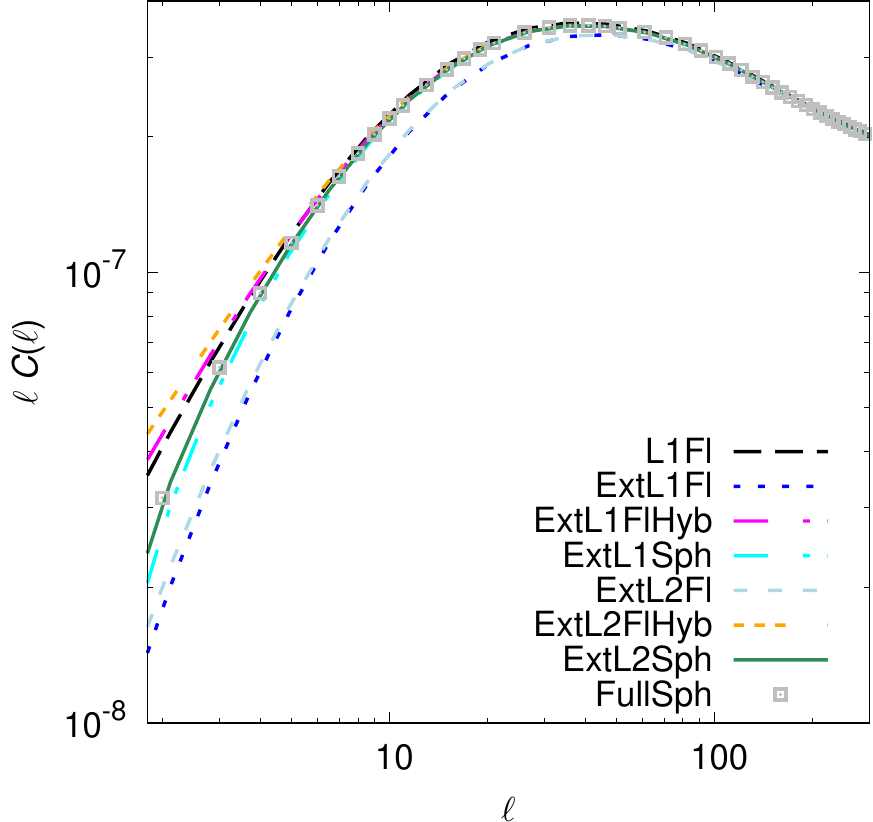}
      \hspace*{2em}
      \includegraphics{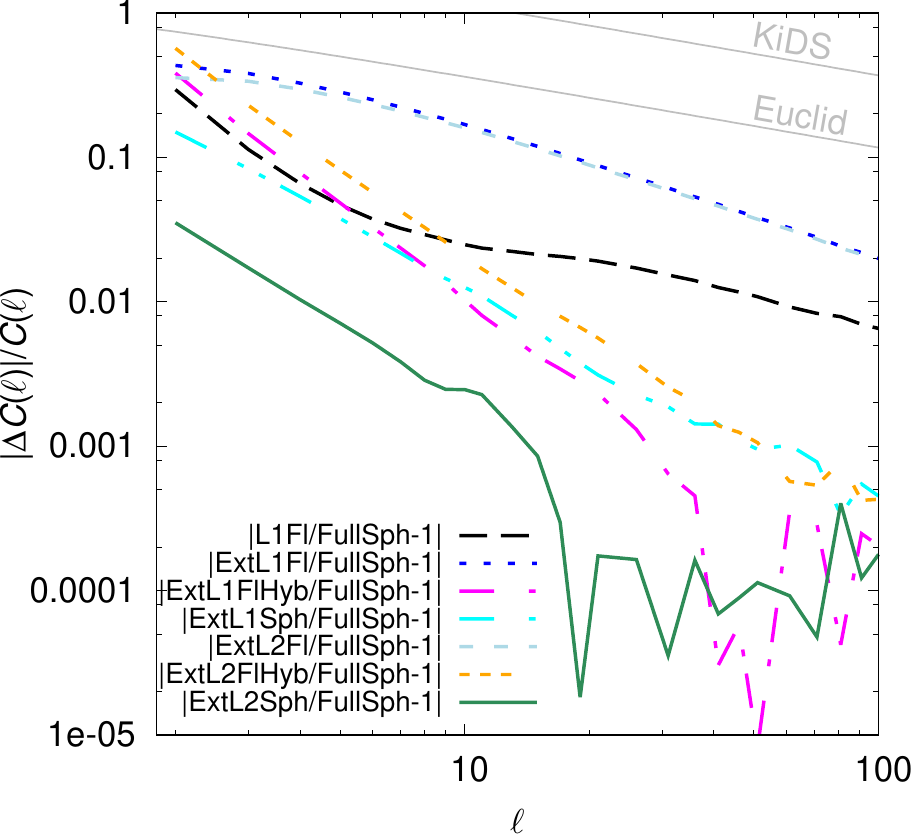}
    }
  \end{center}

  \caption{\label{fig:Cl_cases}%
        The shear power spectrum for different approximations.
        Limber to first order: standard with flat-sky (L1Fl),
        extended for flat sky (ExtL1Fl), extended hybrid for flat sky (ExtL1FlHyb),
        and extended in the spherical expansion (ExtL1Sph);
        second-order Limber approximations: extended flat sky (ExtL2Fl), extended hybrid flat sky (ExtL2FlHyb),
        and extended spherical expansion (ExtL2Sph); full (exact) spherical projection (FullSph).
        The left panel shows the total shear power spectrum.  The right panel shows the fractional difference
        resulting from each approximation,
        relative to the full spherical projection of the shear power spectrum. The two light grey curves on the
        top show the cosmic variance for KiDS- and Euclid-like surveys with areas of $1,500$ and $15,000$
        square degrees, respectively. From \cite{KH17}.
        }
\end{figure}

Many of the above mentioned corrections are more important for third-order
lensing statistics \cite{H02,2005PhRvD..72h3001D,2013arXiv1306.6151V}, which
are presented in Sect.\ \ref{sec:bispectrum}. In \citet{CFHTLenS-2+3pt} we
accounted for source-lens clustering terms contributing to the lensing
bispectrum. Ignoring this contamination, the parameter $\Sigma_8$ (see
eq.~(\ref{eq:Sigma_8})) was biased high by 0.03, which is subdominant compared
to the statistical errors.

% Chapter 4
% shear_corr_estim.tex

%%%%%%%%%%%%%%%%%%%%%%%%%%%%%%%%%%%%%%%%%%%%%%%%%%%%%%%%%%%%
\section{Shear correlation estimators}
\label{sec:corr_estim}
%%%%%%%%%%%%%%%%%%%%%%%%%%%%%%%%%%%%%%%%%%%%%%%%%%%%%%%%%%%%

%%%%%%%%%%%%%%%%%%%%%%%%%%%%%%%%%%%%%%%%%%%%%%%%%%%%%%%%%%%%
\subsection{The shear correlation function}
\label{sec:real_space_2nd}
%%%%%%%%%%%%%%%%%%%%%%%%%%%%%%%%%%%%%%%%%%%%%%%%%%%%%%%%%%%%

The most basic, non-trivial cosmic shear observable is the real-space shear
two-point correlation function (2PCF), since it can be estimated by simply
multiplying the ellipticities of galaxy pairs and averaging.

The two shear components of each galaxy are conveniently decomposed into
\emph{tangential component}, $\gamma_{\rm t}$, and cross-component,
$\gamma_\times$. With respect to a given direction vector $\vec \theta$
whose polar angle is $\phi$,
they are defined as
\begin{equation}
  \gamma_{\rm t} = - \Re \left( \gamma \, {\rm e}^{-2{\rm i}\phi} \right); \quad
  \gamma_\times = - \Im \left( \gamma \, {\rm e}^{-2{\rm i}\phi} \right).
  \label{eq:gamma_tx}
\end{equation}
The minus sign, by convention, results in a positive value of $\gamma_{\rm t}$ for the tangential
alignment around a mass overdensity. Radial alignment around underdensities
have a negative $\gamma_{\rm t}$. A positive cross-component shear is rotated by
$+\pi/4$ with respect to the tangential component.

Three two-point correlators can be formed from the two shear components,
$\langle \gamma_{\rm t} \gamma_{\rm t} \rangle$, $\langle \gamma_\times
\gamma_\times \rangle$ and $\langle \gamma_{\rm t} \gamma_\times \rangle$.
The
latter vanishes in a parity-symmetric universe, where the shear field is
statistically invariant under a mirror transformation. Such a transformation
leaves $\gamma_{\rm t}$ invariant but
changes the sign of $\gamma_\times$.
The two non-zero two-point correlators are
combined into the two components of the shear 2PCF \cite{1991ApJ...380....1M},
\begin{eqnarray}
\eqalign
  \xi_+(\theta) 
  & = \langle \gamma \gamma^\ast \rangle(\theta)  &
  = \langle \gamma_{\rm t} \gamma_{\rm t} \rangle(\theta) + \langle \gamma_\times \gamma_\times \rangle(\theta); \quad
  \nonumber \\
  \xi_-(\theta)
  & = \Re \left[ \langle \gamma \gamma \rangle(\theta) {\rm e}^{-4{\rm i} \phi} \right] &
  = \langle \gamma_{\rm t} \gamma_{\rm t} \rangle(\theta) - \langle \gamma_\times \gamma_\times \rangle(\theta) . 
  \label{eq:xi_pm}
\end{eqnarray}
The two components are plotted in Fig.~\ref{fig:2pcf}. 
We note here that from the equality of the shear and convergence power spectrum and Parseval's theorem, it follows that
$\xi_+$ is identical to the two-point correlation function of $\kappa$.

We defined an estimator of the 2PCF in \citet{SvWKM02} as
\begin{equation}
  \hat \xi_\pm(\theta) = \frac{ \sum_{ij} w_i w_j \left( \varepsilon_{{\rm t}, i} \varepsilon_{{\rm t}, j} \pm
                                \varepsilon_{\times, i} \varepsilon_{\times, j} \right)}{ \sum_{ij} w_i w_j} .
  \label{eq:estim_xi_pm}
\end{equation}
The sum extends over pairs of galaxies ($i, j$) at positions on the sky $\vec
\vartheta_i$ and $\vec \vartheta_j$, respectively,
whose separation $|\vec \vartheta_i - \vec \vartheta_j|$ lies in an angular
distance bin around $\theta$. Each galaxy has a measured ellipticity $\varepsilon_i$, and an
attributed weight $w_i$, which may reflect the measurement uncertainty.
Using the weak-lensing relation (\ref{eq:eps_eps_s_gamma})
and taking the expectation value of (\ref{eq:estim_xi_pm}), we get terms of the following
type, exemplarily stated for $\xi_+$:
\begin{equation}
\langle \varepsilon^{(\rm s)}_{i} {\varepsilon_j^{(\rm s)}}^\ast \rangle;
\langle \varepsilon^{(\rm s)}_i \gamma_j^\ast \rangle;
\langle \gamma_i {\varepsilon_j^{(\rm s)}}^\ast \rangle;
 \quad \mbox{and} \quad \langle \gamma_i \gamma_j^\ast \rangle.
  \label{eq:eps_eps_four_terms}
\end{equation}
We discuss the first three terms in Sect.~\ref{sec:ia}, in the context of intrinsic alignment (IA).
In the absence of IA, those three terms vanish and the last term is equal to
$\xi_+(|\vec \vartheta_i - \vec \vartheta_j|))$. The analogous case holds for $\xi_-$.

The main advantage of the simple estimator (\ref{eq:estim_xi_pm}) is that it
does not require the knowledge of the mask geometry, but only whether a given
galaxy is within the masked area or not. For that reason, many other
second-order estimators that we discuss in the following are based in this one.

The survey and mask geometry is however important to compute the covariance of
(\ref{eq:estim_xi_pm}). This influence was studied in detail in \citet{KS04},
where I developed a Monte-Carlo method to compute the covariance given a galaxy
catalogue. This method was subsequently used for a Principal Component Analysis
\cite{MK05}, and Karhunen-Lo\`eve \cite{KM05} study, to examine the
dependency of various survey properties on the weak-lensing information content.
The same Monte-Carlo method was also used in CFHTLenS to compute the covariance
matrix of the 2PCF \cite{CFHTLenS-2pt-notomo}.

Using (\ref{eq:p_kappa_def}) and (\ref{eq:gamma_kappa_Fourier}), we write the
2PCF in the flat-sky approximation as Hankel transforms of the convergence
power spectrum,
\begin{eqnarray}
  \xi_+(\theta) 
  &
  = \frac 1 {2\pi} \int {\rm d} \ell \, \ell {\rm J}_0(\ell
   \theta)
  [ P_\kappa^{\rm E}(\ell) + P_\kappa^{\rm B}(\ell)];
  \quad
  \nonumber \\
   \xi_-(\theta)
  &
  = \frac 1 {2\pi} \int
   {\rm d} \ell \, \ell {\rm J}_4(\ell \theta)
  [ P_\kappa^{\rm E}(\ell) - P_\kappa^{\rm B}(\ell) ].
   \label{eq:xi_pm_pkappa}
\end{eqnarray}
These expressions can be easily and quickly integrated numerically using fast
Hankel transforms \cite{2000MNRAS.312..257H}.

\begin{figure}

  \begin{minipage}{0.65\textwidth}
  \begin{center}
    \resizebox{1.0\hsize}{!}{
     \includegraphics{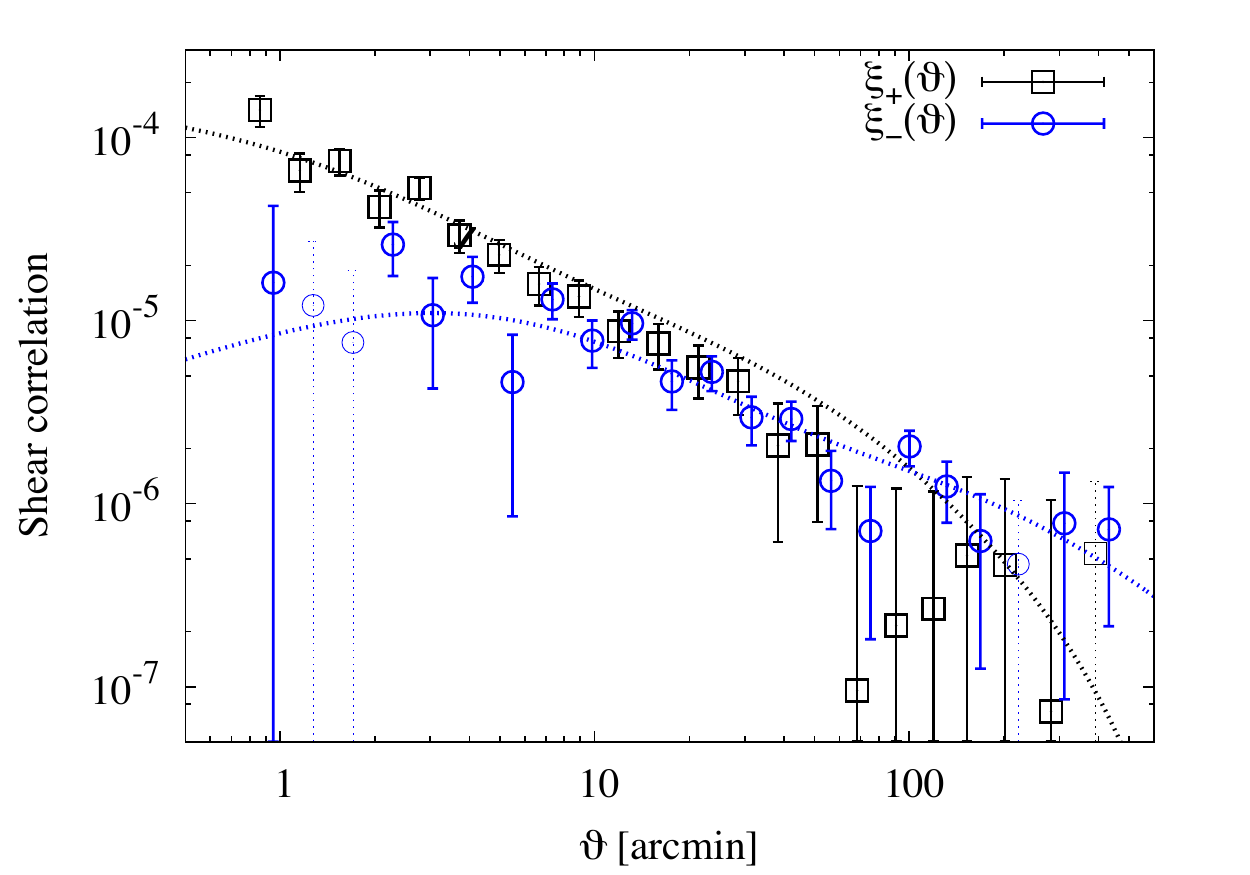}
    }
  \end{center}
  \end{minipage}%
  \hspace*{-0.1\textwidth}%
  \begin{minipage}{0.45\textwidth}

  \caption{2PCF components $\xi_+$ and $\xi_-$ (\ref{eq:xi_pm}) measured in
CFHTLenS. The dotted lines show the WMAP7 model prediction
\cite{2010arXiv1001.4538K}. From \citet{CFHTLenS-2pt-notomo}. }
  \label{fig:2pcf}

  \end{minipage}

\end{figure}

The two 2PCF components mix E- and B-mode power spectra in two different ways.
To separate the two modes, a further filtering of the 2PCF is necessary, which
will be discussed in the following section.

%%%%%%%%%%%%%%%%%%%%%%%%%%%%%%%%%%%%%%%%%%%%%%%%%%%%%%%%%%%%
\subsection{Derived second-order functions}
\label{sec:other_2nd_order}
%%%%%%%%%%%%%%%%%%%%%%%%%%%%%%%%%%%%%%%%%%%%%%%%%%%%%%%%%%%%

Apart from the 2PCF (\ref{eq:xi_pm}), other, derived second-order functions
have been widely used to measure lensing correlations in past and present
cosmic shear surveys. The motivation for derived statistics are to construct
observables that (1) have high signal-to-noise for a given angular scale, (2)
show small correlations between different scales, and (3) separate into E- and
B-modes. In particular the latter property is of interest, since the B-mode can
be used to assess the level of (certain) systematics in the data as we have
seen in Sect.~\ref{sec:E_B_modes}.

All second-order functions can be written as filtered integrals over the
convergence power spectrum, and the corresponding filter functions define their
properties. 

%%%%%%%%%%%%%%%%%%%%%%%%%%%%%%%%%%%%%%%%%%%%%%%%%%%%%%%%%%%%
\subsubsection{Aperture-mass dispersion}
%%%%%%%%%%%%%%%%%%%%%%%%%%%%%%%%%%%%%%%%%%%%%%%%%%%%%%%%%%%%

Another popular statistic is the \emph{aperture-mass dispersion}, denoted as
$\left\langle M_{\rm ap}^2 \right\rangle(\theta)$ (Fig.~\ref{fig:map2}). First,
one defines the \emph{aperture mass} as mean tangential shear with respect to
the centre $\vec \vartheta$ of a circular region, weighted by a filter function
$Q_\theta$ with characteristic scale $\theta$,
\begin{equation}
  M_{\rm ap}(\theta, \vec \vartheta)
  = \int {\rm d}^2 \vartheta^\prime \,
  Q_\theta(|\vec \vartheta - \vec \vartheta^\prime|) \,
  \gamma_{\rm t}(\vec \vartheta^\prime)
  = \int {\rm d}^2 \vartheta^\prime \,
  U_\theta(|\vec \vartheta - \vec \vartheta^\prime|) \,
  \kappa(\vec \vartheta^\prime).
  \label{eq:map}
\end{equation}
The second equality can be derived from the relations between shear and convergence,
which defines the filter function $U_\theta$ in terms of $Q_\theta$ \cite{KSFW94,S96}.  The
aperture mass is therefore closely related to the local projected over-density,
and owes its name to this fact.  The function $U_\theta$ is compensated
(i.e.~the integral over its support vanishes, $\int {\rm d}^2 \vartheta \, U_\theta(\vec \vartheta) = 0$),
and filters out a constant mass sheet $\kappa_0 =
\mbox{const}$, since the monopole mode ($\ell = 0$) is not recoverable from the
shear (\ref{eq:gamma_kappa_Fourier}).  Two choices for the functions
$U_\theta$, and consequently $Q_\theta$, have been widely used for cosmic
shear, a fourth-order polynomial \cite{1998MNRAS.296..873S}, and a Gaussian
function \cite{2002ApJ...568...20C}.

By projecting out the tangential component of the shear, $M_{\rm ap}$ is
sensitive to the E-mode only. One defines $M_\times$ by replacing $\gamma_{\rm
t}$ with $\gamma_\times$ in (\ref{eq:map}) as a probe of the B-mode only.  The
variance of (\ref{eq:map}) between different aperture centres defines the
dispersion $\left\langle M_{\rm ap}^2 \right\rangle(\theta)$, which
can be interpreted as fluctuations of lensing strength between lines of sight,
and therefore have an intuitive connection to fluctuations in the projected
density contrast.

\begin{figure}

  \begin{minipage}{0.65\textwidth}
  \centerline{\resizebox{1.0\hsize}{!}{
    \includegraphics{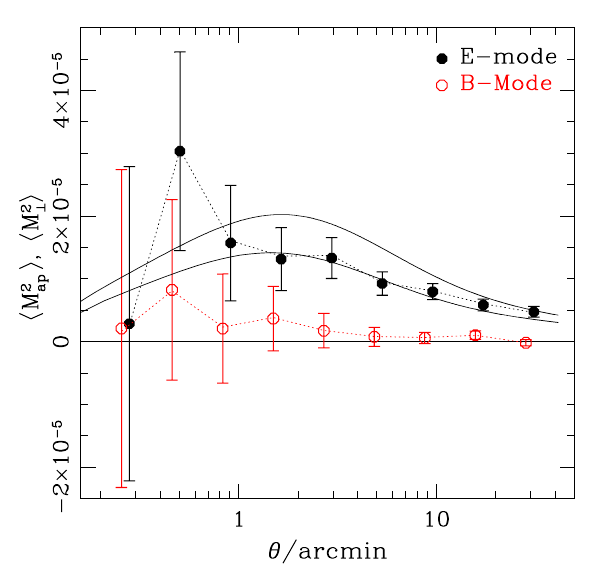}
  }}
  \end{minipage}%
  \hspace*{-0.1\textwidth}%
  \begin{minipage}{0.45\textwidth}

  \caption{Aperture-mass dispersion that we measured in COSMOS. The two solid lines
correspond to predictions with $\sigma_8 = 0.7$ and $0.8$, respectively.
From \citet{SHJKS09}. }
  \label{fig:map2}

  \end{minipage}

\end{figure}

A new $\cal N$ estimator that includes both the aperture-mass dispersion at
various angular scales, and a measure of the 2PCF at one angular scale
$\theta_0$, is discussed in \citet{EKS08}. This new data vector is
$\vec{\mathcal N} = \left(\langle M_{\rm ap}^2\rangle (\theta_1), \ldots,
\langle M_{\rm ap}^2\rangle (\theta_n), \xi_+(\theta_0)\right)$. It is a
compromise between insensitivity to the B-mode (via the aperture-mass
dispersion), and capturing the long-wavelength modes and thus maximizing the
information content (through $\xi_+$). The tightest constraints on cosmological
parameters are obtained with $\theta_0$ of around $10$ arcmin.

%%%%%%%%%%%%%%%%%%%%%%%%%%%%%%%%%%%%%%%%%%%%%%%%%%%%%%%%%%%%
\subsubsection{Practical estimators}
%%%%%%%%%%%%%%%%%%%%%%%%%%%%%%%%%%%%%%%%%%%%%%%%%%%%%%%%%%%%

The aperture-mass dispersion can in principle be
estimated by averaging over many aperture centres $\vec \vartheta$. This is
however not practical: The sky coverage of a galaxy survey is not contiguous,
but has gaps and holes due to masking. Apertures with overlap with masked areas
biases the result, and avoiding overlap results in a substantial area loss.
This is particularly problematic for filter functions whose support extend
beyond the scale $\theta$. One possibility is to fill in the missing data,
e.g.~with inpainting techniques \cite{2009MNRAS.395.1265P}, resulting in a
pixelised, contiguous convergence map on which the convolution (\ref{eq:map})
can be calculated very efficiently \cite{2012MNRAS.423.3405L}. Alternatively,
the dispersion measures can be expressed in terms of the 2PCF, and are
therefore based on the estimator (\ref{eq:estim_xi_pm}) for which the mask
geometry does not play a role.

%%%%%%%%%%%%%%%%%%%%%%%%%%%%%%%%%%%%%%%%%%%%%%%%%%%%%%%%%%%%
\subsubsection{Generalisations}
%%%%%%%%%%%%%%%%%%%%%%%%%%%%%%%%%%%%%%%%%%%%%%%%%%%%%%%%%%%%

In fact, every second-order statistic can be expressed as integrals over the
2PCF because, as mentioned above, all are functions of $P_\kappa$, and the
relation (\ref{eq:xi_pm_pkappa}) can be inverted. In general, they do not
contain the full information about the convergence power spectrum \cite{EKS08},
but separate E- and B-modes.

The general expression for an E-/B-mode separating function $X_{\rm E, B}$ is
\begin{equation}
X_{\rm E, B} = \frac 1 {2\pi} \int_0^\infty {\rm d} \ell \, \ell \, P_\kappa^{\rm E, B}(\ell) \fourier U^2(\ell) .
\label{eq:X_EB_Fourier}
\end{equation}
A practical estimator using (\ref{eq:estim_xi_pm}) is 
\begin{equation} \hat X_{\rm E, B}
= \frac 1 2 \sum_i \vartheta_i \, \Delta \vartheta_i \left[ T_+\left(
{\vartheta_i} \right) \hat \xi_+(\vartheta_i) \pm T_-\left( {\vartheta_i}
\right) \hat \xi_-(\vartheta_i) \right] .
\label{eq:X_EB}
\end{equation}
Here, $\Delta \vartheta_i$ is the bin width, which can vary with $i$, for
example in the case of logarithmic bins.
The filter functions $T_\pm$ and $\fourier U^2$ are
Hankel-transform pairs, given by the integral relation
\cite{2002ApJ...568...20C,2002A&A...389..729S}
\begin{equation}
T_\pm(x) =
\int_0^\infty {\rm d} t \, t \, {\rm J}_{0,4}(x t) \fourier U^2(t).
\label{eq:T_pm}
\end{equation}
This implicit relation between $T_+$ and $T_-$ guarantees the separation into E- and B-modes
of the estimator (\ref{eq:X_EB}).

In some cases of $X_{\rm E, B}$, for example for the aperture mass 
dispersion, the power-spectrum filter $\fourier U$ is explicitely given
as the Fourier transform of a real-space filter function $U$, see
e.g.~(\ref{eq:map}) for the aperture mass.  In other cases the functions
$T_{\pm}$ are constructed first, and $\fourier U$ is calculated by inverting
the relation (\ref{eq:T_pm}). Model predictions of $X_{\rm E}$ can be obtained
from either (\ref{eq:X_EB_Fourier}), or (\ref{eq:X_EB}). For the latter, one
inserts a theoretical model for $\xi_{\pm}$, and does not need to calculate
$\fourier U$.

%%%%%%%%%%%%%%%%%%%%%%%%%%%%%%%%%%%%%%%%%%%%%%%%%%%%%%%%%%%%
\subsubsection{E-/B-mode mixing}
%%%%%%%%%%%%%%%%%%%%%%%%%%%%%%%%%%%%%%%%%%%%%%%%%%%%%%%%%%%%

None of the derived second-order functions introduced so far provide a pure
E-/B-mode separation. They suffer from a leakage between the modes, on small
scales, or large scales, or both. This mode mixing comes from the incomplete
information on the measured shear correlation: On very small scales, up to 10
arc seconds or so, galaxy images are blended, preventing accurate shape
measurements, and thus the shape correlation on those small scales is not
sampled. Large scales, at the order of degrees, are obviously only sampled up
to the survey size. This leakage can be mitigated by (i) extrapolating the
shear correlation to unobserved scales using a theoretical prediction (thereby
potentially biasing the result), or (ii) cutting off small and/or large scales
of the derived functions (thereby loosing information). Figure
\ref{fig:EBmixing} shows the example of the aperture-mass dispersion with a
polynomial filter, see \citet{KSE06}.

\begin{figure}

  \begin{minipage}{0.65\textwidth}
  \centerline{\resizebox{1.0\hsize}{!}{
    \includegraphics{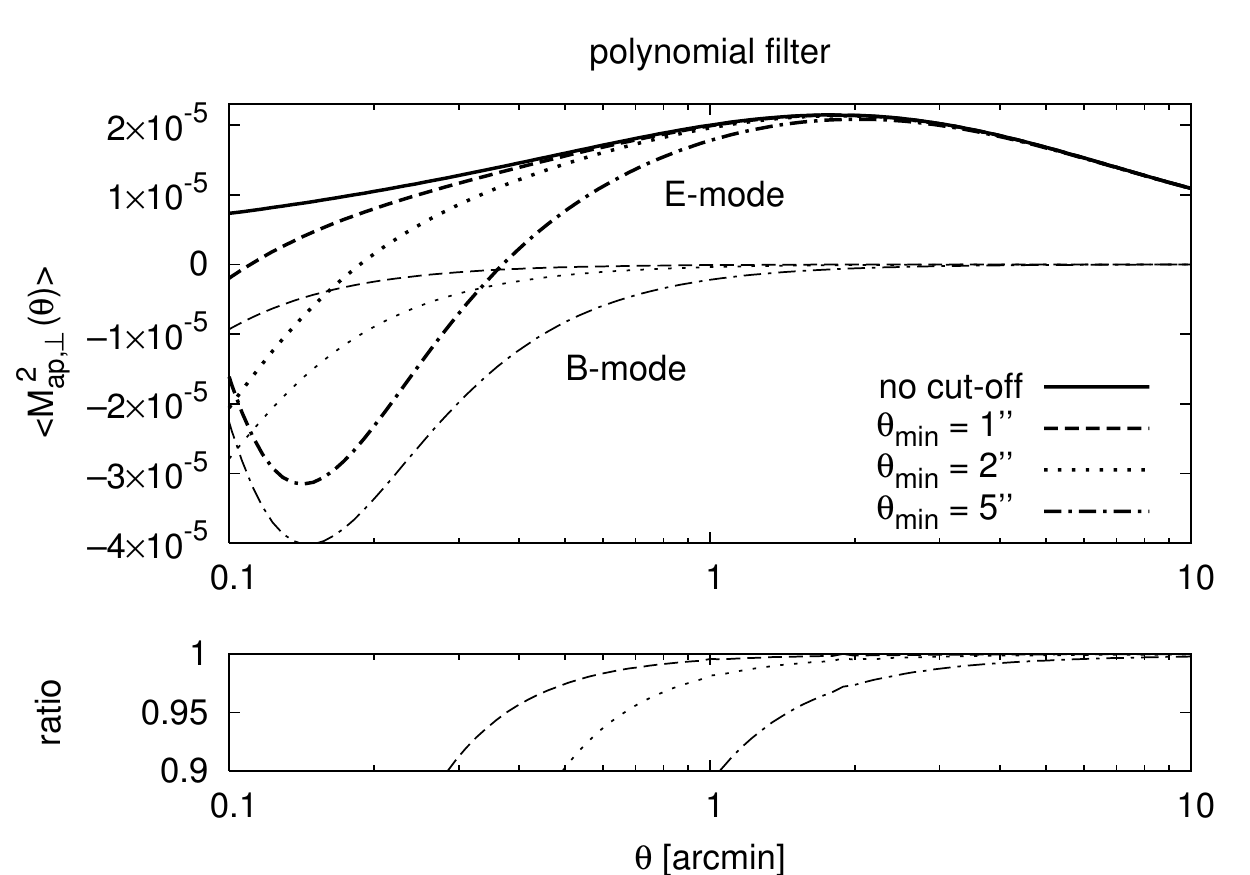}
  }}

  \end{minipage}%
  \hspace*{-0.1\textwidth}%
  \begin{minipage}{0.45\textwidth}

  \caption{\emph{Top panel:} The E-mode aperture-mass dispersion $\langle M_{\rm ap}^2(\theta) \rangle$ and the leakage
    from the E-mode $\langle M_{\rm ap}^2(\theta, \theta_{\rm min}) \rangle$ to
    the B-mode $\langle M_{\perp}^2(\theta, \theta_{\rm min}) \rangle$ due to the
    small-scale cutoff $\theta_{\rm min}$ of the shear correlation function.
    \emph{Bottom panel:} The ratio $\langle
    M_{\rm ap}^2(\theta, \theta_{\rm min}) \rangle/\langle M_{\rm ap}^2(\theta)
      \rangle$. From \citet{KSE06}.
  }
  \label{fig:EBmixing}

  \end{minipage}

\end{figure}

%%%%%%%%%%%%%%%%%%%%%%%%%%%%%%%%%%%%%%%%%%%%%%%%%%%%%%%%%%%%
\subsubsection{E-/B-mode functions from a finite interval}
%%%%%%%%%%%%%%%%%%%%%%%%%%%%%%%%%%%%%%%%%%%%%%%%%%%%%%%%%%%%

E-/B-mode mixing can be avoided altogether by defining derived second-order
statistics via suitable filter functions $T_\pm$ (or, equivalently $U$). For a
pure E-/B-mode separation, those filter functions need to vanish on scales
where the shear correlation is missing.

The first such set of filter functions was derived geometrically, by
defining the E- and B-mode of the shear field on a circle:
\begin{equation}
  {\cal C}(\theta) = {\cal C}_{\rm t}(\theta) + {\cal C}_\times(\theta)
    = \frac 1 {2\pi} \int\limits_0^{2\pi} {\rm d} \phi
      \left( \gamma_{\rm t} + \gamma_\times \right)(\theta, \phi),
\end{equation}
where the tangential and cross-components of the shear are measured with
respect to the center of a circle with radius $\theta$. By construction, ${\cal
C}_{\rm t}$ (${\cal C}_\times$) projects out the E-mode (B-mode).

If we now correlate the field $C$ for two concentric circles with different
radii $\theta_1 < \theta_2$, the resulting second-order E-mode (B-mode)
correlations $\langle {\cal C}_{\rm t}(\theta_1) {\cal C}_{\rm t}(\theta_2)
\rangle$ ($\langle {\cal C}_\times(\theta_1) {\cal C}_\times(\theta_2)$)
correlate the shear at two angular positions with minimum separation $\theta_2
- \theta_1$ and maximum distance $\theta_1 + \theta_2$. This \emph{circle
statistics} thus achieves E-/B-mode separation from shear correlations on a
finite interval.

In practise, the shear cannot be measured on a infinitely thin line, and the
circle is extended to an annulus or ring ${\cal R}$ with finite width. Two
disjoint annuli are correlated to form the \emph{ring statistics} \cite{SK07}.

As all second-order functions, the ring statistics can be written in the forms
(\ref{eq:X_EB_Fourier}) and (\ref{eq:X_EB}). Corresponding filter functions
$U$ and $T_\pm$ have been derived in \cite{SK07}, the latter of which have
finite support.

The exact form of $T_\pm$ is given by the geometrical set-up of the two rings.
This can be generalized, and $T_\pm$ be derived detached from geometry. From the
relations between $T_+$ and $T_-$ and the requirement that both vanish outside 
a finite interval $[\vartheta_{\rm min}; \vartheta_{\rm max}]$, two integral conditions
are sufficient to fulfill these conditions \cite{SK07}:
\begin{equation}
  \int_{\vartheta_{\rm min}}^{\vartheta_{\rm max}}{\rm d}\vartheta\, \vartheta\,T_+(\vartheta)
  =0=\int_{\vartheta_{\rm min}}^{\vartheta_{\rm max}}{\rm d}\vartheta\, \vartheta^3\,T_+(\vartheta) .
  \label{eq:T_p_conditions}
\end{equation}
The corresponding relations for $T_-$ are
\begin{equation}
  \int_{\vartheta_{\rm min}}^{\vartheta_{\rm max}}{{\rm d}\vartheta\over
\vartheta}\,T_-(\vartheta) =0=\int_{\vartheta_{\rm min}}^{\vartheta_{\rm
max}}{{\rm d}\vartheta\over \vartheta^3}\,T_-(\vartheta).
  \label{eq:T_m_conditions}
\end{equation}
The first generalised ring statistics was introduced in
\citet{2010A&A...510A...7E}, who chose the lowest-order polynomials for $T_+$ to
fulfill (\ref{eq:T_p_conditions}) (which is second order).

An optimization scheme for a general ring statistics was developed in
\citet{FK10}. In this work we wrote $T_+$ as linear combination of orthogonal polynomials,
in this case, Chebyshev polynomials of the second kind, up to order $N-1$. The
two integral conditions on $T_\pm$ then become a ($N \times 2$) matrix
equations in the expansion coefficients. It was determined that $N=6$ captures
most of the information of the shear correlation.

Optimisation is then performed by varying the coefficients to maximize two
quantities for a given maximum angular scale $\Psi_{\rm max}$, the $S/N$ of the
ring statistic and the Fisher matrix figure of merit (FoM) for $\Omega_{\rm m}$
and $\sigma_8$. For the Fisher matrix, the (Gaussian) covariance between
different scales was accounted for. Fig.~\ref{fig:FoM_genR} shows the FoM as
function of $\Psi_{\rm max}$ for different statistics. As expected, the FoM
increases with $\Psi_{\rm max}$ as more and more information is included, but
the increase flattens out after around 20 arcmin. Compared to the original ring
statistic from \citet{SK07} (denoted by $Z_+$), the optimised ring statistic
(denoted by $T_+$) achieves two to three times larger FoMs. Depending on the
range of scales $\eta = \vartheta_{\rm min}/\vartheta_{\rm max}$, it even
outperformed the aperture-mass dispersion.

\begin{figure}

    \begin{minipage}{0.55\textwidth}
  \centerline{\resizebox{1.0\hsize}{!}{
    \includegraphics{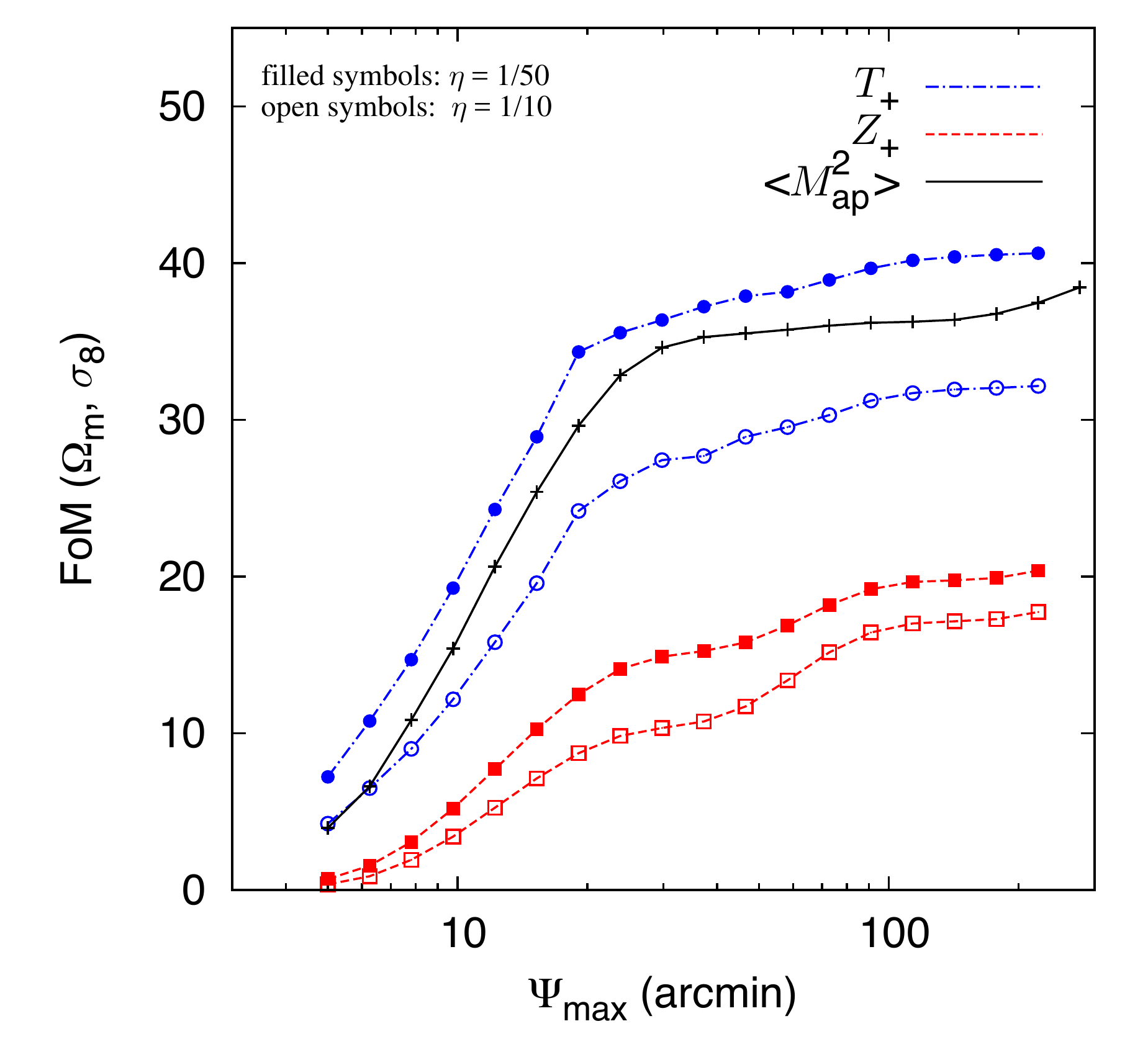}
  }}

  \end{minipage}%
  \hspace*{-0.1\textwidth}%
  \begin{minipage}{0.55\textwidth}

  \caption{%
    Figure of merit (FoM) for the optimised
    filter function $T_+$ (blue
    curves), and the original ring statistic filter function $Z_+$
    from SK07 (in red), as function of the maximum angular scale $\Psi_{\rm max}$.
    Filled and
    open symbols represent
    $\eta \equiv \vartheta_{\rm min}/\vartheta_{\rm max} = 1/50$ and $1/10$, respectively.
    The black curve with crosses corresponds to the
    aperture-mass with aperture diameter $\vartheta_{\rm max}$.
    From \citet{FK10}.
  } 
  \label{fig:FoM_genR}

  \end{minipage}%

\end{figure}

A more general question to ask is, given an interval $[\vartheta_{\rm min};
\vartheta_{\rm max}$], how can we capture all available information of the
E-mode shear correlation on that interval? The information on a subset of
scales, say with $\Psi < \theta_{\rm max}$ should be contained in the entire
interval, since all information on $[\vartheta_{\rm min}; \vartheta_{\rm max}$
contains a signal with a filter function that is zero for $\theta > \Psi$.

Such general E-/B-mode separating second-order quantities from a finite
interval are the so-called COSEBIs \citeaffixed{COSEBIs}{Complete Orthogonal
Sets of E-/B-mode integrals;}. Fig.~\ref{fig:cosebis} shows the COSEBIs
measured with CFHTLenS. COSEBIs do not depend on a continuous angular scale
parameter $\theta$, but are a discrete set of modes $E_n, B_n, n=1, 2 \ldots$
Typically, fewer than 10 COSEBI modes are sufficient to capture all
second-order E-mode information \cite{2012A&A...542A.122A}.

The COSEBI modes are strongly correlated, which makes visual inspection of the
data and comparison to the prediction difficult. Therefore, I compute
uncorrelated data points $E^{\rm ortho}_m$ as orthogonal transformation of the
COSEBIs $E_n$, $E^{\rm ortho}_m = S_{mn} E_n$, where ${\mat S}$ is an
orthogonal matrix, ${\mat S} {\mat S}^{\rm T} = 1$.  The result is presented in
the right panel of Fig.~\ref{fig:cosebis}. Increasing modes $m$ have
larger error bars, which correspond to the elements of the diagonal matrix
${\Sigma}$, obtained by diagonalising the COSEBIs covariance matrix ${\mat C} =
{\mat S} {\Sigma} {\mat S}^{\rm T}$.

\begin{figure}[b]

  \centerline{\resizebox{0.9\hsize}{!}{
    \includegraphics[scale=1.065]{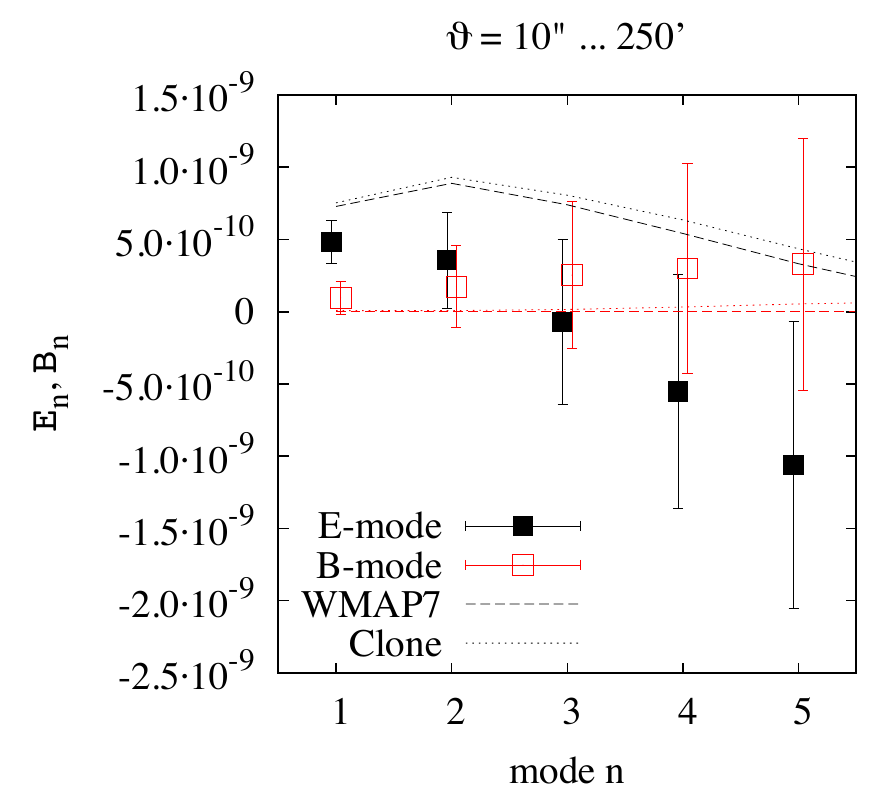}
    \includegraphics{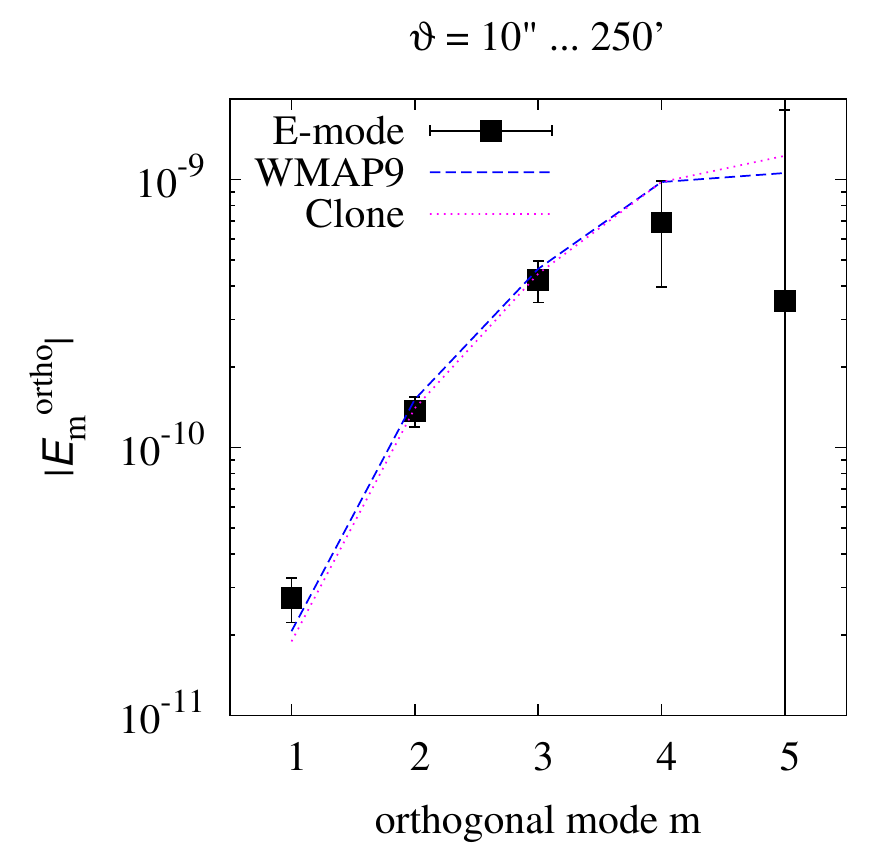}
  }
}

  \caption{The first five COSEBIs modes, measured in CFHTLenS, together with
predictions using a WMAP7 and WMAP9 cosmology, and the CFHTLenS ``Clone''
$N$-body simulation. \emph{Left panel:} The original COSEBI modes, from
\citet{CFHTLenS-2pt-notomo}. \emph{Right panel:} The corresponding tranformed, orthogonal COSEBI modes,
from \citet{CFHTLenS-2+3pt}. }

  \label{fig:cosebis}

\end{figure}

%%%%%%%%%%%%%%%%%%%%%%%%%%%%%%%%%%%%%%%%%%%%%%%%%%%%%%%%%%%%
\subsection{Higher-order correlations}
\label{sec:higher-order}
%%%%%%%%%%%%%%%%%%%%%%%%%%%%%%%%%%%%%%%%%%%%%%%%%%%%%%%%%%%%

%%%%%%%%%%%%%%%%%%%%%%%%%%%%%%%%%%%%%%%%%%%%%%%%%%%%%%%%%%%%
\subsubsection{Third-order correlations}
\label{sec:bispectrum}
%%%%%%%%%%%%%%%%%%%%%%%%%%%%%%%%%%%%%%%%%%%%%%%%%%%%%%%%%%%%

The convergence power spectrum $P_\kappa$ (\ref{eq:p_kappa_limber}) only
captures the Gaussian component of the LSS. There is however substantial
complementary non-Gaussian information in the matter distribution, in particular on
small scales, where the non-linear evolution of structures
creates non-Gaussian weak-lensing correlations. On small and intermediate
scales, these non-linear structures are the dominant contributor to
non-Gaussian lensing signatures, compared to (quasi)-linear perturbations, or
potential primordial non-Gaussianity. Constraints on the latter from cosmic
shear alone can not compete with constraints from other probes such as CMB or
galaxy clustering \cite{2004MNRAS.348..897T,2011MNRAS.411..595P,2012MNRAS.426.2870H}.

To measure these non-Gaussian characteristics, one has to go beyond the second-order
convergence power spectrum. The next-leading order statistic is the bispectrum $B_\kappa$,
which is defined by the
following equation:
\begin{equation}
   \left\langle \fourier \kappa(\vec \ell_1) \fourier \kappa(\vec \ell_3) \fourier \kappa(\vec
      \ell_3) \right\rangle
      = (2 \pi)^2
  \delta_{\rm D}(\vec \ell_1 + \vec \ell_2 + \vec \ell_3)
   \left[ B_\kappa(\vec \ell_1, \vec \ell_2) +
    B_\kappa(\vec \ell_2, \vec \ell_3) + B_\kappa(\vec \ell_3, \vec \ell_1)
  \right].
  \label{eq:b_kappa}
\end{equation}
The bispectrum measures three-point correlations of the convergence defined on
a closed triangle in Fourier space. $B_\kappa$ can be related to the density
bispectrum $B_\delta$ via Limber's equation \cite{2001ApJ...548....7C}. Other
measures of non-Gaussianity are presented in Sect.~\ref{sec:peak_counts}.

The corresponding real-space weak-lensing observable is the shear three-point
correlation function (3PCF)
\cite{2003MNRAS.340..580T,tpcf1,2003ApJ...584..559Z,2006A&A...456..421B}.
Correlating the two-component shear of three galaxies sitting on the vertices
of a triangle, the 3PCF has $2^3 = 8$ components, and depends on three angular
scales. Those eight components can be combined into four complex \emph{natural}
components \cite{tpcf1,SKL05}.

A simple estimator of the 3PCF can be constructed analogous to
(\ref{eq:estim_xi_pm}), by summing up triplets of galaxy ellipticities at
binned triangles. The relations between the 3PCF and the bispectrum are
complex, and it is not straightforward to efficiently evaluate those
numerically. We have derived these equations in \citet{SKL05}, and I computed them numerically
in \citet{PhD}.

%%%%%%%%%%%%%%%%%%%%%%%%%%%%%%%%%%%%%%%%%%%%%%%%%%%%%%%%%%%%
\subsubsection{Generalized aperture-mass skewness}
\label{sec:map3_gen}
%%%%%%%%%%%%%%%%%%%%%%%%%%%%%%%%%%%%%%%%%%%%%%%%%%%%%%%%%%%%

Most measurements and cosmological analyses of higher-order cosmic
shear have been obtained using the aperture-mass skewness $\langle M_{\rm ap}^3
\rangle$ \cite{2003ApJ...592..664P,JBJ04,SKL05}. $\langle M_{\rm ap}^3 \rangle$
is the skewness of (\ref{eq:map}), and can be written as pass-band filter over
the convergence bispectrum. Analogous to the second-order case, relations exist
to represent $\langle M_{\rm ap}^3 \rangle$ as integrals over the 3PCF,
facilitating the estimation from galaxy data without the need to know the mask
geometry. Corresponding filter functions have been found in case of the
Gaussian filter \cite{JBJ04}.

I have contributed to define a generalization of the aperture-mass skewness.
This skewness corresponds to filters with three different
aperture scales, permitting to probe the bispectrum for different $\ell_1,
\ell_2, \ell_3$ in \citet{SKL05}. I have tested the increase of information
from this estimator compared to the pure "diagonal" skewness, and the
aperture-mass dispersion in \citet{KS05}. I showed that the combination of
second- and third-order statistics helps lifting parameter degeneracies, in
particular the one between $\Omega_{\rm m}$ and $\sigma_8$, extending earlier
results 
\cite{1997A&A...322....1B,2004MNRAS.348..897T}
to real-space estimators.

%%%%%%%%%%%%%%%%%%%%%%%%%%%%%%%%%%%%%%%%%%%%%%%%%%%%%%%%%%%%
\subsubsection{Peak counts}
\label{sec:peak_counts}
%%%%%%%%%%%%%%%%%%%%%%%%%%%%%%%%%%%%%%%%%%%%%%%%%%%%%%%%%%%%

In weak-lensing data one can identify projected over-densities by isolating
regions of high convergence, or enhanced tangential shear alignments. The
statistics of such weak-lensing \emph{peaks} are a potentially powerful probe
of cosmology, since peaks are sensitive to the number of halos and therefore
probe the halo mass function, which strongly depends on cosmological
parameters
\citeaffixed{1986MNRAS.222..323K,1989ApJ...347..563P,1989ApJ...341L..71E}{e.g.}.
A \emph{shear-selected} sample of peaks is a tracer of the total mass in halos,
and does not require scaling relations between mass and luminous tracers, such
as optical richness, SZ or X-ray observables.

The relation between peaks and halos is complicated because of projection and
noise. Several small halos in projection, or filaments along the line of sight,
can produce the same lensing alignment as one larger halo. Noise in the form of
intrinsic galaxy ellipticities (see Sect.~\ref{sec:estim_shear})
produces false detections, and alters the
significance of real peaks \cite{2007A&A...462..875S}. Because the number of
halos strongly decreases with mass, noise typically results in an up-scatter of
peak counts towards higher significance, which has to be modeled carefully.

Numerical simulations have shown a large potential of peak counts to constrain
cosmological parameters \cite{2010PhRvD..81d3519K,2012MNRAS.423.1711M}. Shear
peaks single out the high-density regions of the LSS, and therefore probe the
non-Gaussianity of the LSS. Despite peak counts being a non-linear probe of
weak lensing, they require the measurement only to first order in the observed
shear. Thus,
this technique potentially suffers
from less systematics than higher-order shear correlations. This is
similar to galaxy-galaxy lensing, where shapes of background galaxies are correlated
with the position of foreground objects (galaxies, but also groups and clusters).
The decreased sensitivity of galaxy-galaxy and cluster lensing compared to cosmic shear has
been demonstrated with CFTHLenS \cite{2014MNRAS.437.2111V,CSKC14,CFHTLenS-halo-shapes}.

Peak counts are complementary to second-order statistics, and both probes
combined are able to lift parameter degeneracies
\cite{2010MNRAS.402.1049D,2009A&A...505..969P,2011PhRvD..84d3529Y,2012MNRAS.423..983P}.
In addition to peak counts, the two-point correlation function of lensing peaks
carries cosmological information \cite{2013MNRAS.432.1338M}.

Theoretical predictions for peak counts are difficult to obtain, in particular
at high signal-to-noise. Past approaches have been based on Gaussian random fields
\cite{2010ApJ...719.1408F,2010A&A...519A..23M}.
Together with my PhD student Chieh-An Lin, we introduced a new,
flexible model of peak counts is based on samples of halos drawn from the mass
function, which can be generated very quickly without the need to run
time-consuming $N$-body simulations \cite{LK15a}.

\begin{figure}

  \begin{minipage}{0.65\textwidth}
  \centerline{\resizebox{1.0\hsize}{!}{
    \includegraphics{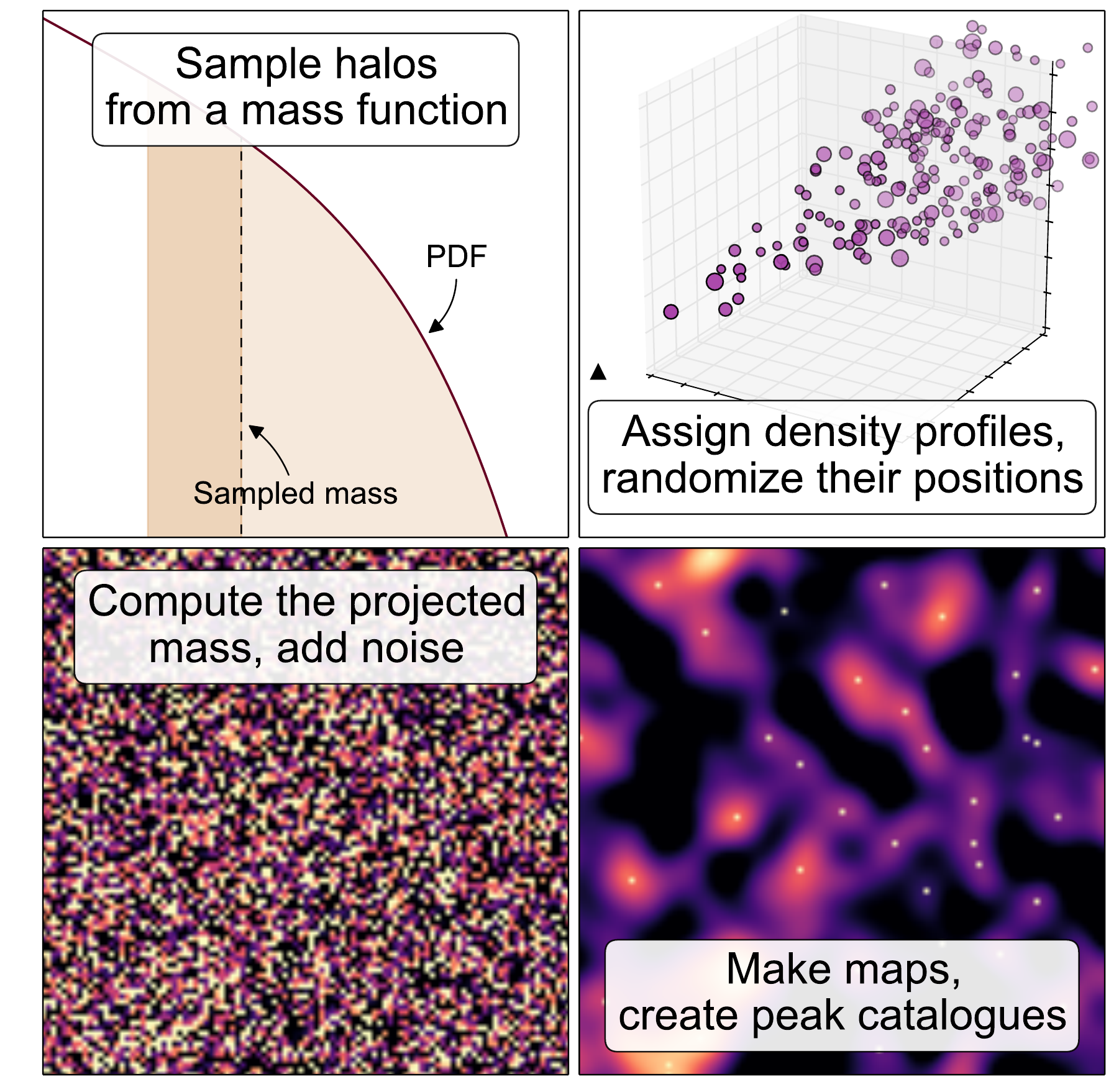}
  }
  }
  \end{minipage}%
  \hspace*{-0.1\textwidth}%
  \begin{minipage}{0.45\textwidth}

  \caption{Illustration of our peak count model from creating fast simulations of uncorrelated
  halos. From \citet{LKS16}.}
  \label{fig:Fig2_LKS16}

  \end{minipage}%

\end{figure}

Fig.~\ref{fig:Fig2_LKS16} illustrates how we generate our peak count model.
First, for a given redshift $z$, halo masses $M$ are randomly drawn from a halo
mass function $n(M, z)$, for which we choose the fitting formula from
\citet{2001MNRAS.321..372J}. The corresponding halos are placed in a comoving
volume of given field of view, where the positions $x$ and $y$ perpendicular to
the line of sigh are uniformly drawn. Halo profiles are attributed to the
halos, in our case the NFW profile following \citet{NFW}, asssuming the
relation between mass and concentration from \cite{2002MNRAS.337..875T}. Since
there is no spatial correlation between halos due to the randomization of their
positions, this corresponds to the 1-halo term in the halo model, see for a
review \cite{2002PhR...372....1C}. Next, lensing convergence and shear are
computed for a given source galaxy redshift distribution, by adding up the
contribution of all halos along the line of sight to a given soure galaxy
redshift. Shape noise is added, if desired the convergence is computed from the
shear, and peaks are counted in the final $\kappa$ map. We characterize the
number of peaks with a histogram of the peak number probability function
(pdf), also peak abundance, or \emph{peak function}, as function of peak
signal to noise ratio $\nu$. Alternatively, we also compute the cumulative pdf
(cdf) following \cite{2010MNRAS.402.1049D} and characterize the peak counts with
the SNR at given percentiles of the cdf.

Our model makes two main assumptions: First, we claim that diffuse, unbound
matter, for example in the form of filaments between halos, does not
significantly contribute to the number of weak-lensing peaks. Similarly,
structures such as voids are not simulated in our model. Note that models based
on the halo model make the same assumptions. Second, we neglect spatial
correlations between halos. Previous work has shown that correlated structure
along the line of sight influences the number of peaks by only a few percent
\cite{2010ApJ...709..286M}. Note that in our model for a given line of sight
more than one halo can contribute to the lensing signal, but in the form of
random, uncorrelated halos.

We test the two hypotheses in \citet{LK15a} by comparing our model predictions
to $N$-body simulations. First, we replace all detected halos in the simulation
by analytical NFW profiles, and remove all remaining dark-matter particles.
This tests our first assumption (together with the universality of the NFW
profile). Next, we randomize the $x$- and $y$-positions of all halos, testing
our second assumption. The result is shown in Fig.~\ref{fig:Fig4_LK15a}. Our
model agrees fairly well with the $N$-body simulation, although the error bars are
large due to the small field of $53.7$ deg$^2$.

\begin{figure}

  \centerline{\resizebox{0.8\hsize}{!}{
    \includegraphics{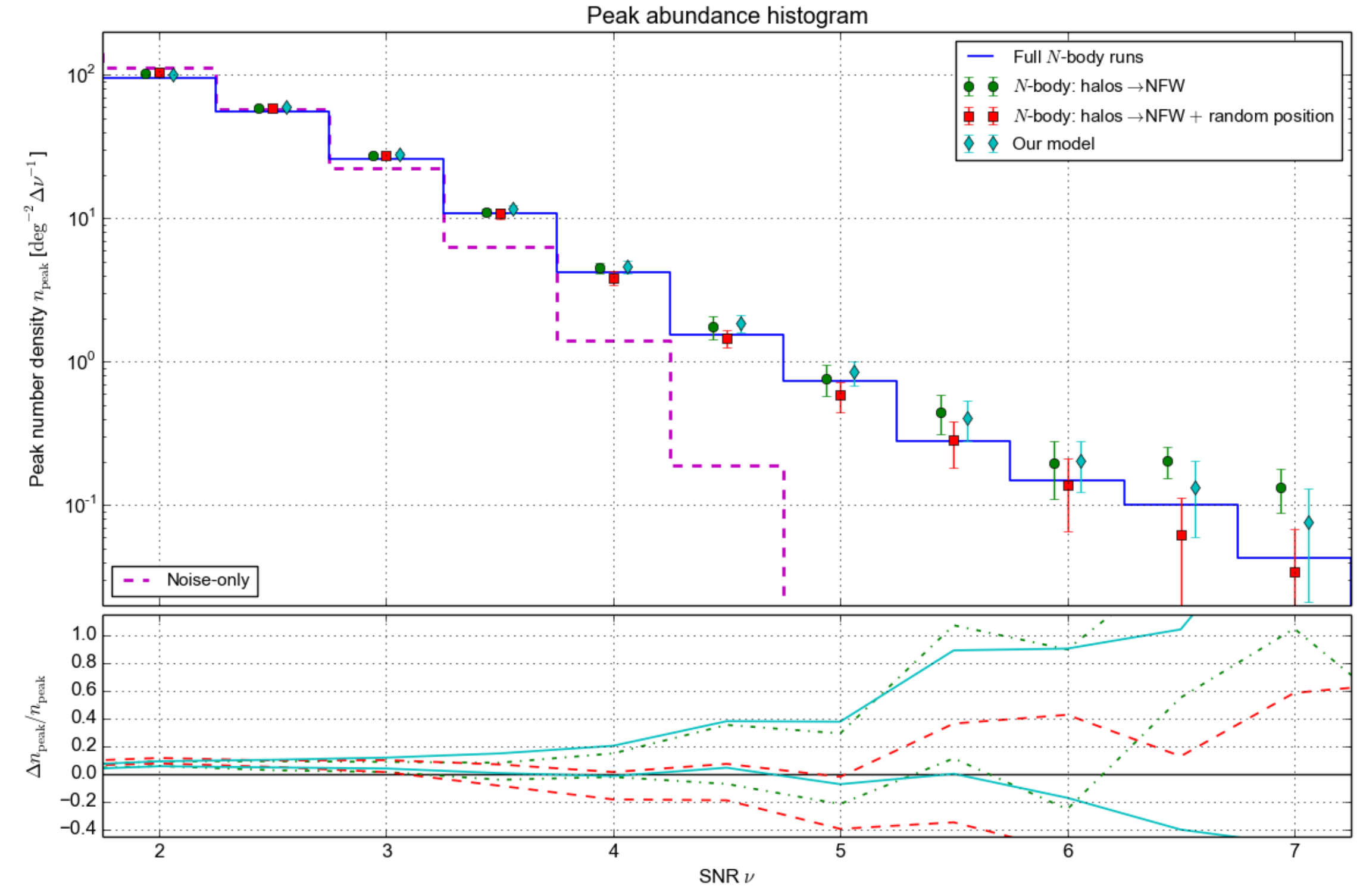}
  }
}

  \caption{Peak counts from different outputs of an $N$-body simulation compared to our model.
   The blue curve shows the result from the full $N$-body simulation. For the green circles we have replaced
   halos with analytical NFW profiles. The green squares correspond to additionally randomized halo positions.
   The cyan diamonds are our model, independent of the simulation. The magenta line corresponds to peaks from noise-only maps.
   The upper panel shows the peak count
   histogram as function of peak SNR $\nu$; the lower plot shows the difference compared to the full $N$-body simulation
  (one standard deviation, where the errors are computed from different realisations of noise maps).
  From \citet{LK15a}.}

  \label{fig:Fig4_LK15a}

\end{figure}

When replacing the $N$-body particles (blue curve) by NFW halos, the peak
counts seem systematically lower (green circles). This might be due to the
missing diffuse matter, or a decreased lensing strength of the NFW profiles
compared to the simulated ones. For example, \citet{2016arXiv161204041L} showed
that the number of peaks depends strongly on the concentration parameter.

We see a further decrease of peak counts when scrambling the halo positions (red
squares), indicating the level of influence of correlated structures. Somewhat
inbetween those two cases is our model, where we replace the simulated halo
numbers and masses by our own draws from the mass function (cyan diamonds). Even though the
number of halos in the simulation agrees well with the analytical mass function
over a large range of redshift and mass \cite{LK15a}, the change of the number
of peaks is visible. Ideally, in a cosmological analysis, one would account for
the uncertainty in mass function, halo profiles, and mass-concentration
relation, and marginalise over it.

The peak count model and data analysis software including parameter inference with
Approximate Bayesian Computation (ABC, see Sect.~\ref{sec:ABC})
is available as the public software \textsc{camelus} \cite{camelus_ascl}.

A similar, stochastical model was proposed by \citet{2009PhRvD..80l3020K}. Our
model extends this earlier work by simulating an entire field of view
corresponding to a real survey. This allows us to include geometrical effects
such as masking, boundary effects, PSF and other spatial systematic residuals,
and finite-field effects due to the conversion from $\kappa$ to $\gamma$. We
test some of these effects in \citet{LKS16} and \citet{LK17}. The constraining
power of peaks compared to the 2PCF was studied in \citet{PL17}. In this paper we
also demonstrated the agreement of our methods with the MICE-simulations
\cite{2015MNRAS.447.1319F}, except for very high SNR $\nu$.

In on-going work, with PhD student Niall Jeffrey (advisor: Filipe Abdala) we
examine fast approximate models such as PINOCCHIO \cite{2002MNRAS.333..623T} to
create a peak count model. This could be a middle way of running simulations
that take significantly less time than a $N$-body simulation, but does contain
spatial correlations between halos to some level.

% Chapter 5
% cosmo_from_shear.tex

%%%%%%%%%%%%%%%%%%%%%%%%%%%%%%%%%%%%%%%%%%%%%%%%%%%%%%%%%%%%
\section{Inference in cosmology}
\label{sec:cosmo_from_shear}
%%%%%%%%%%%%%%%%%%%%%%%%%%%%%%%%%%%%%%%%%%%%%%%%%%%%%%%%%%%%

Much of my work has focused on how to obtain constraints on cosmological parameters
from cosmic shear observations. To that end, I studied the necessary ingredients
for cosmological analyses not only for weak lensing data, but for general problems in cosmology.
This includes the covariance matrix
(Sect.~\ref{sec:cov_estim}), the likelihood function (Sect~.\ref{sec:likelihood},
likelihood sampling techniques for parameter estimation (Sect.~\ref{sec:param_estim}) and
model comparison (Sect.~\ref{sec:model_selection}), and likelihood-free inference methods
(Sect.~\ref{sec:ABC}).

%%%%%%%%%%%%%%%%%%%%%%%%%%%%%%%%%%%%%%%%%%%%%%%%%%%%%%%%%%%%
\subsection{Covariance estimation}
\label{sec:cov_estim}
%%%%%%%%%%%%%%%%%%%%%%%%%%%%%%%%%%%%%%%%%%%%%%%%%%%%%%%%%%%%

The covariance matrix of weak-lensing observables is an essential ingredient
for cosmological analyses of cosmic shear data. Shear correlations at different
scales are not independent but correlated with each other: The cosmic shear
field is non-Gaussian, in particular on small scales, and different Fourier
modes become correlated from the non-linear evolution of the density field.
This mode-coupling leads to an information loss compared to the Gaussian case
(unless higher-order statistics are included). If not taken into account
properly, error bars on cosmological parameters will be underestimated.

Additionally, even in the Gaussian case Fourier modes are spread on a range of
angular scales in real space, causing shear functions to be
correlated across scales. The correlation strength depends on the filter
function relating the power spectrum to the real-space observable
(Sects.~\ref{sec:real_space_2nd}, \ref{sec:other_2nd_order}). The broader the
filter, the stronger is the mixing of scales, and the higher is the
correlation.

For an observed data vector $\vec d = \{ d_i \}, i=1 \ldots m$,
the covariance matrix $\mat C$ is defined as
\begin{equation}
  C_{ij} = \langle \Delta d_i \Delta d_j \rangle =
     \langle d_i d_j \rangle - \langle d_i \rangle \langle d_j \rangle ,
  \label{covariance}
\end{equation}
where the brackets denote ensemble average.

In a typical cosmic shear setting, the data vector $\vec d$ consists of functions of
shear correlations (e.g.~the shear two-point correlation function at $m$ angular
scales $\theta_i$, or band-estimates of the convergence power spectrum
$P_\kappa$ at $m$ Fourier wave bands with centres $\ell_i$). Those functions are quadratic in
the observed galaxy ellipticity $\varepsilon$. The covariance then depends on
fourth-order moments of $\varepsilon$. From (\ref{eq:eps_eps_s_gamma}), one can
see that the covariance can be split into three terms: The shot noise, which is
proportional to $\langle | \varepsilon^{\rm s} |^2 \rangle^2 = \sigma_\varepsilon^4$,
and, in the absence of intrinsic galaxy alignment (Sect.~\ref{sec:ia}), only
contributes to the covariance diagonal; the cosmic variance term, which depends
on fourth moments of the shear; and a mixed term.

In particular the cosmic variance term is difficult to estimate since it
requires the knowledge of the non-Gaussian properties of the shear field.

\subsubsection{The Gaussian approximation}

The covariance of the convergence power spectrum $P_\kappa$ at an individual mode $\ell$
in the Gaussian approximation is the simple
expression \cite{1992ApJ...388..272K,1998ApJ...498...26K,2008A&A...477...43J}
\begin{equation}
  \langle (\Delta P_\kappa)^2 \rangle(\ell)
  = \frac{1}{f_{\rm sky} (2\ell + 1)} \left( \frac{\sigma_\varepsilon^2}{2\bar n} + P_\kappa(\ell) \right)^2 .
  \label{eq:cov_P_kappa}
\end{equation}
Here, the survey observes a fraction of sky $f_{\rm sky}$, with a number
density of lensing galaxies $\bar n$. 
The quadratic expression expands into shot-noise (first term),
cosmic variance (second term), and a mixed term. In this Gaussian
approximation, the fourth-order connected term of $\kappa$ is zero, and the cosmic variance
consists of products of terms second-order in $\kappa$.

Analytical expressions for the Gaussian covariance of real-space second-order
estimators have been obtained in \cite{SvWKM02,KS04,2009MNRAS.397..608S}. The
power-spectrum covariance for shear tomography is easily computed
\cite{2004MNRAS.348..897T}.

\subsubsection{Non-Gaussian contributions}

\Eref{eq:cov_P_kappa} can be extended to the case of a non-Gaussian convergence
field, with the next-leading terms depending on the trispectrum $T_\kappa$
\cite{1999ApJ...527....1S,2004MNRAS.348..897T}. Non-Gaussian evolution leads to
a further coupling of small-scale modes with long wavelength modes that are
larger than the observed survey volume. These super-survey modes were first
introduced as \emph{beat coupling} in \citet{2006MNRAS.371.1188H}, and later
modeled in the halo model framework as \emph{halo sample variance}
\citeaffixed{2009ApJ...701..945S,2013MNRAS.429..344K}{HSV;}. Contrary to the
other terms of the covariance that scale inversely with the survey area $f_{\rm
sky}$, the super-survey covariance decreases faster. Therefore it is important
for small survey areas \cite{2009ApJ...701..945S,2013PhRvD..87l3504T}.

% SSC in simulations http://cdsads.u-strasbg.fr/abs/2014PhRvD..89h3519L
% Minimizing SSC http://cdsads.u-strasbg.fr/abs/2014arXiv1405.2666T
% More SSC http://cdsads.u-strasbg.fr/abs/2012JCAP...04..019D

An alternative, non-analytic path is replacing the ensemble average in
(\ref{covariance}) by spatial averaging, and to estimate the covariance matrix
from a large enough number $n$ of independent $N$-body simulations. To compute
the inverse of this estimator, which is needed in the likelihood function (see
following section), the dimension of the data vector $m$ has to be smaller than
$n$ \cite{andersen03,HSS07}. To reach percent-level precision for the inverse,
$n$ has be much larger than $m$, which for future surveys with many tomographic
bins means that the number of required simulations will be at least a few times
$10^4$ \cite{2013MNRAS.tmp.1312T,2013PhRvD..88f3537D}. This was the path we
chose for CFHTLenS tomography
\cite{CFHTLenS-2pt-tomo,CFHTLenS-mod-grav,CFHTLenS-IA} and higher-order
statistics \cite{CFHTLenS-2+3pt}, and COSMOS \cite{SHJKS09}.

%%%%%%%%%%%%%%%%%%%%%%%%%%%%%%%%%%%%%%%%%%%%%%%%%%%%%%%%%%%%
\subsection{The likelihood function}
\label{sec:likelihood}
%%%%%%%%%%%%%%%%%%%%%%%%%%%%%%%%%%%%%%%%%%%%%%%%%%%%%%%%%%%%

To compare weak-lensing observations to theoretical predictions, one invokes a
likelihood function $L$ as the probability of the observed data $\vec d$ of
length $m$ given a model $M$ with a set of parameters $\vec p$ of dimension
$n$.

For simplicity, in most cases, the likelihood function is modeled as an $m$-dimensional multi-variate
Gaussian distribution,
\begin{equation}
\fl
L(\vec d | \vec p, M) 
=
(2 \pi)^{-m/2} |\mat C(\vec p, M)|^{-1/2}
\exp\left[ - \frac 1 2 \left( \vec d - \vec
    y(\vec p, M) \right)^{\rm t} \mat{C}^{-1}(\vec p, M) \left( \vec d - \vec
    y(\vec p, M) \right) \right].
  \label{eq:likelihood}
\end{equation}
The function $\vec y$ is the model prediction for the data $\vec d$, and depends on the model
$M$ and parameter vector $\vec p$.
This is only an approximation to the true likelihood function, which is unknown, since
shear correlations are non-linear functions of
the shear field, which itself is not Gaussian, in particular on small scales.

The true likelihood function can be estimated by sampling the distribution
using a suite of $N$-body simulations for various cosmological parameters.
Because of the high computation time, this has been done only for a restricted
region in parameter space
\cite{2009A&A...504..689H,2009A&A...505..969P,2011ApJ...742...15T}. For
weak-lensing peak counts (Sect.~\ref{sec:peak_counts}), using our fast
simulations \cite{LK15a}, we have sampled the true likelihood function and
compared this to the Gaussian and the copula likelihood \cite{LK15b}. The
copula is defined by transformed variables for which all one-dimensional pdfs
are Gaussian. This makes the multi-variate likelihood function more Gaussian,
but does not guarantee it \cite{2011PhRvD..83b3501S}.

The log-likelihood function can be approximated by a quadratic form, which is
the inverse parameter covariance at the maximum point, called the \emph{Fisher
matrix} \cite{KS69,TTH97}. The Fisher matrix has become a standard tool to
quickly assess the performance of planned surveys, or to explore the
feasibility of constraining new cosmological models, e.g.~\cite{DETF}. However,
one has to keep in mind that the Fisher matrix is often ill-conditioned, in
particular in the presence of strong parameter degeneracies, and its inversion
requires a very high precision calculating of theoretical cosmological
quantities, as we have shown in \citet{WKWG12}.

In most cases, the parameter-dependence of the covariance in
(\ref{eq:likelihood}) is neglected, since the compuation of the covariance is
very time-consuming, e.g.~when derived from $N$-body simulations. When
estimated from the data themselves, the cosmology-dependence of the covariance
is missing altogether. This is a good approximation, as was shown in
\citet{2009A&A...502..721E} and confirmed in \citet{CFHTLenS-2pt-notomo}, in
particular when only a small region in parameter space is relevant, for example
in the presence of prior information from other cosmological data.

%%%%%%%%%%%%%%%%%%%%%%%%%%%%%%%%%%%%%%%%%%%%%%%%%%%%%%%%%%%%
\subsection{Parameter estimation}
\label{sec:param_estim}
%%%%%%%%%%%%%%%%%%%%%%%%%%%%%%%%%%%%%%%%%%%%%%%%%%%%%%%%%%%%

Theoretical models of cosmic shear observables can depend on a large number of
parameters. Apart from cosmological parameters, a number of additional,
nuisance parameters might be included to characterize systematics, calibration
steps, astrophysical contaminants such as intrinsic alignment, photometric
redshift uncertainties, etc. The number of such additional parameters can get
very large very quickly and reach of the order a few hundred or even thousands, for example if
nuisance parameters are added for each redshift bin \cite{2008arXiv0808.3400B}.

When inferring parameter constraints within the framework of a given
cosmological model, one usually wants to estimate the probability of the
parameter vector $\vec p$ given the data $\vec d$ and model $M$. In a Bayesian
framework, this is the \emph{posterior} probability $\pi$, which is given via
Bayes' theorem as

\begin{equation}
  \pi(\vec p | \vec d, M) = \frac{L(\vec d | \vec p, M) P(\vec p | M)}{E(\vec d | M)},
  \label{eq:Bayes}
\end{equation}
which links the
posterior to the likelihood function (see previous section) via the
\emph{prior} $P$ and the \emph{evidence} $E$. In most cases, one wants to
calculate integrals over the posterior, for example to obtain the mean
parameter vector, its variance, or confidence regions.
Such integrals can be written in general as
\begin{equation}
I(h) = \int {\rm d}^n p \, h(\vec p) \pi(\vec p | \vec d, M),
\label{eq:I_h}
\end{equation}
where $h$ is a function of the parameter $\vec p$. To calculate the mean of the
$\alpha^{\rm th}$ parameter, $I(h) = \bar p_\alpha$, $h(\vec p) =
p_\alpha$. For the variance of $p_\alpha$, set $h(\vec p) = (p_\alpha - \bar
p_\alpha)^2$. For a confidence region ${\cal C}$ (e.g.~the 68\% region
around the maximum) $h$ is the characteristic function $1_{\cal C}$ of the
set ${\cal C}$, that is $h(\vec p) = 1$ if $\vec p$ is in $\cal C$, and 0 else. Note that
this does not uniquely define ${\cal C}$; there are indeed many different ways
to define confidence regions.

In high dimensions, such integrals are most efficiently obtained by
means of Monte-Carlo integration, in which random points are sampled from the
posterior density function. Many different methods exist and have been applied
in astrophysics and cosmology, such as Monte-Carlo Markov Chain \citep[MCMC; ]{cosmomc},
Population Monte Carlo (see Sect.~\ref{sec:PMC}),
or multi-nested sampling
\cite{2008MNRAS.384..449F}. Monte-Carlo sampling allows for very fast
marginalization, for example over nuisance parameters, and projection onto
lower dimensions, e.g.~to produce 1D and 2D marginal posterior constraints.

MCMC provides a chain of $N$ points $\vec p_j$, which under certain conditions
represent a sample from the posterior distribution $\pi$. Using this Markov
chain, integrals of the form (\ref{eq:I_h}) can be estimated as sums over the
$N$ sample points $\vec p_j$, 
\begin{equation}
  \hat I(h) = \frac 1 N \sum_{j=1}^{N} h(\vec p_j).
  \label{eq:I_h_MC}
\end{equation}
One caveat of this estimator is that in general, the samples are actually drawn
from $L(\vec d| \vec p, M) P(\vec p| M)$, or from the unnormalised posterior.
It turns out not to be trivial to estimate the normalisation (evidence) $E$
with MCMC. However, the evidence drops out when using as MCMC
method the very popular
Metropolis-Hasting accept-reject algorithm \cite{Metropolis53,Hastings70}. To
get to the next step in the Markov Chain $\vec p_{i+1}$ from a previous
position $\vec p_i$, only ratios of the posterior $\pi(\vec p_{i+1}| \vec d, M)
/ \pi(\vec p_{i}| \vec d, M)$ are involved. Thus, 
(\ref{eq:I_h_MC}) can be obtained without
the need to compute the parameter-independent normalisation.

Other Monte-Carlo sampling techniques might provide samples under a different
distribution, and (\ref{eq:I_h_MC}) has to be modified accordingly, see for example the following section
with the case of PMC.

Alternatively, in a frequentist framework, one can minimize the function
$\chi^2 = -2 \ln L$. Marginalisation can be performed with the so-called
profile likelihood method \cite{2014A&A...566A..54P}. Frequentist minimisation
is equivalent to Bayesian inference with flat priors on all parameters.

Cosmic shear using current data is sensitive to only a few cosmological
parameters, in particular $\Omega_{\rm m}$ and $\sigma_8$. Shear tomography is
beginning to obtain interesting results on other parameters such as $\Omega_K$,
or $w$. For parameters that are not well constrained by the data, for example
$\Omega_{\rm b}$ or $h$, the (marginal) posterior is basically given by the
prior density. Therefore, the prior should be chosen wide enough to not 
restrict other parameters, and to not result in overly optimistic constraints.

%%%%%%%%%%%%%%%%%%%%%%%%%%%%%%%%%%%%%%%%%%%%%%%%%%%%%%%%%%%%
\subsection{Population Monte Carlo (PMC)}
\label{sec:PMC}
%%%%%%%%%%%%%%%%%%%%%%%%%%%%%%%%%%%%%%%%%%%%%%%%%%%%%%%%%%%%

In \citet{WK09} and \citet{KWR10} we develop for cosmology Population Monte Carlo, a
sampling technique based on iterative importance sampling \cite{cappe:douc:guillin:marin:robert:2007,CGMR03}.
Importance sampling
provides samples under a posterior distribution $\pi$, but samples from a
different distribution, the so-called \emph{proposal} or \emph{importance function} $q$,
choosen to be a simple function from where samples can be generated easily.

To provide samples under $\pi$, we re-write (\ref{eq:I_h}) as
\begin{equation}
  I(h) = \int {\rm d}^n p \, h(\vec p) \pi(\vec p | \vec d, M)
       = \int {\rm d}^n p \, h(\vec p) \frac{\pi(\vec p | \vec d, M)}{q(\vec p)} q(\vec p).
  \label{eq:I_h_IS}
\end{equation}
This identity holds for any $q$ whose support includes the support of $\pi$, and for functions $h$ whose
expectation $I(h)$ is finite.

This expression is estimated by sampling under the proposal $q$.
The Monte-Carlo estimator (\ref{eq:I_h_MC}) is simply modified to account for the additional term in the integral,
\begin{equation}
  \hat I(h) = \frac 1 N \sum_{j=1}^{N} h(\vec p_j) w_j;
  \qquad w_j = \frac{\pi(\vec p_j | \vec d, M)}{q(\vec p_j)}.
  \label{eq:I_h_MC_IS}
\end{equation}
The $w_j$ are called importance importance weights.

To estimate the unnormalised posterior (\ref{eq:I_h_MC_IS}),
can be modified to include the self-normalised importance ratios
\begin{equation}
  \hat I(h) = \frac 1 N \sum_{j=1}^{N} h(\vec p_j) \bar w_j;
  \quad \bar w^t_j = \frac{w^t_j}{\sum_{i=1}^N \bar w^t_i},
  \label{eq:I_h_MC_IS_norm}
\end{equation}
where the $\bar w_j^t$ are the normalised importance weights. This circumvents
the necessity to compute the normalisation $E$. Contrary to the case of MCMC
however, PMC does provide a robust estimate of the evidence that comes at no
extra cost, as I will show below.

The performance of importance sampling depends strongly on the choice of $q$.
If the importance function does not well match $\pi$, many sampled points
receive a very low weight, leading to a very poor efficiency and large variance
of the estimator. PMC proposes to solve this problem by iteratively adapting
the importance function to the posterior: A sequence of importance functions
$q^t, t=1 \ldots T$ aims to approximate the target $\pi$.

The approximation is quantified by the Kullback-Leibler divergence, or relative
entropy $K$,
\begin{equation}
  K(\pi\|q^t) = \int {\rm d}^n p \, \log\left(\frac{\pi(\vec p | \vec d, M)}{q^t(\vec p)}\right) \pi(\vec p),
    \label{eqn:kdiv}
    \label{eq:KL}
\end{equation}
which is an asymmetric measure of the similarity between two distributions. If
$K$ is $0$, the two distributions are identical almost everywhere, whereas for
$K \rightarrow 1$, $\pi$ and $q$ are very different.
 
The PMC algorithm adjusts the density $q^t$ incrementally such that the
divergence decreases with progressive iterations, or $K(\pi\|q^{t+1}) < K(\pi\|q^t)$.

Following \citet{cappe:douc:guillin:marin:robert:2007}, we use a variant of the
expectation-maximization (EM) algorithm to obtain updated parameters that
determine the proposal function, for which we choose a mixture (weighted sum)
of Gaussian or Student-t distributions. The parameters are mixture weights,
mean, and covariance matrix for the Gaussian or Student-t components of the
mixture model. \citet{cappe:douc:guillin:marin:robert:2007} derived closed-form
solutions for these parameters, which are given as integrals over the posterior
$\pi$. In iteration $t+1$, these integrals are approximated by evaluating their
importance sampling estimator using the samples and importance weights from the
previous iteration $t$. The details of the algorithm are summarized in
\citet{WK09}.

\subsubsection{Convergence and effective sample size}

Formally, there is no convergence criterium for PMC.
Eq.~(\ref{eq:I_h_MC_IS_norm}) is an unbiased estimator of $I(h)$ if the support
of the proposal $q$ covers $\pi$. Unlike the case of MCMC, there is no burn-in
phase and no asymptotic convergence of the chain towards the postieror
distribution. However, for a badly adapted proposal, the estimate might be very
noisy. Therefore, a criterium is introduced that monitors the improvement of
the adaption. An estimate of $\exp(-K)$ is the so-called \emph{normalised
perplexity} for iteration $t$, $p^t$,
\begin{equation}
  p^t = \exp(H^t/N),
  \label{eq:perplexity}
\end{equation}
where $H^t$ is the Shannon entropy of the normalised weights,
\begin{equation}
  H^t = - \sum_{j=1}^N \bar w^t_j \log \bar w^t_j.
  \label{eq:H_N}
\end{equation}
The perplexity ranges between $0$ and $1$. Values of $p$ close to unity
indicate good agreement between the importance function and the posterior.

A stopping criterium for a PMC run can be defined to be the iteration for which
$p$ exceeds a pre-determined threshold. Typically, a final importance run with
a higher number of sample points than for each previous iteration is being
carried out using the proposal from the last iteration. This will be the
final sample use for inference, to estimate (\ref{eq:I_h_MC_IS_norm}).

A quantity related to the efficiency of PMC is the \emph{effective sample size}, ESS$^t$,
\begin{equation}
  \mathrm{ESS}^t = \left( \sum_{j=1}^{N}
    \left\{\bar{w}_j^t\right\}^2 \right)^{-1},
  \label{eq:ess}
\end{equation}
where $1 \le \mathrm{ESS}^t \le N$. The effective sample size can be
interpreted as the number of sample points with non-zero weight. The ESS can
directly be compared with the number of sampled points of a Markoc Chain
multiplied with the acceptance rate.

\subsubsection{Performance}

Fig.~\ref{fig:perpl_cosmo_all} shows perplexity and effective sample size as
number of sample points, for a PMC run using CMB anisotropy data from WMAP5
\cite{WMAP5-Hsinshaw08}. We sample seven cosmological parameters of a $w$CDM
model, see \citet{WK09} for details. PMC is run for $10$ iterations, using a
mixture Gaussian importance function. Each point on the figure shows the value
after the corresponding iteration. Each of the first $9$ iterations is performed
with $10,000$ sample points, except for the final one, which has a number five
times larger, to reduce Monte-Carlo noise.

\begin{figure}[!ht]

  \begin{center}
    \resizebox{0.8\hsize}{!}{
      \includegraphics{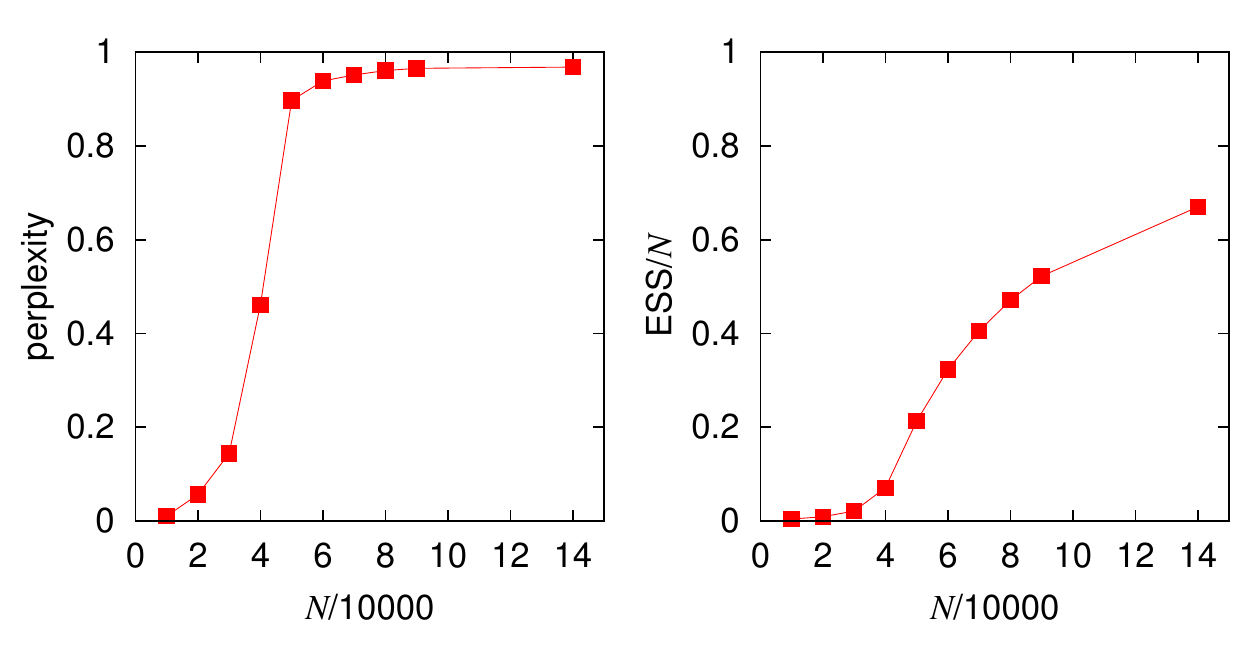}
    }
    \end{center}

    \caption{Perplexity (\emph{left panel}; eq.~(\ref{eq:perplexity})) and effective sample
      size ESS divided by the number of sample points, ESS/$N$ (\emph{right panel}; eq.~(\ref{eq:ess})), as a function of the cumulative
      sample size $N$. The likelihood is WMAP5 for a flat $\Lambda$CDM
      model with six parameters. From \citet{WK09}.  }
  \label{fig:perpl_cosmo_all}
\end{figure}

After five iterations, or $50,000$ sampled points, the perplexity reaches values
of $0.9$ and higher. The normalised ESS increases up to the last iteration and
exceeds $0.6$. This is much higher than the typical MCMC acceptance rate ot
$0.25$, even when taking into account the number of samples from all iterations
($130,000$ in total).

The number of samples needed for PMC is of the same order of magnitude as for
MCMC. The total computing time is therefore similar. However, importance
sampling separates the (typically time-consuming) evaluation of the posterior,
or the likelihood function, from the sampling. Obtaining the sample points (=
draws from Gaussian or Student-t distributions) for each iteration is very
fast. The computation of the importance weights, which
involves evaluation of the likelihood at the sampled points,
can then be performed in parallel. For a
number $n$ of CPUs, the wall-clock time gain is nearly $n$. 

In \citet{WK09} we show for a toy example in 10 dimension that PMC is capable to
well sample the tails of a narrow distribution. The variance of the estimator
(\ref{eq:I_h_MC_IS_norm}) for various functions $h$ such as the mean and
credible regions is typically smaller than for MCMC.

The implementation of Population Monte-Carlo for cosmology is available as the public software
\textsc{CosmoPMC}
\cite{cosmo_pmc_ascl}.

%%%%%%%%%%%%%%%%%%%%%%%%%%%%%%%%%%%%%%%%%%%%%%%%%%%%%%%%%%%%
\subsection{Model selection}
\label{sec:model_selection}
%%%%%%%%%%%%%%%%%%%%%%%%%%%%%%%%%%%%%%%%%%%%%%%%%%%%%%%%%%%%

The previous section discussed estimating parameters within a given
cosmological model, with the goal to measure mean and error bars of model
parameters. Taking one step back, one can ask the more fundamental
question what the best model is that describes the observations. This is a
qualitative different step and needs to compare different models, which is
independent of estimating parameters of those models. Such a comparison has to
account for the ability of models to describe the data, and the model
complexity. This is achieved naturally with posterior probabilities from a
Bayesian analysis.

The Bayesian evidence $E$ is the denominator in Bayes' theorem
(\ref{eq:Bayes}). Since the posterior, being a probability distribution function,
is normalised, this normalization is identical to $E$. Re-writing
(\ref{eq:Bayes}) and integrating yields the evidence as the integral over the
unnormalised posterior (likelihood $\times$ prior)
\begin{equation}
 E(\vec d | M)  \int {\rm d}^n p \, \pi(\vec p| \vec d, M) 
= E(\vec d | M)
 = \int {\rm d}^n p \, L(\vec d | \vec p, M) P(\vec p | M).
  \label{eq:evidence}
\end{equation}
This integral over the entire parameter space can be interpreted as a measure
of the overall model probability $M$ given the data $\vec d$.

The likelihood $L$ accounts for the goodness of the fit with respect to the data,
quantifying the data fidelity of the model. The better the agreement of the
data with the model, the higher the likelihood and thus the larger $E$ will be. 

The Bayesian evidence also crucially depends on the prior $P$. The larger the
parameter space, the smaller the amplitude of $P$ will be in general, since $P$
is a normalised probability distribution, and thus the smaller the evidence
becomes. This penalizes models that have a large parameter space, and that are thus
not very predictive: The predictability of a model $M$ given some data $\vec d$
only makes sense when compared to a prior knowledge. A model has low
predictability if it requires fine-tuning of parameters, i.e.~when the
posterior is very concentrated compared to the prior.

A good model is one for which the prior distribution of parameters closely
matches ``reality'', that is the posterior distribution obtained from a
measurement $\vec d$. This is true in particular when the prior is obtained
from first principles or physical knowledge of the system in question. To use
the Bayesian evidence as model probabilty, this needs to be the case, since the
prior is an integral part of the model, and thus has to be physically
well-motivated instead of choosen in an ad-hoc way \cite{2008MNRAS.388.1314E}.
For arbitrary priors, the results of Bayesian model comparison are arbitrary.

The Bayesian evidence has thus a built-in Occam's razor term, penalizing models
with high complexity (in the form of having many additional parameters, or a
large, ``wasted'' parameter space), see also \citet{berger:jefferys:1992} and
\citet{mackay:2002}.

To compare two models $M_1$ and $M_2$, we use the Bayes factor \cite{jeffreys:1939},
\begin{equation}
	B_{12} = \frac{E_1}{E_2} := \frac{E(\vec d|M_1)}{E(\vec d|M_2)}.
	\label{eq:Bayes_factor}
\end{equation}
If $B_{12}$ is larger (smaller) than unity, the data favour model $M_1$ ($M_2$)
over the alternative model. To quantify the ``strength of evidence'' contained
in the data, \citet{jeffreys:1961} introduced an empirical scale, see Table
\ref{tab:jeffrey}. This is only a rough guideline for decision making and of
course the proposed boundaries are mostly subjective, albeit connected with
information theory. I refer the reader to \citet{2008ConPh..49...71T} for a
pedagogical review about the Bayesian evidence and its use for model
comparison.

\begin{center}\begin{table}% [hb!]
  \caption{Jeffrey's scale to quantify the `strength of evidence' for
    a corresponding range of the Bayes factor $B_{12}$ in (\ref{eq:Bayes_factor}), assuming $B_{12} > 1$. \\}
  \label{tab:jeffrey}
  \begin{tabular}{lll} \hline\hline
    $\ln B_{12}$ & $B_{12}$ & Strength \\ \hline
    $< 1$        & $< 2.7$ & Inconclusive \\
    $1 \ldots 2.5$ & $2.7 \ldots 12$ & Weak \\
    $2.5 \ldots 5$ & $12 \ldots 150$ & Moderate \\
    $>5$           & $>150$ & Strong
  \end{tabular}
\end{table}\end{center}

The integral (\ref{eq:evidence}) is readily approximated with the importance sampling estimator
(\ref{eq:I_h_MC_IS}), where the function $h$ is unity,
\begin{equation}
  \hat E(\vec d, M) =
  \frac 1 N \sum_{j=1}^N w_j.
  \label{eq:ISEstimOfE}
\end{equation}
Note that here the unnormalised importance weights are used.

Below I show an example from \citet{KWR10}, comparing inflatonary models of
the power spectrum of primordial perturbations. I use data from CMB anisotropies
\citep{WMAP5-Hsinshaw08}, SNIa \citep{kowalski-2008} and BAO
\citep{2005ApJ...633..560E}.

I model the primordial scalar perturbations power spectrum as
\begin{equation}
  P_\delta(k) \propto k^{n_{\rm s} + \frac 1 2 \, \alpha_{\rm s} \ln (k/k_0)},
\end{equation}
with the parameters $n_{\rm s}$ being the scalar spectral index, and
$\alpha_{\rm s}$ the ``running'' of the index.
The pivot scale $k_0$ is
fixed to $k_0 = 0.002$
Mpc$^{-1}$. In addition, tensor modes (gravitational waves) have the power
spectrum
\begin{equation}
  P_{\rm t}(k) \propto k^{n_{\rm t}},
\end{equation}
with tensor spectral index $n_{\rm t}$. The ratio between tensor and
scalar perturbation spectra at scale $k_0$ is denoted by $r$. 

The standard $\Lambda$CDM model $M_2$ has $\alpha_{\rm s} = n_{\rm t} = r = 0$,
with only $n_{\rm s}$ being a free parameter. Against $M_2$ we test various
models $M_1$, where we keep combinations of $\alpha_{\rm s}$ and $r$ free. The
tensor index $n_{\rm t}$ is unconstrained by current data and therefore not
included. In addition, we test the Harrison-Zel'dovich model of a scale-free
power spectrum with $n_{\rm s}$. This model has meanwhile been ruled out at the
$5\sigma$ level by Planck \cite{2015arXiv150201589P} for the standard
$\Lambda$CDM model, but this significance decreases for extended models.

To obtain physically motivated priors on the power-spectrum parameter, we
consider the slow-roll approximation of inflation. This approximation provides
an infinite hierarchy of flow equations describing the dynamics of the single
scalar field which drives inflation \cite{2006JCAP...07..002P}. The slow-roll
parameters to first order are $\varepsilon$ and $\eta$, which are given in
terms of the inflaton potential and Hubble parameter during inflation.

Slow-roll conditions are satisfied when $0 \le \varepsilon \le 0.1$ and $|\eta|
\le 0.1$. Although the numerical values are approximate, they are natural
limits for the validity of the Taylor-expansion of the power spectrum $P(k)$ in
$\ln(k/k_0)$, see \citet{2006JCAP...08..009M}. From these limits, and their
relations  to the power-spectrum parameters we derive priors on the latter, see
Table \ref{tab:priors_primordial}. We choose uninformative (flat) priors for
simplicity. With that we have the ingredients for a meaningful Bayesian model
comparison analysis, which are well-defined models with physically motivated
priors.

\begin{table}% [hb!]
  \caption{Prior ranges for primordial model comparison. The prior
    ranges for
    primordial parameters are derived from the slow-roll approximation. \\}
  \label{tab:priors_primordial}
  \begin{tabular}{l|l|l|l}
    Parameter & Description & Min. & Max. \\ \hline
    $n_{\rm s}$  & Scalar spectral index & 0.39   & 1.2 \\
    $\alpha_{\rm s}$ & Running of spectral index & -0.2 & 0.033 \\
    $r$ (lin.~prior) & Tensor-to-scalar ratio & 0 & 1.65 \\
    $\ln r$ (log.~prior)   & Tensor-to-scalar ratio & -80 & 0.50 \\
 \end{tabular}
\end{table}

In Fig.~\ref{fig:evi_primordial} we show the Bayes factor of various models
$M_1$ with respect to the standard, reference model $M_2$ (the flat $\Lambda$CDM
universe with $n_{\rm s}=\mbox{const}$), as function of the number of parameters
$n_{\rm par}$ for each model $M_1$. A running spectral index is favoured
weakly, all other cases are disfavoured. The evidence against the
Harrison-Zel'dovich model ($n_{\rm s}=1$) is only weak, even though the
measured value of $n_{\rm s} = 0.9622 \pm 0.0145$ is inconsistent with unity at
the $2.6\sigma$ level for this data set. This shows the importance of Bayesian
model comparison when compared to the significance of marginalised error bars using a
single model, when making statements about ruling out models.

Tensor perturbations are moderately disfavoured. However, as example of the
influence of the prior choice, we use in addition a model with a flat prior in
$\ln r$ instead of $r$. As lower limit we choose $-80$, corresponding to the energy scale
of Big Bang Nucleosynthesis as a conservative lower limit of the inflation
energy scale \citep{2006PhRvD..73l3523P}. The large prior of the logarithmic
tensor-to-scalar ratio causes this model to be strongly disfavoured.

\begin{figure}% [!tb]

  \begin{minipage}{0.65\textwidth}
  \resizebox{1.0\hsize}{!}{
   \includegraphics[bb=65 0 300 250]{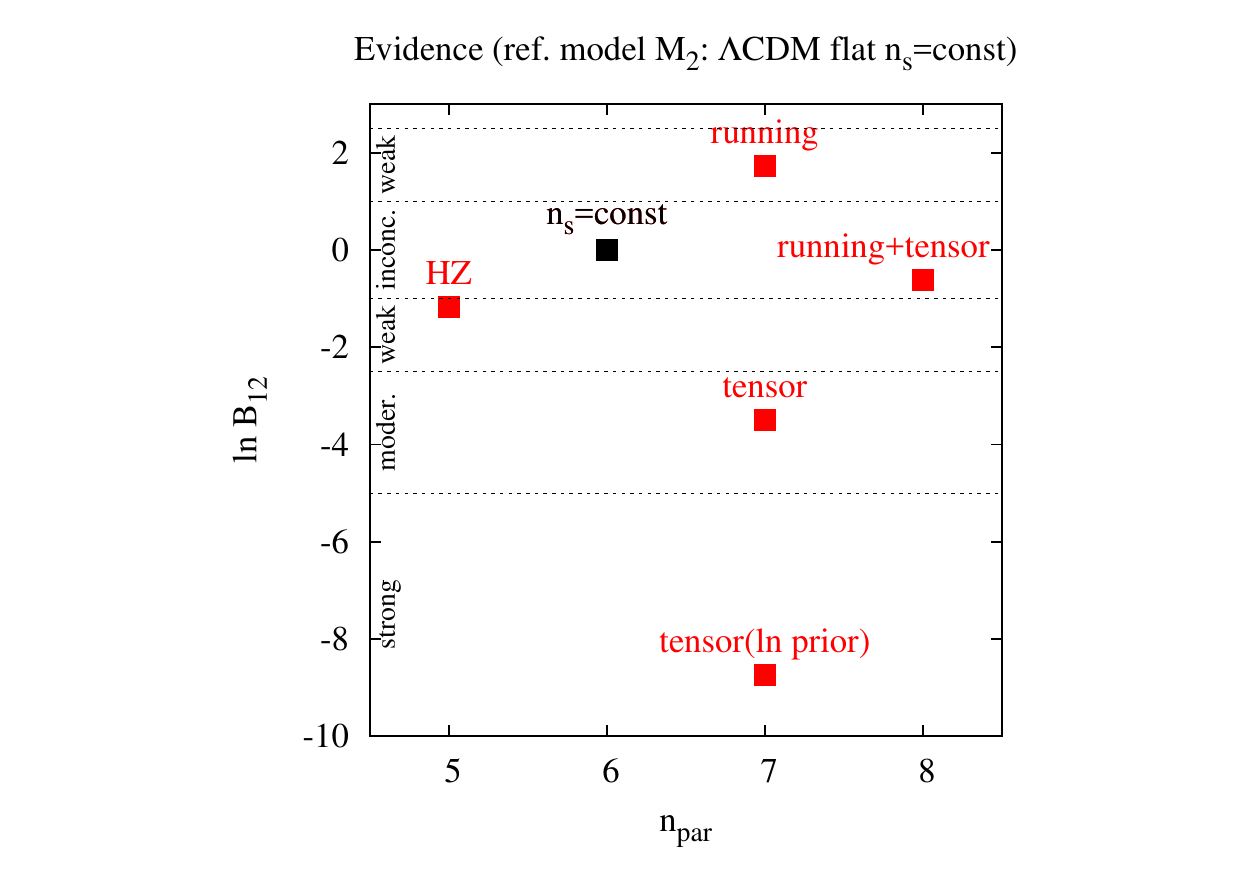}
  }
  \end{minipage}%
  \hspace*{-0.1\textwidth}%
  \begin{minipage}{0.45\textwidth}

  \caption{Ratio of evidences, or Bayes factor, for various models ${M}_1$ with respect to the
    reference model ${M}_2$, a flat $\Lambda$CDM universe with constant
    $n_{\rm s}$. From \citet{KWR10}.}
  \label{fig:evi_primordial}

  \end{minipage}%
\end{figure}

%%%%%%%%%%%%%%%%%%%%%%%%%%%%%%%%%%%%%%%%%%%%%%%%%%%%%%%%%%%%
\subsection{Approximate Bayesian Computation (ABC)}
\label{sec:ABC}
%%%%%%%%%%%%%%%%%%%%%%%%%%%%%%%%%%%%%%%%%%%%%%%%%%%%%%%%%%%%

Approximate Bayesian Computation \cite{2011arXiv1101.0955M,2012MNRAS.425...44C}
is a so-called likelihood-free statistical inference method. That means the
evaluation of a likelihood function is bypassed to obtain samples from the
posterior distribution $\pi$. ABC allows to do parameter inference in cases
where the likelihood function is unknown, or  too expensive (time-consuming)
to evaluate. It is thus ideal for non-Gaussian observables, where the approximation
of a multi-variate Gaussian for the likelihood might be inappropriate. ABC allows
us to test this assumption.

The requisite for ABC is a stochastical process that provides simulated
realisations $\{\vec x_i\}$ of the data $\vec d$ for a
given parameter $\vec p$ and model class $M$. These simulations are the model
prediction for that parameter; the model is here not a single, deterministic
vector $\vec y(\vec p|M)$ as before, but a stochastic random variable drawn
from a distribution.

ABC requires that this distribution is the true underlying probability
distribution, or likelihood $L$, of the observable $\vec d$. Then, the
simulations $\vec x_i$ are a sample under $L$, and the density of the sample
points represent an estimate of $L$ of a given parameter $\vec p$ and model
$M$. For parameter inference, we need to know the probability of the data,
$L(\vec d| \vec p, M)$ given $\vec p$ and $M$. that we observed $\vec d$. How
do we get this probabilty without calculating $L(\vec d)$?

This is best illustrated for discrete data: If we have $N$ model prediction
realisations $\{
\vec x_i \}$ for a parameter $p$ of some discrete data $\vec d$, the
probability of observing $\vec d$ is the number of realisations $r$ for which
$\vec x_i = \vec d$, divided by $N$. Interestingly, this is a frequentist
interpretation of probability, which is the number of matches over the total
number of events.

We can write this relative frequency of matches as sum over the distribution
(the likelihood $L$) of $\vec x_i$ times the Kronecker delta to only select matches.
If in addition the parameter $\vec p$ is drawn from the prior distribution $P(\vec p|M)$, the
the distribution of matching events, which we call $\hat \pi_{\rm ABC}$, is
\begin{equation}
  \hat \pi_{\rm ABC}(\vec p | \vec d, M) := \sum_{i=1}^N L(\vec x_i| \vec p, M) P(\vec p|M)
  \delta_{\vec x_i, \vec d} = L(\vec d| \vec p, M) P(\vec p|M).
  \label{eq:pi_ABC_discr}
\end{equation}
This justifies thus the use of the symbol $\pi$, since the result is an unbiased estimator of the
(unnormalised) posterior of $\vec p$!

An additional step is performed with ABC: The number of realisations $N$ can be
as small as one. This might sound surprising at first, since a single
realisation $\vec x$ is not likely to match $\vec d$, so $\hat \pi_{\rm
ABC}$ can be zero. But the probability of a match is equal to $L(\vec x| \vec
p, M)$, independent of the number of realizations $N$,
so the expectation value of (\ref{eq:pi_ABC_discr}) is still an
unbiased estimator of $\pi$ even for $N=1$. The large variance of this one-sample
estimator is
compensated by the large number of sampled parameters $\vec p$ that are typically
explored in sampling of the posterior. It turns out that the overall sampling
(or Monte-Carlo) noise does not increase for $N=1$. This one-sample test
leads to an \emph{accept-reject} algorithm.

The application of ABC to continuous data requires further adaptations. Since
strict equality between two continuous variables are practically not possible,
sampled points $\vec x_i$ are accepted when they fall within some
\emph{tolerance level} $\varepsilon$ of $\vec d$. For multi-variate data, this
also requires a metric $D$ that can be compared to $\varepsilon$. In addition,
the complexity of high-dimensional data is typically reduced to a
lower-dimensional space using a summary statistic $\vec s$ of the data. Thus, a
model $\vec x$ is accepted if
\begin{equation}
  D[\vec s(\vec d), \vec s(\vec x)] < \varepsilon; \quad \mbox{or equivalently} \quad \vec x \in {\cal D}_\varepsilon[\vec s(\vec d)],
  \label{eq:ABC_acceptance}
\end{equation}
where ${\cal D}_\varepsilon(\vec z)$ is the $q$-dimensional ball with radius $\varepsilon$ centred on $\vec z$.

The accepted points follow a distribution that is a modified version of (\ref{eq:pi_ABC_discr}),
\begin{equation}
    \pi_{{\rm ABC}, \varepsilon}(\vec p| \vec d, M) = L_\varepsilon(\vec d| \vec p, M) P(\vec p|M),
  \label{eq:pi_ABC_eps}
\end{equation}
where $L_\varepsilon(\vec d|M)$ is the probability that a proposed parameter
  $\vec p$ passes the one-sample tolerance test (\ref{eq:ABC_acceptance}),
\begin{equation}
  L_\varepsilon(\vec d | \vec p, M) \equiv \int {\rm d}^n x \, L(\vec x | \vec p, M) \mathbf{1}_{{\cal D}_\varepsilon[\vec s(\vec d)]}[\vec s(\vec x)] .
  \label{eq:ABC_accept}
\end{equation}
The sum over discrete events $\vec x_i$ is now an integral over models, and the Kronecker delta has been replaced with the indicator function $\mathbf{1}_A(\vec x)$,
which is unity if $\vec x \in A$, and zero otherwise.
This accept-reject algorithm is illustrated in Fig.~\ref{fig:ABC_scheme}.

\begin{figure}% [!tb] 

  \centerline{
  \resizebox{0.8\hsize}{!}{
   \includegraphics{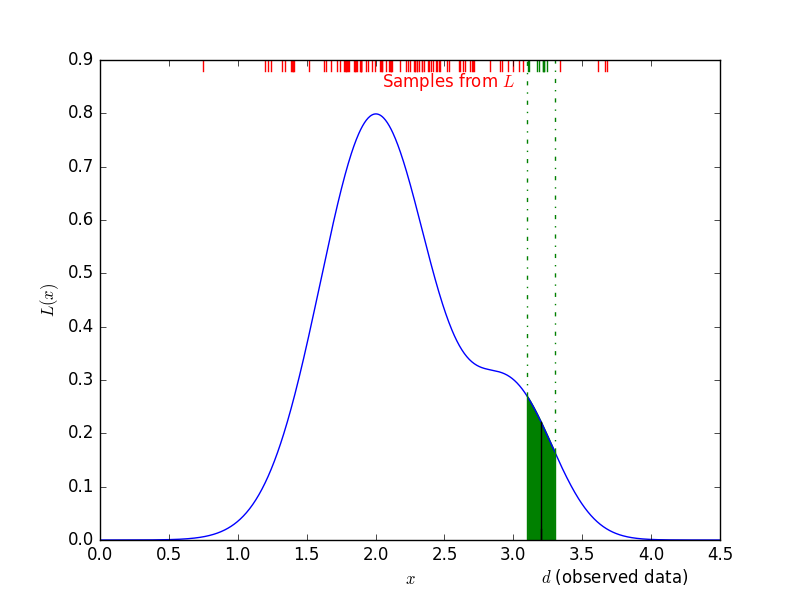}
  }
  }

  \caption{Illustration of the ABC accept-reject method in one dimension. Sample points (with positions indicated at the top)
					 are sampled from the underlying likelihood function $L$ of some observable $x$. The observed data
					 point marked at $d$. We want to obtain the likelihood function at the position of the data, $L(x=d)$. A Monte-Carlo
					 estimate of $L(d)$ is given by the density of sample points at $d$. ABC approximates this density by defining a tolerance
					 $\varepsilon$ around $d$, and counting the number of points within this limit, normalised by the total number of samples.
					 In a one-sample test limit, this frequency is the acceptance probability, which is equal to the green area divided by
					 the total area under the $L$ curve.
					 }
  \label{fig:ABC_scheme}
\end{figure}

The main assumption of ABC is now that the probability distribution function
(\ref{eq:ABC_accept}) is a good approximation of the true, underlying likelihood
function $L$ of the data $\vec d$,
\begin{equation}
  L_\varepsilon(\vec d | \vec p, M) \approx L(\vec d | \vec p, M).
\end{equation}
and consequently the ABC posterior (\ref{eq:pi_ABC_eps}) an approximation of the true posterior $\pi$.

This assumption relies on the three ingredients for ABC sampling: the summary
statistic $\vec s$, the distance function $D$, and the tolerance $\varepsilon$.
Note that traditional Monte-Carlo likelihood sampling approaches also depend on
the choice of a summary statistic: In most cases the size of the observed data
$\vec d$ is huge (e.g.~CMB pixelised maps or time series, weak-lensing galaxy
shape catalogues, SNIa lightcurve time measurements), and is typically reduced
to a much smaller observable (e.g.~power spectrum, correlation function,
magnitude + stretch + color), by a mapping $\vec s(\vec d)$.

\citet{McKinley09} showed that the choice of a summary statistic and distance
function are of great importance for the ABC performance. In \citet{LK15b} we
only used one summary statistic, namely $\vec s(\vec x) = \vec x$, where $\vec
x$ is the (already reduced) data vector. In our case of weak-lensing peak
counts (Sect.~\ref{sec:peak_counts}), $\vec x$ was chosen to be the number of
peaks $\vec n = (n_1, n_2, \ldots)$ (peak abundance or pdf) as function of peak
SNR $\nu_i$.

\citet{LKS16} compared two different distance functions $D$, which are the
square root of scalar products of the two vectors $\vec x$ and $\vec d$. One
distance, $D_2$, uses the full data covariance matrix to compute the scalar
product, the other one, $D_1$, only weighs the data by the variance (diagonal
of the covariance). Note that ABC with the distance $D_1$ does not necessarily
neglect all correlation between data points --- these are included
automatically in the simulated model predictions under the joint, multi-variate
likelihood function $L$. These correlations are only neglected under $D_1$ when
deciding whether a model $\vec x$ for a parameter $\vec p$ is close enough to
$\vec d$ for $\vec p$ to be accepted.

We found that $D_1$ and $D_2$ give very similar results when data points are
weakly correlated, but $D_1$ provided overly tight constraints on parameters for
highly correlated data, with under-estimation of the error of $\Sigma_8$ by 40\%.

The tolerance $\varepsilon$ is best set in an iterative approach. This
automatically solves the problem of fixing the tolerance a priori: If
$\varepsilon$ is too large, too many points are accepted. In that case,
$L_\varepsilon$ (\ref{eq:ABC_accept}) approaches unity, and ABC effectively
provides samples from the prior $P$. If $\varepsilon$ is too close to zero, so
is $L_\varepsilon$, and sampling becomes very inefficient, with the
overwhelming majority of points being rejected. In \cite{LK15b} we use an
iterative importance-sampling algorithm together with ABC. This sequential
Monte-Carlo (SMC) ABC method has similarities with population Monte Carlo
(Sect.~\ref{sec:PMC}), and is therefore also calles PMC ABC.

% but we used
%three different choices for the (already reduced) data vector $\vec x$, and
%compared the results. In our case of weak-lensing peak counts
%(Sect.~\ref{sec:peak_counts}) $\vec x$ was chosen to be (i) the number of peaks
%$n_i$ (peak abundance or pdf) as function of peak SNR $\nu_i$; (ii) the SNR
%values $\nu_i$ at given percentiles of the cumulative peak number (cdf)
%$\sum_i=0^j n_j$; (iii) the same as (ii) but with a minimal SNR of $\nu_{\rm
%min} = 3.5$.

% Chapter 6
% measurement.tex

%%%%%%%%%%%%%%%%%%%%%%%%%%%%%%%%%%%%%%%%%%%%%%%%%%%%%%%%%%%%
\section{Measuring weak lensing}
\label{sec:measuring_wl}
%%%%%%%%%%%%%%%%%%%%%%%%%%%%%%%%%%%%%%%%%%%%%%%%%%%%%%%%%%%%

%%%%%%%%%%%%%%%%%%%%%%%%%%%%%%%%%%%%%%%%%%%%%%%%%%%%%%%%%%%%
%\subsection{Data analysis methods for weak lensing}
%\label{sec:data_analysis}
%%%%%%%%%%%%%%%%%%%%%%%%%%%%%%%%%%%%%%%%%%%%%%%%%%%%%%%%%%%%

The cosmological interpretation of cosmic shear measurements requires the observation
of large and deep sky areas in superb
image quality, together with sophisticated image analysis methods.
A significant detection of cosmic shear requires a very large number of
galaxies to high redshifts (of order unity) and low signal-to-noise ratios (down to $10$ - $15$).
The shapes of those faint galaxies
have to be measured with high accuracy.
Galaxy images have to be corrected for the \emph{point-spread function}
(PSF). The PSF is the combined effect of the imaging system consisting of the
atmosphere (for ground-based surveys), telescope optics, and detector. To
estimate the PSF, a very pure sample of stars, uncontaminated by small
galaxies, has to be selected.
Shape measurements typically have to be
calibrated. To ensure measurement biases small enough compared to the
statistical errors, large sets of realistic image simulations need to be used.
In addition, the interpretation of shape correlations depends crucially on the
redshift distribution of the lensed galaxy sample, see (\ref{eq:kappa}).
Multiple optical band observations have to be used to estimate photometric
redshifts.
This section gives a brief overview of the methods for weak lensing measurements I have used in my work.

%%%%%%%%%%%%%%%%%%%%%%%%%%%%%%%%%%%%%%%%%%%%%%%%%%%%%%%%%%%%
\subsection{Galaxy shape measurement}
\label{sec:shapes}
%%%%%%%%%%%%%%%%%%%%%%%%%%%%%%%%%%%%%%%%%%%%%%%%%%%%%%%%%%%%

%%%%%%%%%%%%%%%%%%%%%%%%%%%%%%%%%
%\subsubsection{Direct estimation}
%\label{sec:shapes_direct}
%%%%%%%%%%%%%%%%%%%%%%%%%%%%%%%%%

The first family of shape measurement methods is moment-based direct estimation.
One of the most widely used methods of this type is \emph{KSB}
\nocite{1995ApJ...449..460K} (Kaiser, Squires \& Broadhurst 1995). KSB measures
ellipticity directly on an image using weighted second-order moments of the
galaxy brightness distribution. The convolution of the image with the PSF is
approximated by linear operations on the ellipticities; it is therefore a
perturbative method.
Alternatively, instead of correcting the ellipticities as in KSB, the
PSF deconvolution can be done directly on the moments of the galaxy light
distribution \cite{2000ApJ...536...79R}. This is more rigorously explored 
by \emph{DEIMOS} (deconvolution of image moments), a truncated hierarchy of
higher-order moments in \citet{2011MNRAS.412.1552M}.

KSB was used in our cosmic-shear measurements \cite{FSHK08,SHJKS09}.
Deconvolution in moment space, inspired by DEIMOS, but using unweighted moments
on denoised images, was one of my contributions to the GREAT3 challenge.

%%%%%%%%%%%%%%%%%%%%%%%%%%%%%%%%%
%\subsubsection{Model fitting}
%\label{sec:shapes_model}
%%%%%%%%%%%%%%%%%%%%%%%%%%%%%%%%%

A second large family of shape measurement methods is based on model fitting.
These indirect methods assume a model for the surface brightness $I$, including
ellipticity parameters, and fit the model to the observed image. One of the
advantages of such \emph{forward-fitting} methods is the straightforward
treatment of the PSF: The model is more easily and robustly convolved with the PSF 
than the observed (pixellised, noisy, maybe partially masked) image is deconvolved.
A fully Bayesian forward-fitting method is
\emph{lens}fit, which measures the posterior distribution of ellipticity for
galaxies on individual exposures, and combines the results in a Bayesian way
without information loss
\cite{2007MNRAS.382..315M,2008MNRAS.390..149K,CFHTLenS-shapes}. A further
notable model-fitting method that I have used is \emph{gfit}
\cite{2012arXiv1211.4847G}. \emph{lens}fit turned out to be the superior method
for CFHTLenS. \emph{gfit} was used as main method by the CEA-EPFL team in the
GREAT3 challenge, and is used in on-going work on shear calibration
\cite{pujol_shear_bias_17}, and for the analysis of the Canada-France Imaging Survey (CFIS),
see Sect.~\ref{sec:upcoming_surveys}.

%%%%%%%%%%%%%%%%%%%%%%%%%%%%%%%%%%%%%%%%%%%%%%%%%%%%%%%%%%%%
\subsubsection{Shape measurement biases}
\label{sec:shape_biases}
%%%%%%%%%%%%%%%%%%%%%%%%%%%%%%%%%%%%%%%%%%%%%%%%%%%%%%%%%%%%

One can make the very general statement that the non-linear dependence of
ellipticity estimators on the light distribution in the presence of noise
creates a bias, the so-called \emph{noise bias}. This bias has been
investigated for methods based on moments \cite{2004MNRAS.353..529H} and model
fitting \cite{2012MNRAS.425.1951R,2014MNRAS.441.2528K}.

A further source of bias is related to incorrect assumptions about the light
distribution, the so-called \emph{model bias} \cite{2010MNRAS.404..458V}. Not
only model-fitting techniques, but also direct methods are still not free from
such assumptions: For example, moment-based methods require weight functions
whose profile and size should match the observed images, and in fact, direct
and indirect methods show very similar biases related to galaxy morphology
\cite{pujol_shear_bias_17}.

Shape biases can be characterised to first approximation by a multiplicative
component $\vec m$, and and additive term $\vec c$. These bias parameters are
given by the relation between observed and true ellipticity
\cite{2006MNRAS.366..101H,STEP1},
\begin{equation}
  \varepsilon_i^{\rm obs} = (1 + m_i) \varepsilon_i^{\rm true} + c_i; \quad i = 1, 2.
  \label{eq:ell_bias}
\end{equation}
The shear biases $\vec m$ and $\vec c$ are generally functions of galaxy
properties and redshift. Current shape measurement methods provide shear
estimates with resudial (after calibration) $m$ at the $1$ to few percent
level, and $c$ between $10^{-3}$ and $10^{-2}$. Typically, the calibration of
measured shears is performed using large image simulations
(Sect.~\ref{sec:image_sims}). Recent work has been looking into calibration by
using the data themselves, the so-called \emph{meta-calibration} approach
\cite{2017arXiv170202600H,2017ApJ...841...24S}.
Future surveys require the accuracy of calibrated
shapes to be on the order of $0.1$\%
\cite{2006MNRAS.366..101H,2013MNRAS.429..661M,2013MNRAS.431.3103C}, see
Sect.~\ref{sec:upcoming_surveys}.

% Second effect: effective weight function W changes with ellipticity: W is in parts given by PSF size.
% Elliptical galaxies might have information on smaller scales than round ones, which are then weighted
% differently by the PSF [B10].

% Also: need to know for calibration dP(e)/dg, how shear transforms ellipticity
% measurements.

%The galaxy shape measurement is still a very active, albeit very technical,
%area of research since our ability to analyze future lensing surveys strongly
%depend on it.

%%%%%%%%%%%%%%%%%%%%%%%%%%%%%%%%%%%%%%%%%%%%%%%%%%%%%%%%%%%%
\subsection{PSF correction}
\label{sec:PSF}
%%%%%%%%%%%%%%%%%%%%%%%%%%%%%%%%%%%%%%%%%%%%%%%%%%%%%%%%%%%%

To estimate the PSF at the position of a galaxy, one has to select stars on the
image, measure their shape, and interpolate the resulting PSF to the position
of the galaxy. This requires a sample of suitable stars,
i.e.~without saturated pixels, not hit by cosmic rays, and uncontaminated by
galaxies. A common selection criteria is the identification of the
\emph{stellar locus} in a size-magnitude diagram. This is a region of bright
and small objects that is relatively well isolated from resolved galaxies and
unresolved, dim objects such as very faint galaxies and detection artefacts.
Additionally, colour information can be added to classify stars and galaxies.

The PSF (in form of parameters, a pixellised vignette, or a high-resolution
model), is then interpolated onto the galaxy position. For ground-based
observations in the past, this has usually been done with a two-dimensional
polynomial or a rational function. For mosaic multi-CCD cameras,
discontinuities between chips are common and have to be accounted for in the
PSF model, for example by performing fits on each chip individually
\cite{CFHTLenS-shapes}.

%%%%%%%%%%%%%%%%%%%%%%%%%%%%%%%%%%%%%%%%%%%%%%%%%%%%%%%%%%%%
\subsection{Error modelling and residual systematics}
\label{sec:error-model}
%%%%%%%%%%%%%%%%%%%%%%%%%%%%%%%%%%%%%%%%%%%%%%%%%%%%%%%%%%%%

Any weak lensing data analysis must be completed with a robust error modeling.
This step is necessary to quantify any residual systematics caused by an
imperfect PSF correction, since those residuals can mimic a cosmological
signal.

%%%%%%%%%%%%%%%%%%%%%%%%%%%%%%%%%%%%%%%%%%%%%%%%%%%%%%%%%%%%
\subsubsection{Star-galaxy correlation}
\label{sec:star_gal_corr}
%%%%%%%%%%%%%%%%%%%%%%%%%%%%%%%%%%%%%%%%%%%%%%%%%%%%%%%%%%%%

The most commonly used approach is a null test of the correlation
between the stellar ellipticities $\varepsilon^\star$ (before PSF correction)
and the corrected galaxy shapes $\varepsilon$.  This star-galaxy ellipticity
correlation function is defined as
\begin{equation}
  \xi_{\rm sys}=\langle \varepsilon^\star \varepsilon\rangle .
  \label{eq:xi_sys}
\end{equation}
Note that a non-zero signal on a small region on the sky could come from chance
alignments between PSF pattern and a coherent shear from large-scale structure.
To interpret measurements of $\xi_{\rm sys}$, this cosmic variance contribution
needs to be accounted for \cite{CFHTLenS-sys}.

%%%%%%%%%%%%%%%%%%%%%%%%%%%%%%%%%%%%%%%%%%%%%%%%%%%%%%%%%%%%
\subsubsection{PSF model - residual correlation}
\label{sec:PSF_model_res_corr}
%%%%%%%%%%%%%%%%%%%%%%%%%%%%%%%%%%%%%%%%%%%%%%%%%%%%%%%%%%%%

Two correlation functions quantify the PSF model. These are the auto-correlation
of the PSF residuals, $D_1$, and the cross-correlation between PSF and PSF residuals,
$D_2$ \cite{2010MNRAS.404..350R}. These two correlation functions are defined as
\begin{eqnarray}
  D_1(\theta) & = \left\langle \left( \varepsilon - \varepsilon_{\rm m} \right)^\ast
                                 \left( \varepsilon - \varepsilon_{\rm m} \right) \right\rangle(\theta) ;
    \nonumber \\
  D_2(\theta) & = \left\langle \varepsilon^\ast \left( \varepsilon - \varepsilon_{\rm m} \right) +
                                 \left( \varepsilon - \varepsilon_{\rm m}\right)^\ast \varepsilon 
                                 \right\rangle(\theta) .
  \label{eq:D1D2}
\end{eqnarray}
Here, $\varepsilon$ is the observed ellipticity of a star, and
$\varepsilon_{\rm m}$ the PSF model ellipticity\footnote{In this sub-section, we
drop the super-script `$^\star$' to detnote ellipticty measured for stars; I
remind the reader that the different symbol `$^\ast$' denotes complex
conjugation.}. These functions do not only measure the amount of residuals
$\left( \varepsilon - \varepsilon_{\rm m} \right)$, but the spatial correlation
of residuals. This can be introduced by a PSF model that does not well
represent the spatial variation of the PSF over the detector. In case of a
perfect PSF model, both correlation functions are expected to vanish
identically, $\langle D_1 \rangle = \langle D_2 \rangle = 0$.

\citet{2010MNRAS.404..350R} writes the observed ellipticity $\varepsilon$ of a
galaxy as the sum of true ellipticity $\varepsilon_{\rm t}$
and noise $N$,
\begin{equation}
  \varepsilon = \varepsilon_{\rm t} + N,
\end{equation}
and the model ellipticity $\varepsilon_{\rm m}$ as sum of true ellipticity
$\varepsilon_{\rm t}$ and model uncertainty $m$,
\begin{equation}
  \varepsilon_{\rm m} = \varepsilon_{\rm t} + m .
\end{equation}
The two correlation functions can then be written as
\begin{eqnarray}
  D_1(\theta) & = & - \left\langle m^\ast N + N^\ast m \right\rangle(\theta)
                    + \left\langle m^\ast m \right\rangle(\theta);
    \nonumber \\
  D_2(\theta) & = & - \left\langle m^\ast N + N^\ast m \right\rangle(\theta)
                    -  \left\langle m^\ast \varepsilon_{\rm t} + \varepsilon_{\rm t}^\ast m
                    \right\rangle(\theta) .
\end{eqnarray}
In particular the second function $D_2$ is a useful diagnostic since it is
negative definite. In case of over-fitting, the model tends to fit the noise
rather the true ellipticity, which creates correlations between the noise $N$
and the model uncertainty $m$, and $D_2$ becomes significantly negative.

I illustrate such a case in Fig.~\ref{fig:D1_D2_DES}. Non-zero $D_1$ and $D_2$
indicate PSF residual correlations, in particular on small scales, where the
spatial PSF model, a bi-variate polynomial with varying degree,
seems to not well fit data. The increasing negative $D_2$ with
increasing polynomial degree shows cases of over-fitting for the more complex
models, indicating that the true PSF variation shows less degrees of freedom.

A second example is discussed below in Sect.~\ref{sec:image_sims_great10}.

\stoptwocol
\begin{figure}
  \begin{center}
    \resizebox{0.9\hsize}{!}{
      \includegraphics{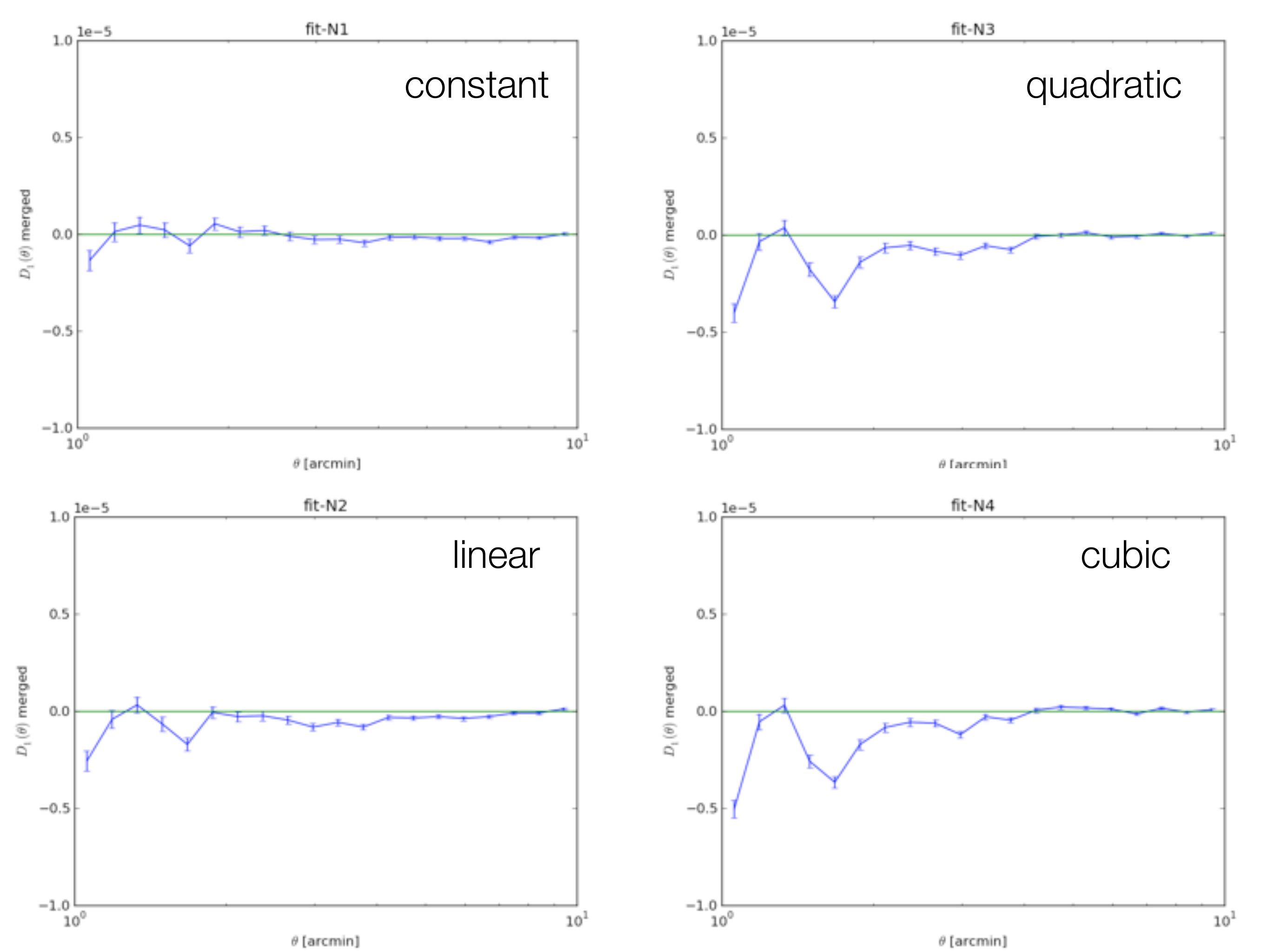}
    }
    \resizebox{0.9\hsize}{!}{
      \includegraphics{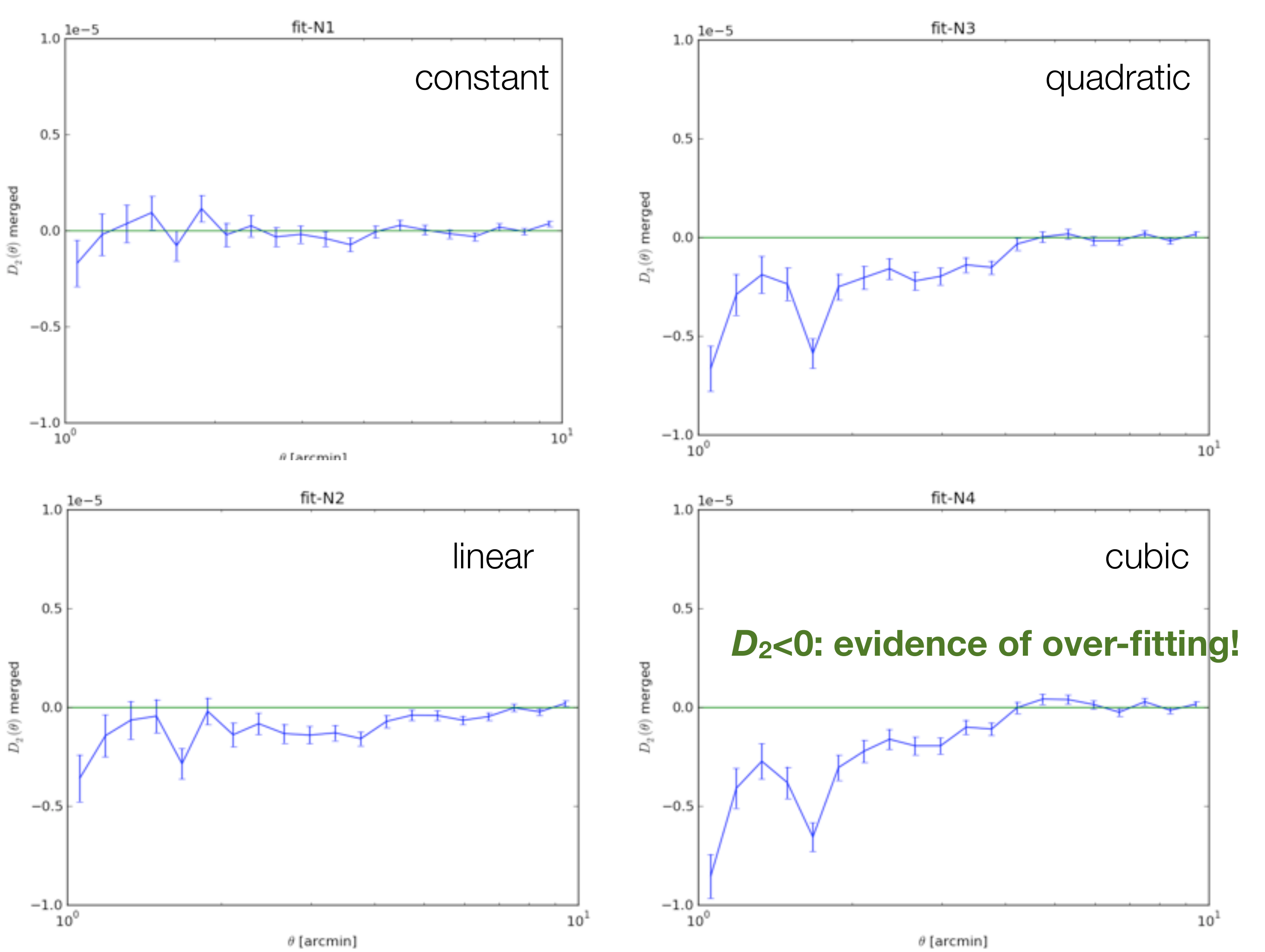}
    }
  \end{center}

  \caption{The PSF diagnostic correlation functions $D_1$ (\emph{upper four panels}) and $D_2$ (\emph{lower four panels}) as function
    of angular scale $\theta$, for the Dark Energy Survey (DES) data challenge \#5 simulations.
    The four sub-panels show PSF polynomial models of increasing degree, from 0 (constant) to 3 (cubic).
    See Sect.~\ref{sec:PSF_model_res_corr}.
  }

  \label{fig:D1_D2_DES}.

\end{figure} 
\begtwocol

%%%%%%%%%%%%%%%%%%%%%%%%%%%%%%%%%%%%%%%%%%%%%%%%%%%%%%%%%%%%
\subsection{Image simulations}
\label{sec:image_sims}
%%%%%%%%%%%%%%%%%%%%%%%%%%%%%%%%%%%%%%%%%%%%%%%%%%%%%%%%%%%%

Image simulations have been created as collaborative projects within the
weak-lensing community, such as the Shear TEsting Project (STEP) with the two
consecutive blind tests STEP1 \cite{STEP1} and STEP2 \cite{STEP2}. Public
challenges like the GRavitational lEnsing Accuracy Testing (GREAT) projects
have been launched to reach out to a larger community, in particular computer
science, to invite more ideas to tackle the problem of galaxy shape
measurement. This contains GREAT08
\cite{2009AnApS...3....6B,2010MNRAS.405.2044B}, GREAT10
\cite{2010arXiv1009.0779K,2012MNRAS.423.3163K,Kitching:2012fj}, and GREAT3
\cite{2013arXiv1308.4982M}.

Those collaborative image simulation projects typically started under simple,
well-controlled conditions, for example, a constant PSF, constant shear over
the field, and analytical galaxy light distributions with high signal-to-noise.
They then progressed to more complex and more realistic images, for example
galaxy images based on observed HST deep fields. The purpose of those
simulations is to test estimates of shear with amplitudes of a few percent
to an accuracy at also the percent level. This is typically quantified in
terms of multiplicative and additive bias (\ref{eq:ell_bias}). The number
of simulated images is necessarily very large, producing hundreds of gigabytes
of data.

In the following sections, I briefly discuss some of my past work on galaxy
shape measurement, bias quantification, and PSF modelling with image simulations.

%%%%%%%%%%%%%%%%%%%%%%%%%%%%%%%%%%%%%%%%%%%%%%%%%%%%%%%%%%%%
\subsubsection{GREAT10}
\label{sec:image_sims_great10}
%%%%%%%%%%%%%%%%%%%%%%%%%%%%%%%%%%%%%%%%%%%%%%%%%%%%%%%%%%%%

Fig.~\ref{fig:great10_PSF} shows the PSF pattern of a GREAT10 star
challenge image. Together with Bernhard Riedl, Diploma student under my and
Jochen Weller's supervision (Ludwigs-Maximilians-Universit\"at M\"unchen), we
fit this pattern with bi-variate polynomials in the pixel coordinates $x$ and
$y$ of varying degree. The diagnostic functions $D_1$ and $D_2$ are then
calculated using the public software \textsc{athena} \cite{athena_ascl}.

The diagnostic correlation functions are plotted in Fig.~\ref{fig:great10_D12}
for increasing polynomial degrees 1 (bi-linear), 3 (bi-cubic) and 5. First,
below 50 pixels the correlations cannot be measured since there are no pairs of
stars due to the finite size postage stamps. Second, there is a very
significant correlation of residuals up to a few hundred pixels. Clearly, the
polynomial is not a good fit to capture those small-scale PSF variations. Third,
on scales above 500 pixels, the linear function still displays strong
correlations, but the third-order polynomial best fits the data showing the
smallest correlations. Increasing the polynomial order to 5 re-introduces
correlations at around 500 pixels, hinting to an over-fitting problem.

\begin{figure}

  \begin{minipage}{0.65\textwidth}
  \begin{center}
    \resizebox{1.0\hsize}{!}{
      \includegraphics{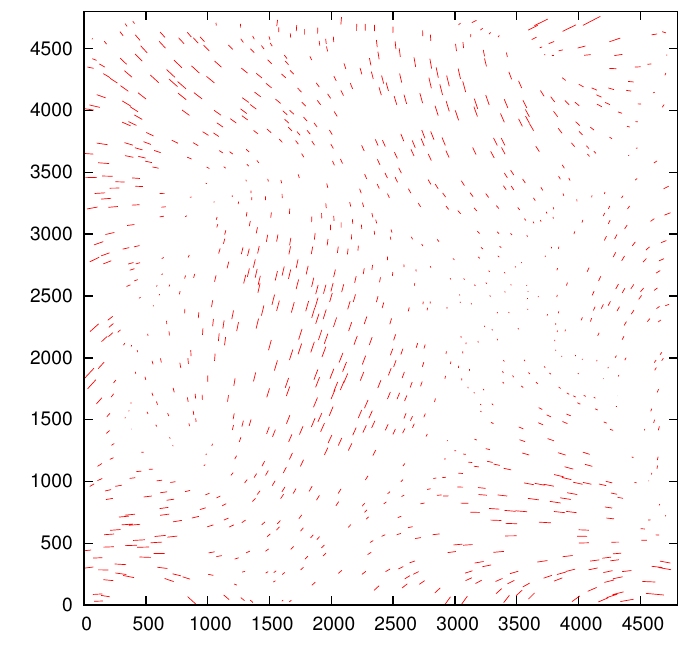}
    }
  \end{center}
  \end{minipage}%
  \hspace*{-0.1\textwidth}%
  \begin{minipage}{0.45\textwidth}

  \caption{PSF pattern of a starfield from the GREAT10 challenge. The $x$-
    and $y$-axis are pixel units.}
  \label{fig:great10_PSF}

  \end{minipage}%

\end{figure}

\begin{figure}
  \begin{center}
    \resizebox{1.0\hsize}{!}{
      \includegraphics{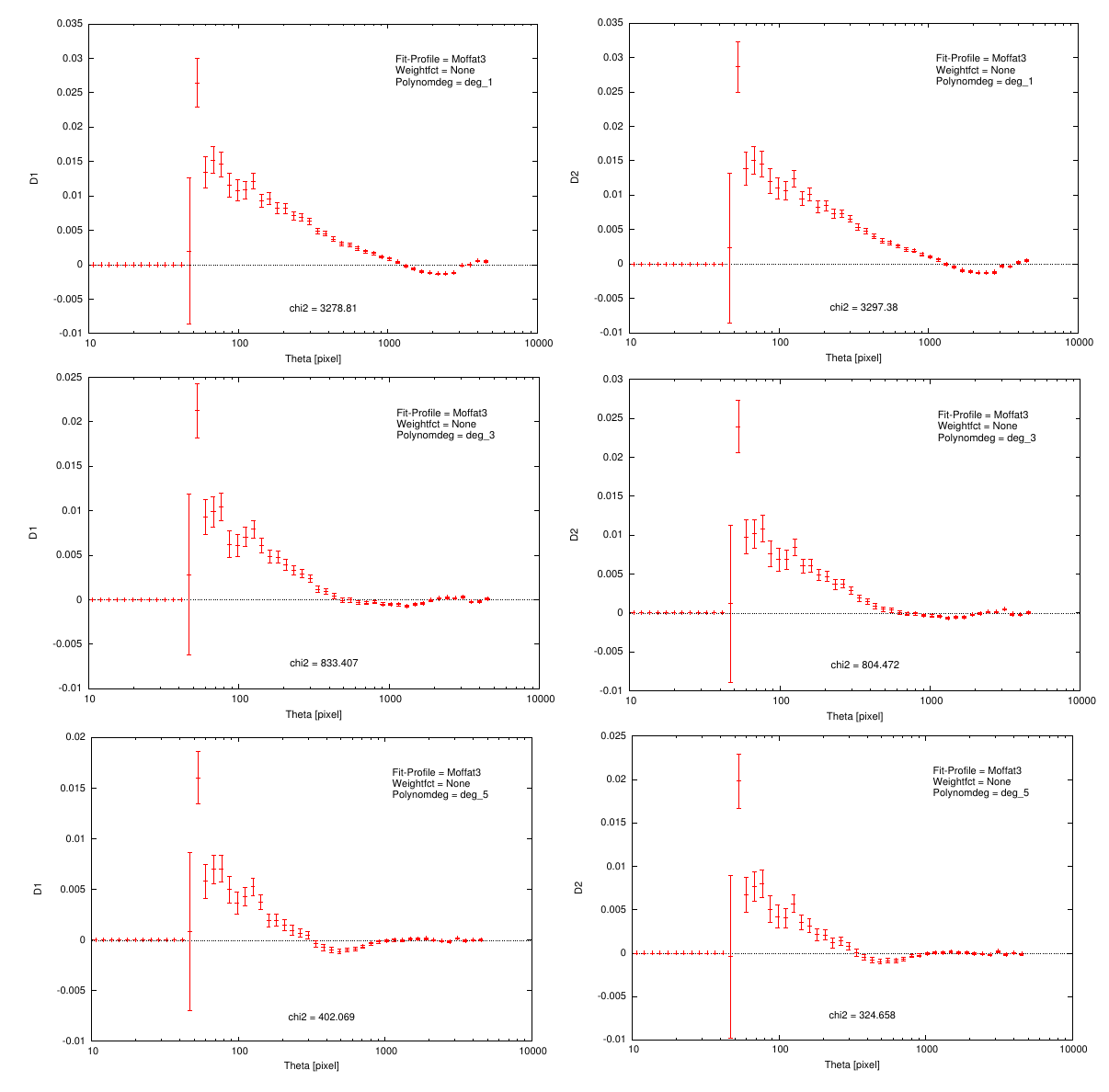}
    }
  \end{center}

  \caption{The diagnostic functions $D_1$ (\emph{left column}) and
    $D_2$ (\emph{right column}) as function of pixel units.
    The three rows from top to bottom show the case of a PSF interpolation
    model as bi-variate polynomial in $x$ and $y$ of order 1, 3, and 5, respectively.
    Figure from Bernhard Riedl's master thesis (unpublished).
  }

  \label{fig:great10_D12}

\end{figure}

%%%%%%%%%%%%%%%%%%%%%%%%%%%%%%%%%%%%%%%%%%%%%%%%%%%%%%%%%%%%
\subsubsection{CFHTLenS}
\label{sec:image_sims_CFHTLenS}
%%%%%%%%%%%%%%%%%%%%%%%%%%%%%%%%%%%%%%%%%%%%%%%%%%%%%%%%%%%%

The CFHTlenS collaboration created image simulations (1) to calibrate for
shear biases (Sect.~\ref{sec:shape_biases}), and (2) to model residual
systematics correlations (Sect.~\ref{sec:error-model}).

As a cross-check, two sets of simulations were created, using the (modified)
code from the GREAT08 and GREAT10 challenges as described in
\citet{2010MNRAS.405.2044B}, and the SkyMaker package
\citep{2009MmSAI..80..422B}, respectively. Several important features were
added to the simulations compared to the GREAT challenge:

First, the galaxies were modeled as disk+bulge as fitted by \emph{lens}fit,
with ellipticity and size distribution matching the observed data. A mismatch
would result in a wrong bias calibration; for example, the STEP and GREAT08/10
simulations did not include the large observed number of small galaxies, and
thus the bias of those objects could not be calibrated correctly.

The additive and multiplicative biases were fitted to all simulated galaxies as
functions of size and SNR. 
I propagated the uncertainties of $m$ from the fits to the correlation function
covariance. Their contribution turned out to be negligeable, see
Fig.~\ref{fig:CFHTLenS-cov_m}.

\stoptwocol
\begin{figure}
  \begin{center}
    \resizebox{1.0\hsize}{!}{
      \includegraphics{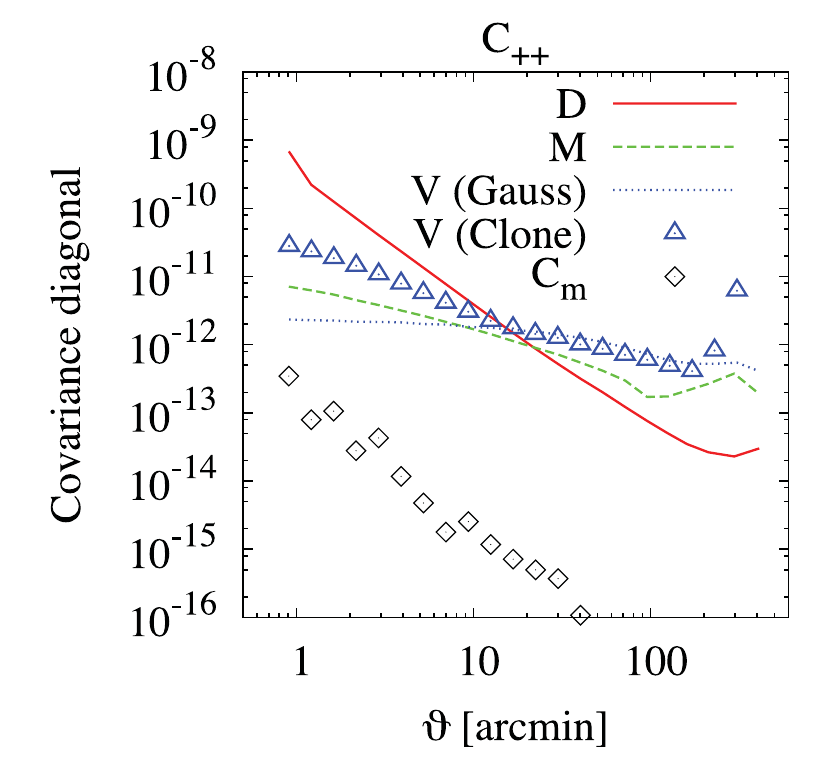}
      \includegraphics{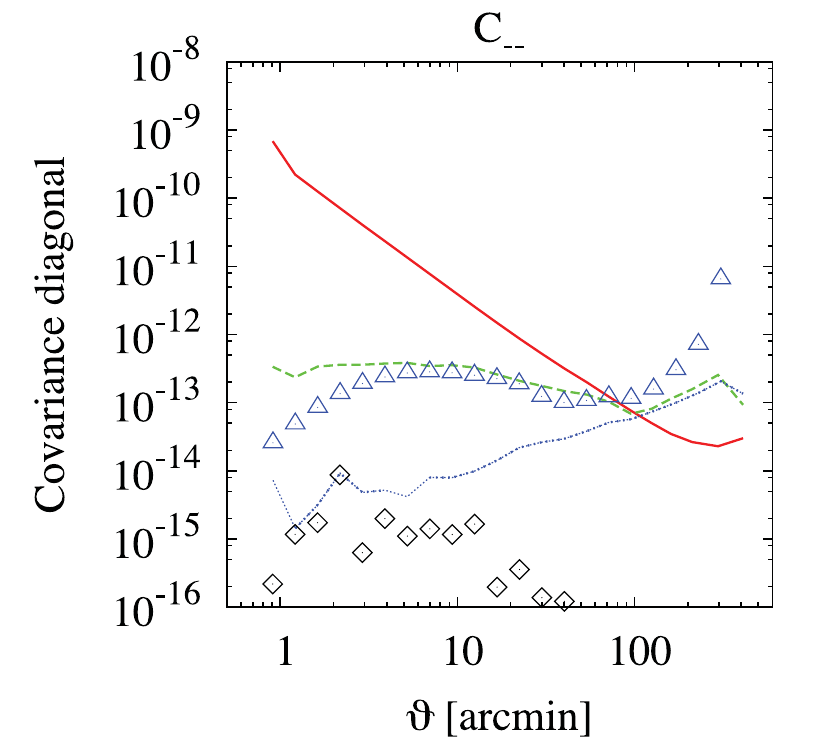}
    }
  \end{center}

  \caption{Diagonal of the covariances $\mat C_{++}$ (\emph{left panel}) and
  $\mat C_{--}$ (\emph{right panel}). The variance $\mat C_m$ due to the shear calibration
  of $m$ (black diamonds) is sub-dominant compared to the shot noise $\mat D$ (solid red line),
  mixed term $\mat M$ (green dashes), and cosmic variance $\mat V$ (blue dashes: Gaussian approximation;
  blue triangles: non-Gaussian covariance estimated from the CFHTLenS Clone simulations). From
  \citet{CFHTLenS-2pt-notomo}.}

  \label{fig:CFHTLenS-cov_m}

\end{figure}
\begtwocol

Second, each galaxy got assigned a shear from the CFHTLenS ``Clone'' $N$-body
simulation \citep{CFHTLenS-Clone}. A realistic cosmological shear component is
important in the quantitative analysis of systematics \citep{CFHTLenS-sys}.

%%%%%%%%%%%%%%%%%%%%%%%%%%%%%%%%%%%%%%%%%%%%%%%%%%%%%%%%%%%%
\subsubsection{GREAT3}
\label{sec:image_sims_great3}
%%%%%%%%%%%%%%%%%%%%%%%%%%%%%%%%%%%%%%%%%%%%%%%%%%%%%%%%%%%%

The GREAT3 weak-lensing image challenge \cite{2013arXiv1308.4982M} was
organised by R.~Mandelbaum and B.~Rowe, and ran from mid-2013 to April 2014.
Results were presented and discussed at the final meeting at CMU in Pittsburgh
in May 2014 \cite{great3-I}. I was part of the ``CEA--EPFL'' team including from
CEA Florent Sureau, Jean-Luc Starck, Fred Maurice Ngol\'e Mboula, St\'ephane
Paulin-Henriksson, and from EPFL Marc Gentile and Fr\'ed\'eric Courbin.

I also submitted results under ``CEA\_denoise'', for which I first denoised the
galaxy images by applying the \textsc{mr\_filter} multi-scale wavelet filter to
the images \cite{2006A&A...451.1139S}. I then used \textsc{SExtractor}
\cite{1996A&AS..117..393B} to measure unweighted second moments of the galaxy
light distribution, and corrected for the PSF in moment space, following
\citet{2000ApJ...536...79R}, as mentioned earlier. The results were however not
great and could not compete with the leading group of best methods.

%%%%%%%%%%%%%%%%%%%%%%%%%%%%%%%%%%%%%%%%%%%%%%%%%%%%%%%%%%%%
%\paragraph{\textsc{gfit}}
\label{sec:image_sims_great3_gfit}
%%%%%%%%%%%%%%%%%%%%%%%%%%%%%%%%%%%%%%%%%%%%%%%%%%%%%%%%%%%%

The main shape measurement method of our team was the forward-fitting
maximum-likelihood method \textsc{gfit}. This algorithm is a new version
(rewritten in \textsc{python} from scratch) of \textsc{gfit} presented in
\citet{2012arXiv1211.4847G}, which was used in the GREAT10 galaxy challenge.
Galaxies are detected by a \textsc{SExtractor} run,
whose output parameters of centroid position, size, and ellipticity are used as
first-guess starting point of the model fitting procedure. Galaxy light
profiles are modeled as bulge (S\'ersic index $n=1$ + exponential (S\'ersic
$n=4$) disk, which are concentric and aligned with identical intrinsic
ellipticity. The eight parameters centroid, flux, disc flux fraction, bulge
radius, disk radius, and ellipticity were fitted. Several minimizers were
explored.

\paragraph{Weighing and filtering}

Often, measured galaxy shapes are weighted depending on various quantities: the
galaxy S/R, size, or the best-fit $\chi^2$, fit error bar, or confidence in
the result. This down-weighs galaxies with uncertain or biased shape estimates
and generally improves the results.

Each of the GREAT3 images had $10,000$ galaxies, where two pairwise galaxies
had the same intrinsic ellipticity roated by $90\deg$ to reduce shape noise. To
fully benefit from this noise cancelling scheme, one has an interest to weigh
all galaxies equally. However, for some objects no shape can be attributed, for
example if the minimizer does not converge. Thus, at the end of the day, we
decided to introduce weights for the galaxies with the hope to reduce the biases.

A first, simple weighing scheme was to eliminiate galaxies with large fit
residuals by setting their weight to zero. I then developed an improved weighting
scheme, which I describe in the following.

For a given GREAT3 image, the \textsc{gfit} output parameters for each galaxies
were used to create new simulations of that image, with properties of noise,
PSF, pixellation etc.\ similar to the input image. \textsc{gfit} was run on
this second simulated image, and a PCA decomposition of measured galaxy
parameters was performed on the rms ellipticity between input and ouput, $|
\Delta e|$. In other words, the galaxies were classified according to
ellipticity bias. Using PCA I could identify combinations $v_i$ of galaxy
parameters that contribute most to this bias, and to devise a weighting scheme
to downweigh objects for which those parameters correlated with a large bias.
To keep the number of parameters reasonably small, I split them up into several
sets of parameters, and studied one set at a time. Here, I will quote results
from one such set of parameters (case `a' in \cite{great3-I}), which are
%
%\begin{itemize}
%
%\item
the \textsc{gfit} flux $\ln(\ln I_0)$, disc radius $\ln r_{\rm disk}$,
bulge radius $\ln r_{\rm bulge}$, disc flux fraction  $f_{\rm disk}$.
%
%\item[(b)] full-width half maximum $\ln$ FWHM, \textsc{SExtractor} size $\ln
%R_{\rm F}$, S/N $\ln \nu$, \textsc{SExtractor} flux $\ln F$, disc radius $\ln
%r_{\rm disk}$, bulge radius $\ln r_{\rm bulge}$.
%
%\end{itemize}

\begin{figure}

  \begin{minipage}{0.55\textwidth}
  \begin{center}
    \resizebox{1.0\hsize}{!}{
      \includegraphics{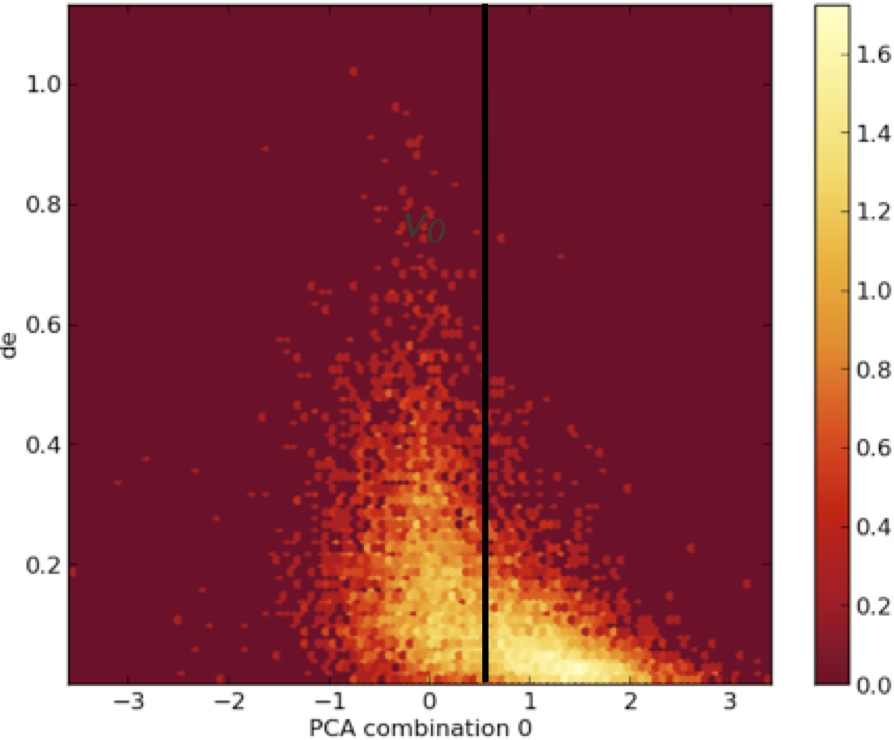}
    }
  \end{center}
  \end{minipage}%
  \hspace*{-0.1\textwidth}%
  \begin{minipage}{0.55\textwidth}

  \caption{Log-density of objects as function of PCA component $v_0$ ($x$-axis) and
    ellipticity bias $|\Delta e|$ ($y$-axis). The horizontal line is the cut-off value
    $v_{\rm c} = 0.6$.
  }
  \label{fig:great3_PCA_v40}.

  \end{minipage}%

\end{figure}

The dependence of the parameters (linear, logarithmic or double-logarithmic)
being most significant to quantify large residuals
was established by trying out various variants.

The weights were determined by ploting $|\Delta e|$ versus various PCA
components $v_i$ to select a cutoff value $v_{\rm c}$ that provided a good
separation of objects with low and high bias $|\Delta e|$.

%For case (a),
A density plot in $v_0$ (the $0^{\rm th}$ PCA component) and
the ellipticity bias $| \Delta e|$ is show in Fig.~\ref{fig:great3_PCA_v40}.
There is a clear tail of high-$| \Delta e|$ objects for small values of $v_0$,
whereas there is a high density of objects with low $|\Delta e|$ at $v_0 = 1
\ldots 2$. The cut-off value $v_{\rm c} = 0.6$ is shown by the horizontal line.
Thus, objects with PCA component $v_0 < v_{\rm c}$ ($v_0 > v_{\rm c}$) show 
generally a strong (weak) bias $|\Delta e|$. The first category was assigned
the weight $w_{\rm low} = 0.2$. Objects in the low-bias category kept their
weight $w_{\rm high} = 1$.

The $0^{\rm th}$ PCA component is
\begin{eqnarray}
  v_{0} & = 0.38 \ln(\ln I_0) + 0.59 r_{\rm disk} + 0.44 \ln r_{\rm bulge} - 0.28 f_{\rm disk}
  \nonumber \\
        & = \ln \left[ (\ln I_0)^{0.38} \, r_{\rm disk}^{0.59} \,
            r_{\rm bulge}^{0.44} \, {\rm e}^{-0.28 f_{\rm disk}}
            \right] .
\end{eqnarray}
This means that the bias is smaller for objects with larger flux, disk and
bulge radius, but smaller disk fraction. The improvement on the overall GREAT3
of this weighting scheme compared to uniform weights was around $15$ - $20\%$.

\begin{figure}
  \begin{center}
    \resizebox{0.7\hsize}{!}{
      \includegraphics{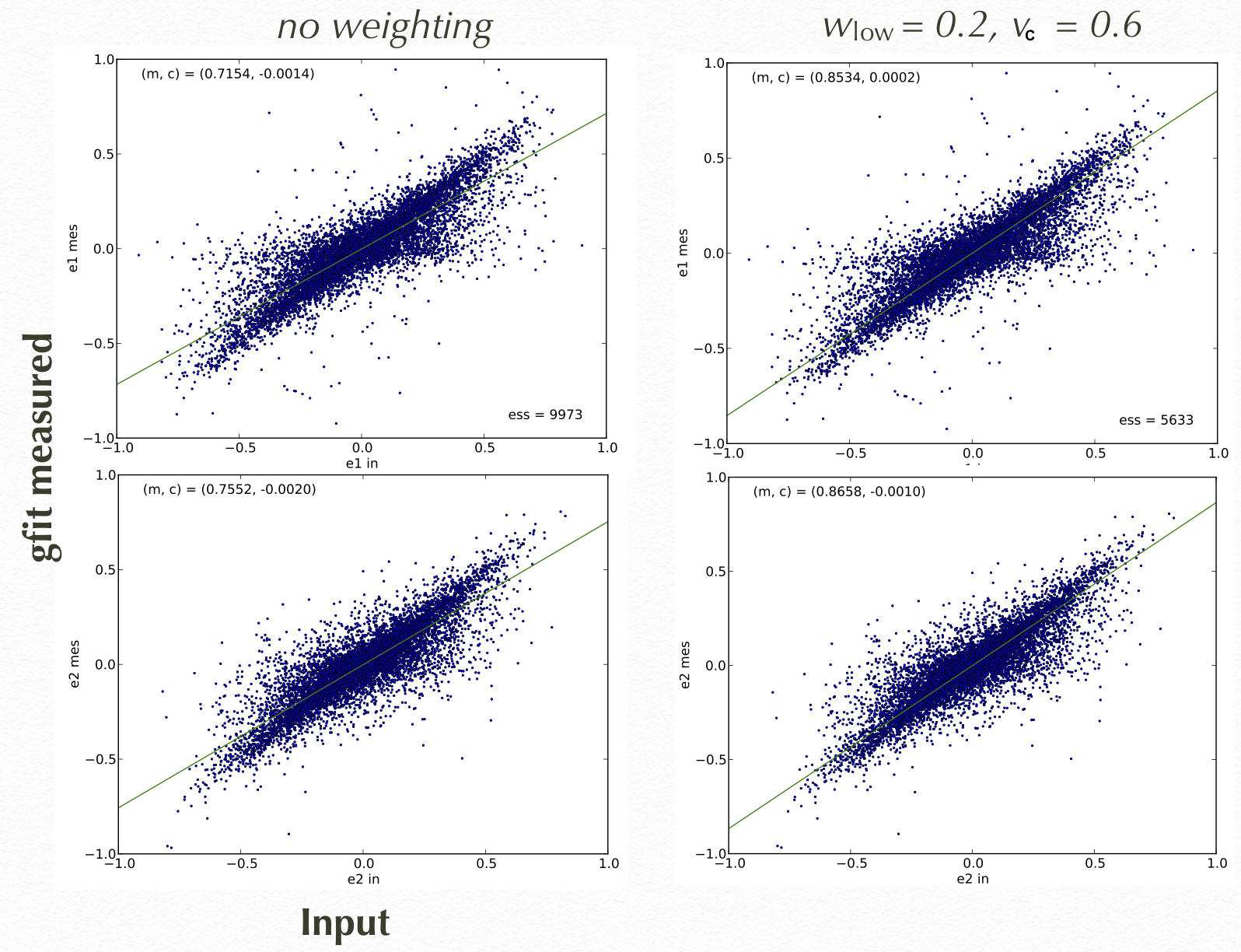}
    }
  \end{center}

  \caption{Input ($x$-axes) versus output ($y$-axes) ellipticity
    ellipticity bias. The upper row is $e_1$, the lower row $e_2$.
  Left panels show the no-weighting (uniform weighting) case, the left column
  is for case (a) with a weight $w_{\rm low} = 0.2$ of high-bias objects whose
  $0^{\rm th}$ PCA component is smaller than $v_{\rm c} = 0.6$.
  The lines are fits of multiplicative ($m$) and additive ($c$) bias with values
  indicated in the panels.
  }

  \label{fig:great3_PCA_v40_m_c}.

\end{figure}

%%%%%%%%%%%%%%%%%%%%%%%%%%%%%%%%%%%%%%%%%%%%%%%%%%%%%%%%%%%%
\subsection{Redshift estimation}
\label{sec:redshift_estim}
%%%%%%%%%%%%%%%%%%%%%%%%%%%%%%%%%%%%%%%%%%%%%%%%%%%%%%%%%%%%

%%%%%%%%%%%%%%%%%%%%%%%%%%%%%%%%%%%%%%%%%%%%%%%%%%%%%%%%%%%%
\subsubsection{Photometric redshifts}
\label{sec:photo-z}
%%%%%%%%%%%%%%%%%%%%%%%%%%%%%%%%%%%%%%%%%%%%%%%%%%%%%%%%%%%%

Weak lensing observables, being integrals along the line of sight weighted by
the source galaxy distribution $n(z)$ (\ref{eq:lens_efficiency}), require knowledge of
the latter if they are to be interpreted cosmologically. To first order, the mean redshift $\bar z$
has to be determined, but also the shape of $n(z)$ plays an important role.
\citet{2006MNRAS.366..101H} find a rough estimate of $P_\kappa(\ell \sim 1000)
\propto \Omega_{\rm de}^{-3.5} \sigma_8^{2.9} {\bar z}^{1.6} |w|^{0.31}$.
Clearly, for a desired accuracy on cosmological parameters, the mean redshift
of sources has to be known to at least that accuracy, and to a much higher
accuracy in the case of parameter on which the power spectrum has a weaker
dependence such as $w$.

Spectroscopy of all the faint galaxies used for a typical weak-lensing survey
is too costly, and redshifts have to be estimated from broad-band photometry,
using the technique of \emph{photometric redshifts}, or photo-$z$s. There are
various methods to measure photometric redshifts, of which template-based
approaches are one of the most popular. They perform $\chi^2$-type fits of
(redshifted) template SEDs to the flux in the observed bands. Exemplary methods
that have been used in a weak-lensing context include \emph{LePhare}
\cite{2006A&A...457..841I} and \emph{Bayesian Photometric redshift estimation}
\citeaffixed{2000ApJ...536..571B}{BPZ; }. LePhare photo-$z$'s for CFHTLS were
published in \cite{CIK09}; BPZ was the method of choice for CFHTLenS
\cite{CFHTLenS-photoz}. Both methods do not only performe a point-estimate
$\hat z$ but provide the full pdf of the redshift.

%The currently-reached amplitude of dispersion $\sigma_z$ is sufficient for
%future surveys. However, they require $\sigma_z$ to be known to sub-percent
%accuracy \cite{2006ApJ...636...21M}. Catastrophic outliers can strongly bias
%tomographic shear power spectra, and their rate has to be lower than a percent
%\cite{2010MNRAS.401.1399B,2010ApJ...720.1351H}. This rate is applied to the
%galaxy sample after possible rejection of likely outliers by the photometric
%redshift code. The required precision of photometric redshifts necessitates a
%very large spectroscopic calibration set. For current surveys, this number is
%on the order of ten thousand, and a magnitude larger for future surveys
%\cite{2006ApJ...636...21M}. The spectroscopic survey has to be a representative
%sample of the lensed galaxy population, covering all possible types and
%redshifts. In most cases however, the spectroscopic surveys are too shallow to
%be complete down to the limiting magnitude of weak-lensing galaxies, and suffer
%from a non-zero spectroscopic redshift failure rate. Several calibration
%methods that cope with these issues have been applied to cosmic shear results
%in \cite{KiDS-450}.

%%%%%%%%%%%%%%%%%%%%%%%%%%%%%%%%%%%%%%%%%%%%%%%%%%%%%%%%%%%%
\subsubsection{Clustering-based redshift estimation}
\label{sec:cluster_z}
%%%%%%%%%%%%%%%%%%%%%%%%%%%%%%%%%%%%%%%%%%%%%%%%%%%%%%%%%%%%

To assess the quality of photometric redshifts and to recover the true redshift
distribution, one can make use of the spatial clustering of galaxies. From the
amount of cross-correlation of photometric samples between different redshift
bins one can deduce the amount of redshift outliers. In \citet{BvWMK10} we
introduced this method, and applied it to photometric clustering
\cite{CK11,MWCK15} and weak-lensing \cite{CFHTLenS-2pt-tomo}.

Fig.~\ref{fig:CK12-Fig6} is an example from \citet{CK11}. For the full galaxy
sample of the CFHTLS-Wide T0006 release, it shows the angular auto-correlation
functions of redshift bin \#1, $0.4 < z < 0.6$, and the two auto-correlation
functions of that bin with galaxies at higher redshift bins \#3 ($0.8 < z < 1$)
and \#4 ($1 < z < 1.2$), respectively. From these measurement we estimate the
contamination $f_{ij}$ between redhift bin pairs $i$ and $j$ due to photometric
mis-identification. We carry out the so-called \emph{global pair-wide
analysis}, considering only two bins at a time while in turn setting to zero
contaminations of other bins, and also neglecting higher-order effects such as
magnification bias. The resulting constraints of the contamination fractions,
$f_{ij}$, $i,j=1,3,4$, are shown in the in the middle and right panels. Around
three percent of galaxies in bin \#3 are mis-identified into bin \#1, which
leads to the non-zero cross-correlation between the two bins (dotted line in
the left panel). The mixing between bins \#1 and \#4 is consistent with zero,
as is the cross-correlation function (dashed line in the left panel).

\begin{figure}
  \begin{center}
    \resizebox{1.0\hsize}{!}{
      \includegraphics[scale=1.65]{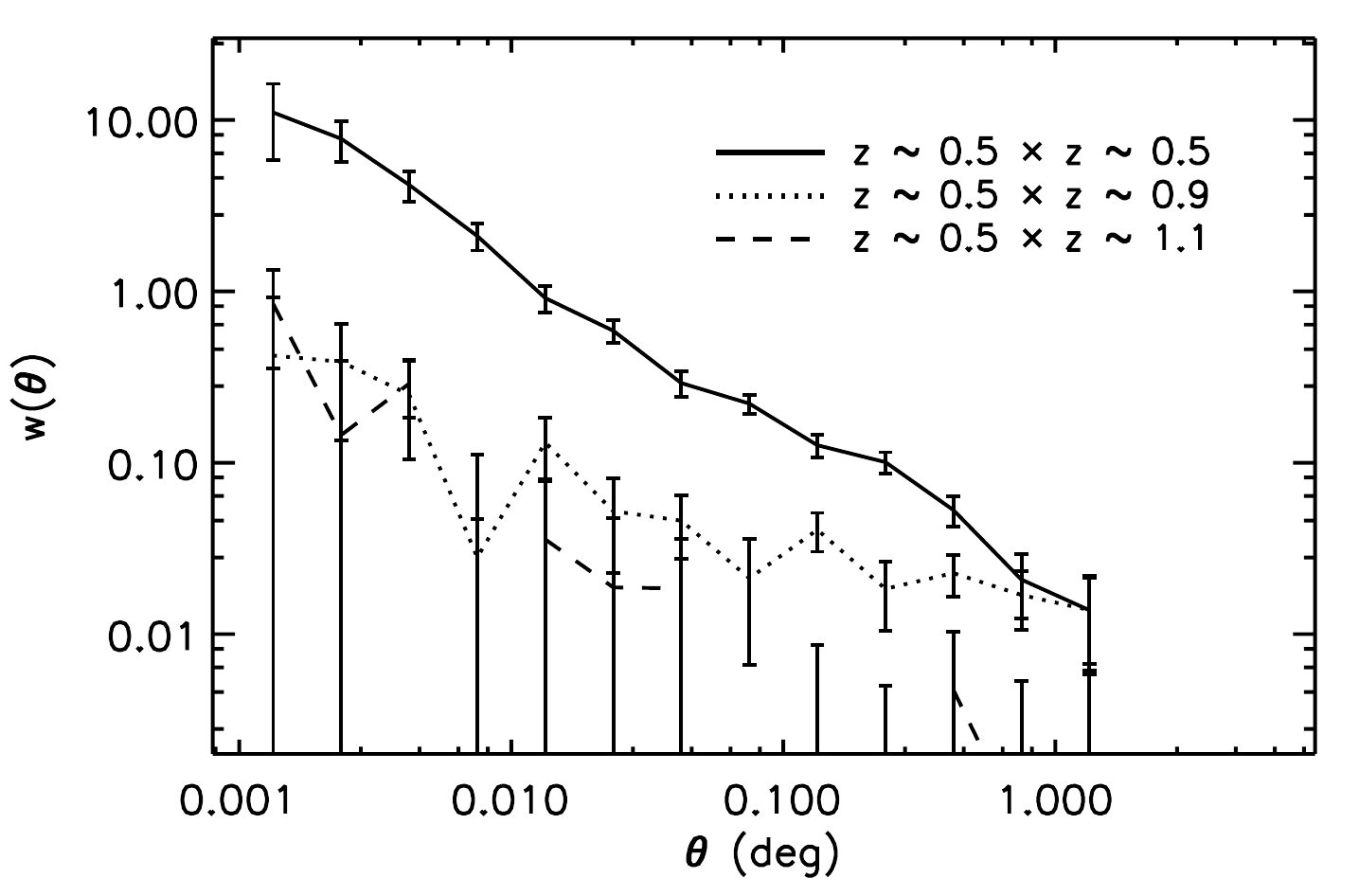}
      \includegraphics{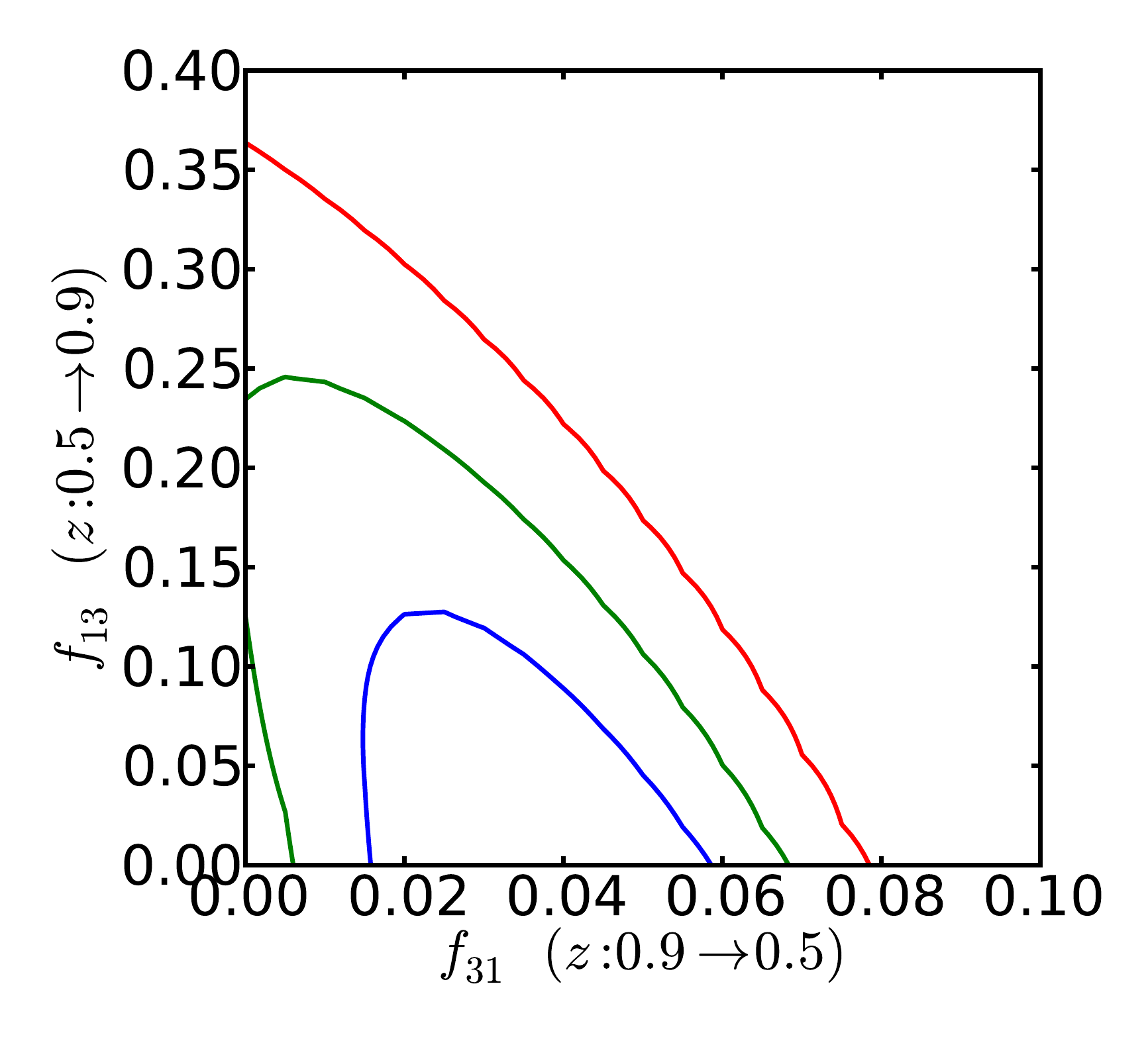}
      \includegraphics{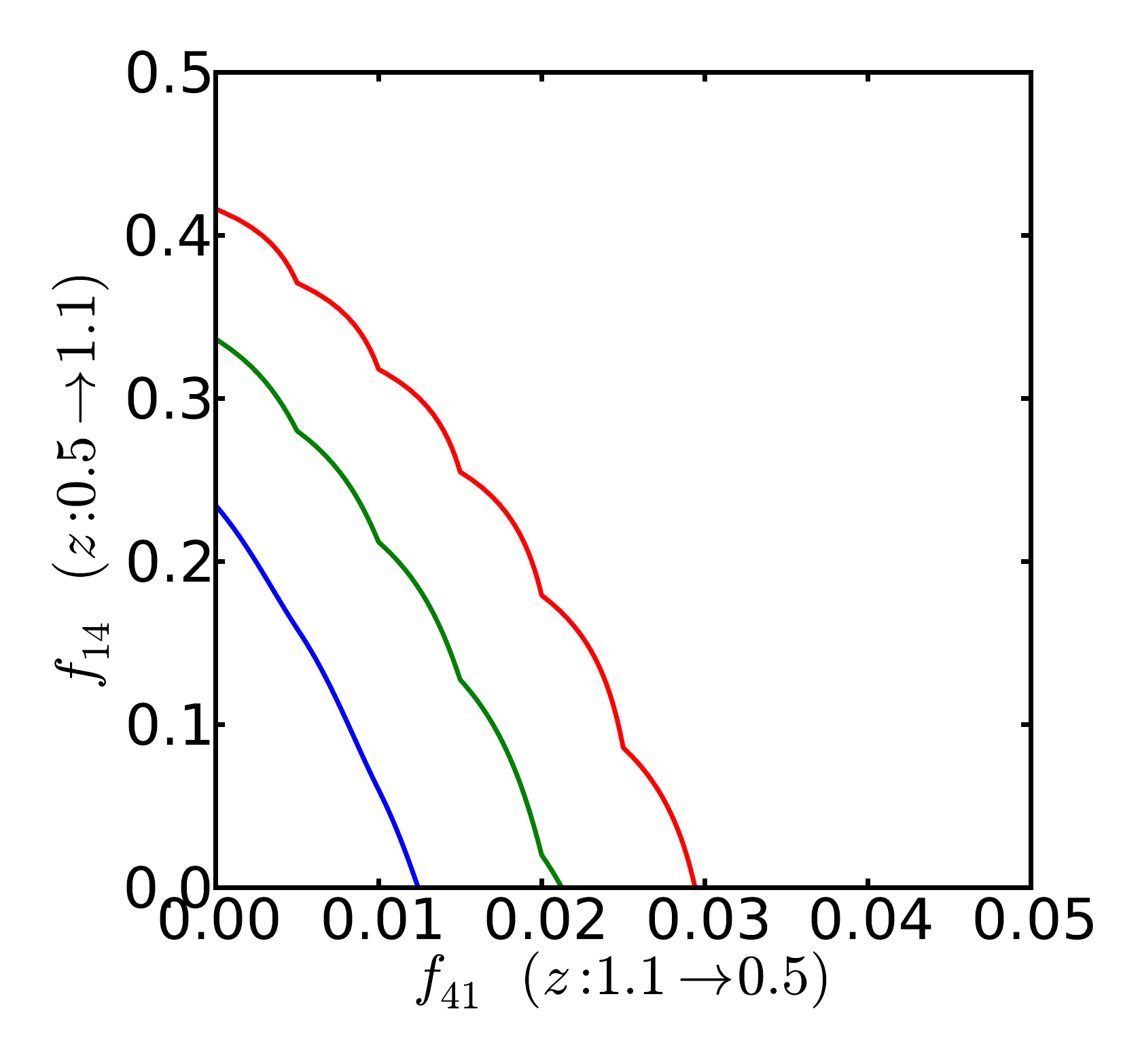}
    } 
  \end{center} 

  \caption{Cross correlation analysis between redshift bins for the full
    sample. \emph{Left}: angular auto-correlation function in the redshift bin $0.4 < z < 0.6$
    (straight line) and cross-correlation between the bins  $0.4 < z
    < 0.6$ and $0.8 < z < 1.0$  (dotted line) and between $0.4 < z <
    0.6$ and $1.0  < z < 1.2$ (dashed line).
    \emph{Middle and right}: quantitative estimates of the contamination (percentage
    of galaxies scattered) from a pairwise
    analysis between redshift bins $0.4 < z < 0.6$ and $0.8 < z < 1.0$
    (\emph{middle}) and  between $0.4 < z < 0.6$ and $1.0 < z < 1.2$ (\emph{right}).
    The contours show the 68.3 (blue), 95.5 (green) and 99.7 (red) confidence regions.
    From \citet{CK11}.
    }

  \label{fig:CK12-Fig6}

\end{figure}

Similarly, the cross-correlation of the photometric with a spectroscopic sample
can reveal the true redshift distribution \cite{Newman08}, although this
reconstruction is hampered by a possible redshift-dependent bias of the
photometric galaxy sample \cite{2010ApJ...724.1305S}. In \citet{Scottez16} we
applied the cross-correlation method to CFHTLS/VIPERS, and showed that it can
yield redshifts for individual galaxies if color information is present.

% Chapter 7
% results.tex

%%%%%%%%%%%%%%%%%%%%%%%%%%%%%%%%%%%%%%%%%%%%%%%%%%%%%%%%%%%%
\section{Observational results and cosmological constraints}
\label{sec:obs_results}
%%%%%%%%%%%%%%%%%%%%%%%%%%%%%%%%%%%%%%%%%%%%%%%%%%%%%%%%%%%%

This section highlights some of the observational results for cosmological parameter
constraints from cosmic shear to which I contributed.

%%%%%%%%%%%%%%%%%%%%%%%%%%%%%%%%%%%%%%%%%%%%%%%%%%%%%%%%%%%%
\subsection{Second-order statistics}
\label{sec:basic_results}
%%%%%%%%%%%%%%%%%%%%%%%%%%%%%%%%%%%%%%%%%%%%%%%%%%%%%%%%%%%%

In the following sub-sections I discuss the basic observational results from
second-order cosmic shear statistics. The parameter combination that cosmic
shear (including non-linear scales) is most sensitive to the parameter $\Sigma_8$, defined
by
\begin{equation}
 \Sigma_8 = \sigma_8 \left(\frac{\Omega_{\rm m}}{\Omega_{\rm m, 0}}\right)^\alpha,
  \label{eq:Sigma_8}
\end{equation}
with typical values of $\alpha \approx 0.5$ - $0.7$. The pivot value $\Omega_{{\rm m}, 0}$ can
be chosen freely. Fig.~\ref{fig:Sigma} and Table \ref{tab:Sigma}
shows this combination measured in recent years for $\Omega_{\rm m, 0} = 0.3$. If the original
measurement (indicated in the table) corresponds to a different pivot value, I transform to
$\Omega_{\rm m, 0} = 0.3$ including a simple error propagagtion computation.

%%%%%%%%%%%%%%%%%%%%%%%%%%%%%%%%%%%%%%%%%%%%%%%%%%%%%%%%%%%%
\subsubsection{CFHTLS-T0003, 2007}
\label{sec:consol_era}
%%%%%%%%%%%%%%%%%%%%%%%%%%%%%%%%%%%%%%%%%%%%%%%%%%%%%%%%%%%%

The third data release (T0003) of the wide part of the Canada-France-Hawaii
Telescope Legacy Survey (\survey{CFHTLS}) with an observed area of $53$ deg$^2$
provided 2D cosmic shear results out to very large, linear scales
\citeaffixed{FSHK08}{$7.7$ degrees, corresponding to $170$ Mpc at the mean lens
redshift of $0.5$; }. This enabled us to infer cosmological constraints
using large scales only, thereby reducing uncertainties from non-linear and
baryonic physics on small scales. Using $\langle M_{\rm ap}^2 \rangle(\theta)$
on scales $\theta > 85^\prime$ we obtained $\sigma_8 (\Omega_{\rm m} /
0.25)^{0.53} = 0.837 \pm 0.084$. This study used
the photometric redshifts from the 4 square degree deep part of \survey{CFHTLS}
\cite{2006A&A...457..841I}, taking into account sampling variance. The deep
fields have an area of $4$ square degrees, an increase of nearly a factor
$2500$ over the \survey{HDF}. Constraints on neutrino masses were obtained using
this data release in \citet{TSUK09}.

By that time, ground-based surveys had become large enough to enable detailed
residual systematics tests. For \survey{CFHTLS}, \citet{KB09} as well as first
multi-colour observations \cite{FuPhD} revealed an anomalous shear amplitude
scaling with source redshift and a variance between \instrument{MegaCam}
pointings larger than expected. \citet{KB09} quantified their influence on
cosmological parameters from a joint analysis. However, it took three more
years of work by the CFHTLenS team (Sect.~\ref{survey_era}) of around 20 members and
a complete re-analysis of CFHTLS images to finally obtain a robust shear
catalogue. The origin of those systematics was never found, and the price to
pay for a systematic-free data set was to reject $25\%$ of the \instrument{MegaCam}
pointings that were plagued with PSF residuals.

%%%%%%%%%%%%%%%%%%%%%%%%%%%%%%%%%%%%%%%%%%%%%%%%%%%%%%%%%%%%
\subsubsection{Re-analysis of COSMOS, 2009}
\label{sec:cosmos}
%%%%%%%%%%%%%%%%%%%%%%%%%%%%%%%%%%%%%%%%%%%%%%%%%%%%%%%%%%%%

I contributed to a re-analysis of the weak-lensing data \cite{SHJKS09},
independent from the first series of papers from that survey
\cite{2007ApJS..172..219L,2007ApJS..172..239M}. We obtained improved
photo-$z$'s from twice the number of bands \cite{2009ApJ...690.1236I}. Due to
the low number of high-$S/N$ stars in \instrument{ACS} fields, and temporal
instabilities of \instrument{HST}, the PSF model was obtained by PCA of the PSF
pattern from dense stellar fields. In this work we presented a five-bin tomographic
analysis, leading to constraints on the deceleration parameter $q_0 = - \ddot a
a / \dot a^2 = \Omega_{\rm m}/2 - \Omega_\Lambda$, with a detection of
acceleration ($q_0 < 0$) at 94.3\% confidence, including additional priors on
$h$ and $\Omega_{\rm b}$.

% Howto create this: see notes

\stoptwocol
\begin{figure}
  \begin{center}
    \resizebox{1.05\hsize}{!}{
      \includegraphics{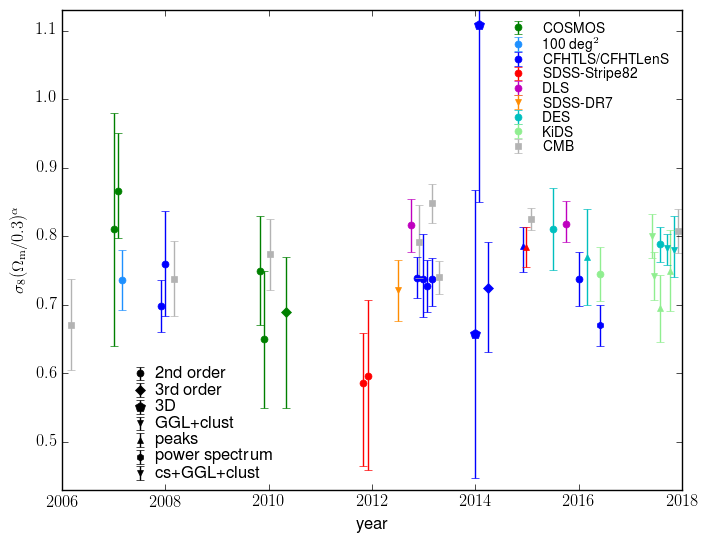}
    }
  \end{center}

    \caption{Mean and 68\% error bars for the parameter $\sigma_8 \left(\Omega_{\rm m}/0.3\right)^\alpha$,
    for various cosmic shear observations, plotted as function of their publication date
    (first arXiv submission). All parameter values are given in Table \ref{tab:Sigma}.
    Different surveys are distinguished by colour as indicated in the figure. Data points are shown
    for second-order statistics (circles), third-order (diamonds), 3D lensing (pentagons), galaxy-galaxy
    lensing (+ galaxy clustering; triangle), and CMB (squares). This plot is an updated version of Fig.~7 from
    \citet{K15}.
    }

    \label{fig:Sigma}

  \end{figure}
\begtwocol

    \bigskip\bigskip
    {\footnotesize
    \renewcommand{\arraystretch}{1.25}
\begin{longtable}{|l|l|r@{ = }l|l|p{0.25\textwidth}|}
\caption{Parameter values of $\sigma_8 \Omega_{\rm m}^\alpha$
used for Fig.~\ref{fig:Sigma}. For different surveys (first column)
            the second column is the value rescaled to $\Omega_{\rm m} = 0.3$ which is plotted in the figure,
            obtained from
            the original measurement (third column). The fourth column indicates
            the number of redshift bins, the column 5 is the reference.
                This table is an updated version
            of Table 1 from Kilbinger (2015). \\ \\
    {\scriptsize $^a$ 5 narrow photo-$z$ bins and one wide bin of faint galaxies.\\[0.1em]
    $^b$ 3D lensing, no $z$-binning.\\[0.1em]
    $^c$ The index $\alpha = 0.5$ is adopted
            from the WMAP9 measurement, published on \texttt{http://lambda.gsfc.nasa.gov}.
            The resulting values and errors are therefore only illustrative.}
}
\label{tab:Sigma}\\
\hline\hline
\rule[-3mm]{0em}{8mm}Survey	 &$\sigma_8(\Omega_{\rm{m}}/0.3)^\alpha$	 &\multicolumn{2}{c|}{original\phantom{x}measurement}	 &$N_z$	 &Reference	\\\cline{1-1}\cline{2-2}\cline{3-3}\cline{4-4}\cline{5-5}\cline{6-6}\hline
COSMOS	 &$0.810\pm{0.17}$	 &$\sigma_8(\Omega_{\rm{m}}/0.3)^{0.48}$ & $0.81\pm{0.17}$	 &1	 &\citet{2007ApJS..172..239M}	\\\cline{1-1}\cline{2-2}\cline{3-3}\cline{4-4}\cline{5-5}\cline{6-6}
COSMOS	 &$0.866^{+0.085}_{-0.068}$	 &$\sigma_8(\Omega_{\rm{m}}/0.3)^{0.44}$ & $0.866^{+0.085}_{-0.068}$	 &3	 &\citet{2007ApJS..172..239M}	\\\cline{1-1}\cline{2-2}\cline{3-3}\cline{4-4}\cline{5-5}\cline{6-6}
100\phantom{{x}}deg$^2$	 &$0.736\pm{0.0438}$	 &$\sigma_8(\Omega_{\rm{m}}/0.24)^{0.59}$ & $0.84\pm{0.05}$	 &1	 &\citet{JonBen07}	\\\cline{1-1}\cline{2-2}\cline{3-3}\cline{4-4}\cline{5-5}\cline{6-6}
CFHTLS	 &$0.699\pm{0.0383}$	 &$\sigma_8(\Omega_{\rm{m}}/0.25)^{0.64}$ & $0.785\pm{0.043}$	 &1	 &\citet{FSHK08}	\\\cline{1-1}\cline{2-2}\cline{3-3}\cline{4-4}\cline{5-5}\cline{6-6}
CFHTLS	 &$0.760\pm{0.0763}$	 &$\sigma_8(\Omega_{\rm{m}}/0.25)^{0.53}$ & $0.837\pm{0.084}$	 &1	 &\citet{FSHK08},\phantom{x}large\phantom{{x}}scales	\\\cline{1-1}\cline{2-2}\cline{3-3}\cline{4-4}\cline{5-5}\cline{6-6}
COSMOS	 &$0.750\pm{0.08}$	 &$\sigma_8(\Omega_{\rm{m}}/0.3)^{0.51}$ & $0.75\pm{0.08}$	 &$5\!+\!1^a$	 &\citet{SHJKS09}	\\\cline{1-1}\cline{2-2}\cline{3-3}\cline{4-4}\cline{5-5}\cline{6-6}
COSMOS	 &$0.650\pm{0.1}$	 &$\sigma_8(\Omega_{\rm{m}}/0.3)^{0.62}$ & $0.65\pm{0.1}$	 &1	 &\citet{SHJKS09}	\\\cline{1-1}\cline{2-2}\cline{3-3}\cline{4-4}\cline{5-5}\cline{6-6}
COSMOS	 &$0.69^{+0.08}_{-0.14}$	 &$\sigma_8(\Omega_{\rm{m}}/0.3)^{0.46}$ & $0.69^{+0.08}_{-0.14}$	 &1	 &\citet{2011MNRAS.410..143S},\phantom{x}$+$IA	\\\cline{1-1}\cline{2-2}\cline{3-3}\cline{4-4}\cline{5-5}\cline{6-6}
SDSS-Stripe82	 &$0.585^{+0.0743}_{-0.121}$	 &$\sigma_8(\Omega_{\rm{m}}/1.0)^{0.7}$ & $0.252^{+0.032}_{-0.052}$	 &1	 &\citet{2012ApJ...761...15L}	\\\cline{1-1}\cline{2-2}\cline{3-3}\cline{4-4}\cline{5-5}\cline{6-6}
SDSS-Stripe82	 &$0.597^{+0.11}_{-0.138}$	 &$\sigma_8(\Omega_{\rm{m}}/0.264)^{0.67}$ & $0.65^{+0.12}_{-0.15}$	 &1	 &\citet{2011arXiv1112.3143H}	\\\cline{1-1}\cline{2-2}\cline{3-3}\cline{4-4}\cline{5-5}\cline{6-6}
SDSS-DR7	 &$0.721\pm{0.0451}$	 &$\sigma_8(\Omega_{\rm{m}}/0.25)^{0.57}$ & $0.8\pm{0.05}$	 &1	 &\citet{2013MNRAS.432.1544M}	\\\cline{1-1}\cline{2-2}\cline{3-3}\cline{4-4}\cline{5-5}\cline{6-6}
DLS	 &$0.816\pm{0.0385}$	 &$\sigma_8(\Omega_{\rm{m}}/0.265)^{0.5}$ & $0.868\pm{0.041}$	 &1	 &\citet{2012arXiv1210.2732J},\phantom{x}priv.~comm.	\\\cline{1-1}\cline{2-2}\cline{3-3}\cline{4-4}\cline{5-5}\cline{6-6}
CFHTLenS	 &$0.740\pm{0.0301}$	 &$\sigma_8(\Omega_{\rm{m}}/0.27)^{0.59}$ & $0.787\pm{0.032}$	 &1	 &\citet{CFHTLenS-2pt-notomo}	\\\cline{1-1}\cline{2-2}\cline{3-3}\cline{4-4}\cline{5-5}\cline{6-6}
CFHTLenS	 &$0.738^{+0.0654}_{-0.056}$	 &$\sigma_8(\Omega_{\rm{m}}/0.27)^{0.65}$ & $0.79^{+0.07}_{-0.06}$	 &1	 &\citet{CFHTLenS-2pt-notomo},\phantom{x}large\phantom{{x}}scales	\\\cline{1-1}\cline{2-2}\cline{3-3}\cline{4-4}\cline{5-5}\cline{6-6}
CFHTLenS	 &$0.728\pm{0.0377}$	 &$\sigma_8(\Omega_{\rm{m}}/0.27)^{0.55}$ & $0.771\pm{0.04}$	 &2	 &\citet{CFHTLenS-2pt-tomo}	\\\cline{1-1}\cline{2-2}\cline{3-3}\cline{4-4}\cline{5-5}\cline{6-6}
CFHTLenS	 &$0.737^{+0.0305}_{-0.0391}$	 &$\sigma_8(\Omega_{\rm{m}}/0.27)^{0.46}$ & $0.774^{+0.032}_{-0.041}$	 &6	 &\citet{CFHTLenS-IA},\phantom{x}$+$IA	\\\cline{1-1}\cline{2-2}\cline{3-3}\cline{4-4}\cline{5-5}\cline{6-6}
CFHTLenS	 &$0.657\pm{0.21}$	 &$\sigma_8(\Omega_{\rm{m}}/0.27)^{0.46}$ & $0.69\pm{0.22}$	 &$\infty^b$	 &\citet{CFHTLenS-3d},\phantom{x}$+$bary	\\\cline{1-1}\cline{2-2}\cline{3-3}\cline{4-4}\cline{5-5}\cline{6-6}
CFHTLenS	 &$1.107\pm{0.258}$	 &$\sigma_8(\Omega_{\rm{m}}/0.27)^{0.44}$ & $1.16\pm{0.27}$	 &$\infty^b$	 &\citet{CFHTLenS-3d},\phantom{x}large\phantom{{x}}scales,$+$bary	\\\cline{1-1}\cline{2-2}\cline{3-3}\cline{4-4}\cline{5-5}\cline{6-6}
CFHTLenS	 &$0.725^{+0.0659}_{-0.0942}$	 &$\sigma_8(\Omega_{\rm{m}}/0.27)^{0.57}$ & $0.77^{+0.07}_{-0.1}$	 &1	 &\citet{CFHTLenS-2+3pt},\phantom{x}$+$IA	\\\cline{1-1}\cline{2-2}\cline{3-3}\cline{4-4}\cline{5-5}\cline{6-6}
CFHTLenS	 &$0.785^{+0.028}_{-0.0374}$	 &$\sigma_8(\Omega_{\rm{m}}/0.27)^{0.64}$ & $0.84^{+0.03}_{-0.04}$	 &1	 &\citet{2014arXiv1412.0757L}	\\\cline{1-1}\cline{2-2}\cline{3-3}\cline{4-4}\cline{5-5}\cline{6-6}
SDSS-Stripe82	 &$0.785\pm{0.0287}$	 &$\sigma_8(\Omega_{\rm{m}}/0.27)^{0.42}$ & $0.82\pm{0.03}$	 &1	 &\citet{2014arXiv1412.3683L}	\\\cline{1-1}\cline{2-2}\cline{3-3}\cline{4-4}\cline{5-5}\cline{6-6}
DES-SV	 &$0.811^{+0.059}_{-0.06}$	 &$\sigma_8(\Omega_{\rm{m}}/0.3)^{0.478}$ & $0.811^{+0.059}_{-0.06}$	 &3	 &\citet{2015arXiv150705552T}	\\\cline{1-1}\cline{2-2}\cline{3-3}\cline{4-4}\cline{5-5}\cline{6-6}
DLS	 &$0.818^{+0.034}_{-0.026}$	 &$\sigma_8(\Omega_{\rm{m}}/0.3)^{0.5}$ & $0.818^{+0.034}_{-0.026}$	 &5	 &\citet{2015arXiv151003962J}	\\\cline{1-1}\cline{2-2}\cline{3-3}\cline{4-4}\cline{5-5}\cline{6-6}
CFHTLenS	 &$0.738\pm{0.0393}$	 &$\sigma_8(\Omega_{\rm{m}}/1.0)^{0.5}$ & $0.404\pm{0.0215}$	 &7	 &\citet{2016arXiv160105786J},\phantom{x}$+$IA,bary,$z$-sys	\\\cline{1-1}\cline{2-2}\cline{3-3}\cline{4-4}\cline{5-5}\cline{6-6}
DES-SV	 &$0.770\pm{0.07}$	 &$\sigma_8(\Omega_{\rm{m}}/0.3)^{0.6}$ & $0.77\pm{0.07}$	 &1	 &\citet{2016arXiv160305040K}	\\\cline{1-1}\cline{2-2}\cline{3-3}\cline{4-4}\cline{5-5}\cline{6-6}
CFHTLenS	 &$0.670\pm{0.03}$	 &$\sigma_8(\Omega_{\rm{m}}/0.3)^{0.5}$ & $0.67\pm{0.03}$	 &2	 &\citet{2016arXiv160700008A},\phantom{x}$z$-sys	\\\cline{1-1}\cline{2-2}\cline{3-3}\cline{4-4}\cline{5-5}\cline{6-6}
KiDS	 &$0.745\pm{0.039}$	 &$\sigma_8(\Omega_{\rm{m}}/0.3)^{0.5}$ & $0.745\pm{0.039}$	 &4	 &\citet{KiDS-450},\phantom{x}$+$IA,bary,$\Delta{n(z)}$	\\\cline{1-1}\cline{2-2}\cline{3-3}\cline{4-4}\cline{5-5}\cline{6-6}
KiDS	 &$0.801\pm{0.032}$	 &$\sigma_8(\Omega_{\rm{m}}/0.3)^{0.5}$ & $0.801\pm{0.032}$	 &4	 &\citet{2017arXiv170605004V},\phantom{x}cs+ggl+clust,band-power,$+$IA,bary,$\Delta{n(z)}$	\\\cline{1-1}\cline{2-2}\cline{3-3}\cline{4-4}\cline{5-5}\cline{6-6}
KiDS	 &$0.742\pm{0.035}$	 &$\sigma_8(\Omega_{\rm{m}}/0.3)^{0.5}$ & $0.742\pm{0.035}$	 &4	 &\citet{2017arXiv170706627J},\phantom{x}cs+ggl+RSD,$+$I-A,bary,$\Delta{n(z)}$	\\\cline{1-1}\cline{2-2}\cline{3-3}\cline{4-4}\cline{5-5}\cline{6-6}
KiDS	 &$0.696^{+0.048}_{-0.05}$	 &$\sigma_8(\Omega_{\rm{m}}/0.3)^{0.38}$ & $0.696^{+0.048}_{-0.05}$	 &1	 &\citet{2018MNRAS.474.1116S},\phantom{x}high-SNR-peaks	\\\cline{1-1}\cline{2-2}\cline{3-3}\cline{4-4}\cline{5-5}\cline{6-6}
DES-Y1	 &$0.789^{+0.024}_{-0.026}$	 &$\sigma_8(\Omega_{\rm{m}}/0.3)^{0.5}$ & $0.789^{+0.024}_{-0.026}$	 &4	 &\citet{2017arXiv170801538T}	\\\cline{1-1}\cline{2-2}\cline{3-3}\cline{4-4}\cline{5-5}\cline{6-6}
DES-Y1	 &$0.783^{+0.021}_{-0.025}$	 &$\sigma_8(\Omega_{\rm{m}}/0.3)^{0.5}$ & $0.783^{+0.021}_{-0.025}$	 &4	 &\citet{2017arXiv170801530D}	\\\cline{1-1}\cline{2-2}\cline{3-3}\cline{4-4}\cline{5-5}\cline{6-6}
KiDS	 &$0.750\pm{0.059}$	 &$\sigma_8(\Omega_{\rm{m}}/0.3)^{0.5}$ & $0.75\pm{0.059}$	 &1	 &\citet{2018MNRAS.474..712M},\phantom{x}low-SNR-peaks	\\\cline{1-1}\cline{2-2}\cline{3-3}\cline{4-4}\cline{5-5}\cline{6-6}
DES-Y1	 &$0.78^{+0.05}_{-0.04}$	 &$\sigma_8(\Omega_{\rm{m}}/0.3)^{0.5}$ & $0.78^{+0.05}_{-0.04}$	 &4-5	 &\citet{2017arXiv171005045G},\phantom{x}density-split-statistic	\\\cline{1-1}\cline{2-2}\cline{3-3}\cline{4-4}\cline{5-5}\cline{6-6}
WMAP3$^c$	 &$0.671\pm{0.0669}$	 &$\sigma_8(\Omega_{\rm{m}}/0.234)^{0.5}$ & $0.76\pm{0.05}$	 &-	 &\citet{2007ApJS..170..377S}	\\\cline{1-1}\cline{2-2}\cline{3-3}\cline{4-4}\cline{5-5}\cline{6-6}
WMAP5$^c$	 &$0.738\pm{0.0544}$	 &$\sigma_8(\Omega_{\rm{m}}/0.258)^{0.5}$ & $0.796\pm{0.036}$	 &-	 &\citet{WMAP5-Komatsu08}	\\\cline{1-1}\cline{2-2}\cline{3-3}\cline{4-4}\cline{5-5}\cline{6-6}
WMAP7$^c$	 &$0.774^{+0.0512}_{-0.0518}$	 &$\sigma_8(\Omega_{\rm{m}}/0.273)^{0.5}$ & $0.811^{+0.03}_{-0.031}$	 &-	 &\citet{2010arXiv1001.4538K}	\\\cline{1-1}\cline{2-2}\cline{3-3}\cline{4-4}\cline{5-5}\cline{6-6}
WMAP9	 &$0.792\pm{0.0529}$	 &$\sigma_8(\Omega_{\rm{m}}/1.0)^{0.5}$ & $0.434\pm{0.029}$	 &-	 &\citet{2012arXiv1212.5226H}	\\\cline{1-1}\cline{2-2}\cline{3-3}\cline{4-4}\cline{5-5}\cline{6-6}
Planck2013	 &$0.848\pm{0.0286}$	 &$\sigma_8(\Omega_{\rm{m}}/0.27)^{0.46}$ & $0.89\pm{0.03}$	 &-	 &\citet{2013arXiv1303.5076P},\phantom{x}$C(\ell)$	\\\cline{1-1}\cline{2-2}\cline{3-3}\cline{4-4}\cline{5-5}\cline{6-6}
Planck2013	 &$0.740\pm{0.0242}$	 &$\sigma_8(\Omega_{\rm{m}}/0.27)^{0.3}$ & $0.764\pm{0.025}$	 &-	 &\citet{2013arXiv1303.5080P},\phantom{x}SZ	\\\cline{1-1}\cline{2-2}\cline{3-3}\cline{4-4}\cline{5-5}\cline{6-6}
Planck2015	 &$0.825\pm{0.0161}$	 &$\sigma_8(\Omega_{\rm{m}}/1.0)^{0.5}$ & $0.4521\pm{0.0088}$	 &-	 &\citet{2015arXiv150201589P},\phantom{x}$C(\ell)$	\\\cline{1-1}\cline{2-2}\cline{3-3}\cline{4-4}\cline{5-5}\cline{6-6}
Planck2015+SPT	 &$0.808\pm{0.0324}$	 &$\sigma_8(\Omega_{\rm{m}}/1.0)^{0.25}$ & $0.598\pm{0.024}$	 &-	 &\citet{2017arXiv171207541S}	\\\cline{1-1}\cline{2-2}\cline{3-3}\cline{4-4}\cline{5-5}\cline{6-6}
\end{longtable}

    }

%%%%%%%%%%%%%%%%%%%%%%%%%%%%%%%%%%%%%%%%%%%%%%%%%%%%%%%%%%%%
\subsubsection{CFHTLenS, 2012 - 2015}
\label{survey_era}
%%%%%%%%%%%%%%%%%%%%%%%%%%%%%%%%%%%%%%%%%%%%%%%%%%%%%%%%%%%%

A milestone for cosmic shear represented the CFHT lensing survey
\citeaffixed{CFHTLenS-data}{\survey{CFHTLenS}; }. With CFHTLenS we provided
measurements that relied on independently cross-checked photometric redshifts,
and a robust estimate of residual systematics on weak-lensing shear
correlations (Sect.~\ref{sec:error-model}). In these studies, the residual
sytematics analysis was done completely independently from the cosmological
parameters analysis, in order not to bias the cosmological results.

\stoptwocol
  \begin{figure}

     \centerline{Flat $\Lambda$CDM}

        \resizebox{1.0\hsize}{!}{	
          \includegraphics{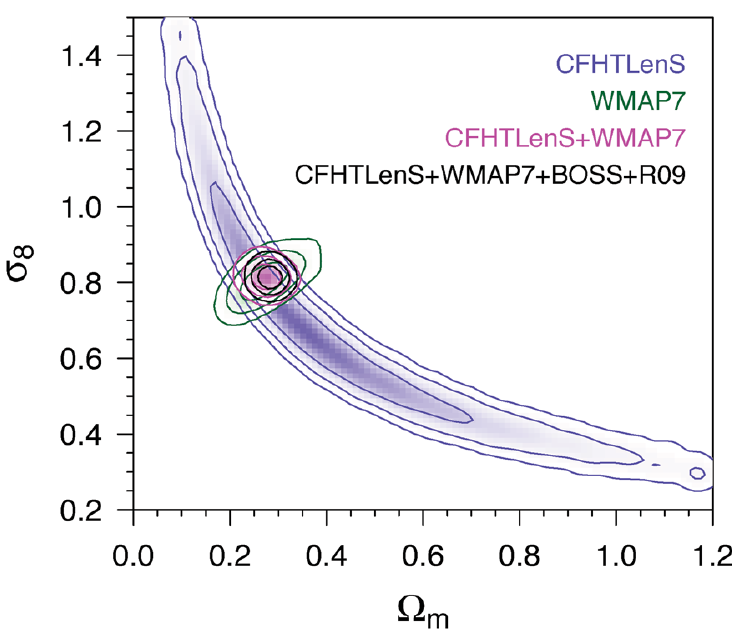}
          \includegraphics{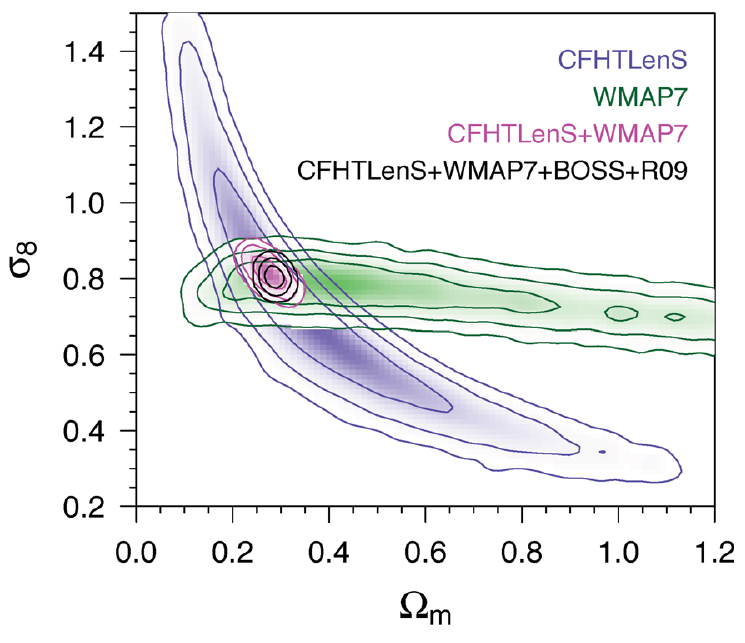}
        }

     \caption{The %near-orthogonality of $\Omega_{\rm m}$ and $\sigma_8$ constraints from
     2D cosmic shear and CMB.
     CFHTLenS, WMAP7, BAO from BOSS \cite{2012arXiv1203.6594A}, and a HST $H_0$ prior
     \cite[`R09']{2009ApJ...699..539R}. From \citet{CFHTLenS-2pt-notomo}.
     }

     \label{fig:Om_s8_flatLCDM}

  \end{figure}
\begtwocol

Photometric redshifts for each source galaxy were obtained in
\citet{CFHTLenS-photoz}, the robustness of which was verified using
spectroscopic redshifts, \survey{COSMOS} 30-band photo-$z$s, and a
cross-correlation analysis \cite{CFHTLenS-2pt-tomo}. Galaxy shapes were
measured on individual exposures with \emph{lens}fit and calibrated using two
independent suites of image simulations \cite{CFHTLenS-shapes}. An excess
correlation between star and galaxy shapes (\ref{eq:xi_sys}) was found on
$25\%$ of the observed fields, which in turn were discarded from the
cosmological analysis \cite{CFHTLenS-sys}. Two-dimensional cosmic shear
correlation functions from \survey{CFHTLenS} were presented in
\citet{CFHTLenS-2pt-notomo}. A two-bin tomographic analysis was performed by
\citet{CFHTLenS-2pt-tomo}. The same tomographic data were used to place
constraints on modified gravity \cite{CFHTLenS-mod-grav}. Further, a six-bin
tomographic analysis was performed where cosmological and intrinsic-alignment
parameters were constrained simultaneously \cite{CFHTLenS-IA}. Late-type
galaxies were found to not show any significant intrinsic alignment, while for
early type galaxies IA was detected at about $2\sigma$.

\stoptwocol
\begin{figure}

   \begin{center}
      %\resizebox{0.8\hsize}{!}{	
      \resizebox{1.0\hsize}{!}{	
        \includegraphics{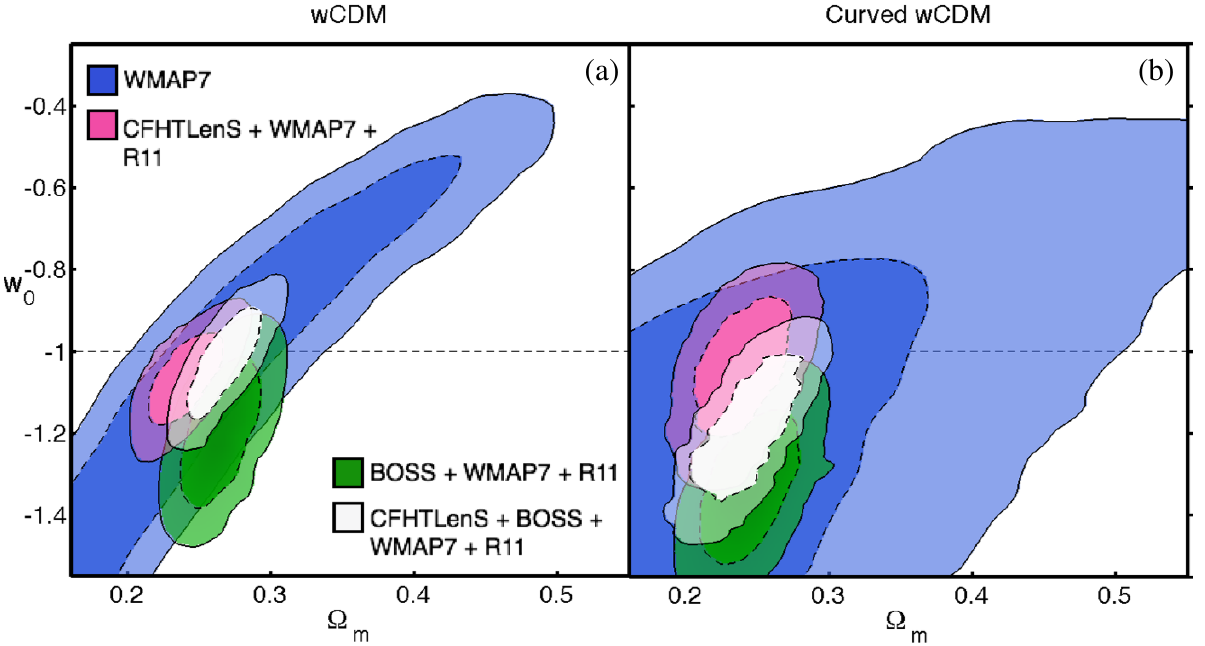}
      }
   \end{center}

   \caption{Combined constraints on $\Omega_{\rm m}$ and $w_0$ from cosmic shear, CMB, and BAO.
	The model is a $w$CDM universe with flat (free) curvature in 
	panel a (b). Cosmic shear is six-bin tomography from CFHTLenS. The CMB and BAO
	data are the same as in Fig.~\ref{fig:Om_s8_flatLCDM}. The HST $H_0$ prior is replaced with an 
	updated version \cite{2011ApJ...730..119R}. From \citet{CFHTLenS-IA}.
  }

   \label{fig:Om_w0}

\end{figure}
\begtwocol

For a $\Lambda$CDM cosmology, cosmic shear constrains a combination of
$\Omega_{\rm m}$ and $\sigma_8$ that is perpendicular to the one obtained from
CMB \cite{Contaldi03}. Adding cosmic shear to \survey{WMAP (Wilkinson Microwave
Anisotropy Probe)} results in typical reduction of error bars on $\Omega_{\rm
m}$ and $\sigma_8$ of up to $50\%$, similar to other low-$z$ cosmological
probes such as Baryonic Acoustic Oscillations (BAO). For example, the
\survey{WMAP7} constraints of $\Omega_{\rm m} = 0.273 \pm 0.03$ and $\sigma_8 =
0.811 \pm 0.031$ \cite{2010arXiv1001.4538K} get tightened when adding
\survey{CFHTLenS}, resulting in $\Omega_{\rm m} = 0.274 \pm 0.013$ and
$\sigma_8 = 0.815 \pm 0.016$ \cite{CFHTLenS-2pt-notomo}, see
Fig.~\ref{fig:Om_s8_flatLCDM}. Planck's cosmological findings from temperature
anisotropies (together with CMB lensing and WMAP polarization) correspond to a
higher matter density and normalization compared to most previous pobes, with
$\Omega_{\rm m} = 0.315 \pm 0.017$ and $\sigma_8 = 0.829 \pm 0.012$, or
$\sigma_8 (\Omega_{\rm m} / 0.27)^{0.46} = 0.89 \pm 0.03$
\cite{2013arXiv1303.5076P}. This is consistent with CFHTLenS at the $2\sigma$
level, see Fig.~\ref{fig:Om_s8_Planck}. Further, Planck's counts of
Sunyaev-Zel'dovich (SZ) clusters results in a lower normalization of $\sigma_8
(\Omega_{\rm m} / 0.27)^{0.3} = 0.78 \pm 0.01$ \cite{2013arXiv1303.5080P}.
Sect.~\ref{sec:further_results_CFHTLenS} discusses whether adding
extra-parameters such as massive neutrinos are needed to reconcile recent high-
and low-$z$ data.

A model with variable curvature does not change the cosmic-shear constraints
on $\Omega_{\rm m}$ and $\sigma_8$ by a lot. Pre-\survey{Planck} CMB data alone
cannot constrain the curvature of the Universe, and adding other probes such as
measurements of $H_0$ or weak lensing are required. \survey{Planck} and
high-resolution ground-based millimetre-wavelength radio telescopes of similar
sensitivity and resolution such as \instrument{SPT (South Pole Telescope)} and
\instrument{ACT (Atacama Cosmology Telescope)} have detected weak-lensing of
the CMB by large-scale structures (\emph{CMB lensing}), which helps to break
the geometrical degeneracy. This results in tight constraints on $\Omega_{K}$
from CMB alone
\cite{2011PhRvL.107b1302S,2012ApJ...756..142V,2013arXiv1303.5076P}.
Fig.~\ref{fig:Om_s8_Planck} shows joint cosmic shear and CMB constraints for a
free-curvature model.

Since the effect of dark energy on the supression of the growth of structure is
relatively small, 2D weak lensing is not very sensitive to dark energy.
Tomographic weak lensing can place interesting constraints on the dark-energy
equation of state parameter. Fig.~\ref{fig:Om_w0} shows how CMB constraints
from \survey{WMAP7} --- with an additional prior on $H_0$ from
\citet{2011ApJ...730..119R} --- are reduced by \survey{CFHTLenS} six-bin
tomography. The parameters $\Omega_{\rm m}$ and $w_0$ are measured to better
than $10\%$ accuracy, for both a flat and free-curvature $w$CDM model. The
improvement is similar to adding Baryonic Acoustic Oscillation (BAO) data from
the \survey{SDSS-III Baryon Oscillation Spectroscopic Survey}
\citeaffixed{2012arXiv1203.6594A}{\survey{BOSS}; } to CMB data.

Constraints on modified gravity using the parametrization in
eqs.~(\ref{eq:Poisson_mod_Psi}, \ref{eq:Poisson_mod_Phi_plus_Psi}) showed
consistency with GR \cite{CFHTLenS-mod-grav}. A simple model was considered
where $\Sigma$ and $\mu$ did not vary spatially, and at early times tend
towards GR, so that deviations of GR are allowed at late times where the
accelerated expansion happens. The present-day values of those two parameters
were measured to be $\Sigma_0 = 0.00 \pm 0.14$, and $\mu_0 = 0.05 \pm 0.25$,
combining \survey{CFHTLenS} weak-lensing tomographic data
\cite{CFHTLenS-2pt-tomo}, redshift-space distortions from \survey{WiggleZ}
\cite{2012MNRAS.425..405B} and \survey{6dFGS} \cite{2012MNRAS.423.3430B},
\survey{WMAP7} CMB anisotropies from small scales, $\ell \ge 100$
\cite{2011ApJS..192...16L}, and the \citet{2011ApJ...730..119R} $H_0$ prior
(see Fig.~\ref{fig:modgrav}).

\stoptwocol
  \begin{figure}

    \resizebox{1.0\hsize}{!}{
        \includegraphics{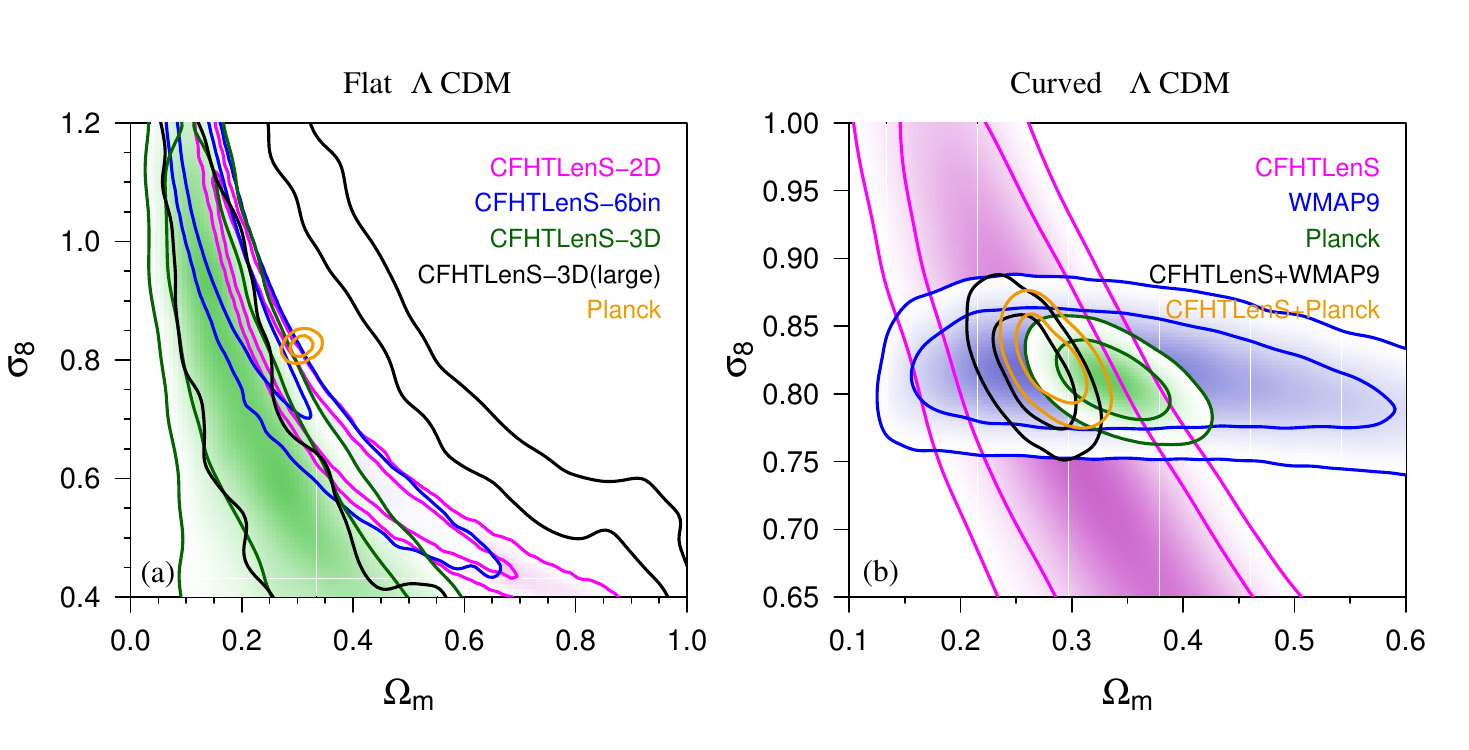}
      }

   \caption{Cosmic shear and CMB $68.3\%$ and $95.5\%$ confidence levels for $\Omega_{\rm m}$ and $\sigma_8$
  in a $\Lambda$CDM universe. (a) Assuming flatness.
  CFHTLenS 2D, 6-bin tomography, 3D, and 3D from large scales only are compared to Planck constraints.
  (b) With free curvature, showing CFHTLenS (joint second- and third-order), WMAP9, Planck,
  CFHTLenS $+$ WMAP9, and CFHTLenS $+$ Planck constraints,
	from \citet{CFHTLenS-2+3pt}.
  }

   \label{fig:Om_s8_Planck}

\end{figure}
\begtwocol

%%%%%%%%%%%%%%%%%%%%%%%%%%%%%%%
\subsubsection{Further results from CFHTLenS}
\label{sec:further_results_CFHTLenS}
%%%%%%%%%%%%%%%%%%%%%%%%%%%%%%%

Other groups different from the CFHTLenS collaboration have used those data for further, extended analysis.
The cosmological constraints from CMB temperature anisotropies measured by the
Planck satellite \cite{2013arXiv1303.5076P} seem to be in slight tension with
other probes. In particular, Planck found a higher power-spectrum normalisation
$\sigma_8$. Several works proposed massive neutrinos to alleviate the tension
with low-$z$ probes such as weak lensing: Massive neutrinos are still
relativistic at recombination and do not significantly influence the CMB
anisotropies. They become however non-relativistic at late time, and dampen the
growth of structure, therefore reducing the low-$z$ clustering power. Joint
analyses including massive neutrinos from Planck and our CFHTLenS weak-lensing data of
\citet{CFHTLenS-2pt-notomo} were found to improve parameter constraints with
the inclusion of non-zero neutrino masses
\cite{2014PhRvL.112e1303B,2014arXiv1403.4599B}, but the evidence still favors a
$\Lambda$CDM model without additional parameters for massive neutrinos
\cite{2014arXiv1404.5950L}.

\begin{figure}

  \begin{minipage}{0.6\textwidth}
   \centerline{Flat $\Lambda$CDM}

   \begin{center}
      \resizebox{1.0\hsize}{!}{ 
        \includegraphics{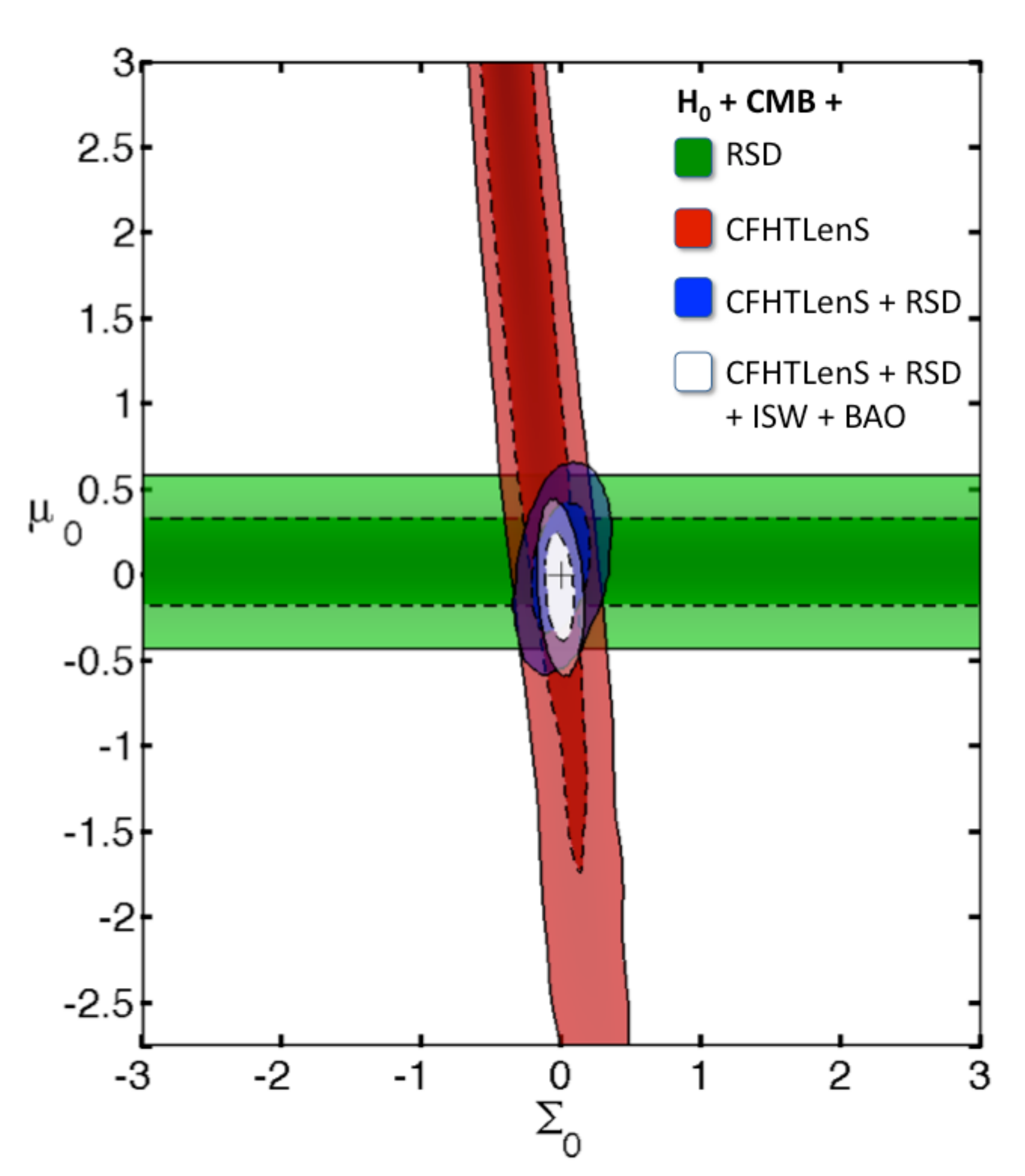}
      }
   \end{center}
  \end{minipage}%
  \hspace*{-0.1\textwidth}%
  \begin{minipage}{0.5\textwidth}

   \caption{Combined constraints on the present-day modified-gravity parameters
    $\Sigma_0$ and $\mu_0$, from redshift-space distortions (RSD), cosmic shear
    (CFHTLenS), and their combination, including the case of additional BAO
    \cite{2012arXiv1203.6594A} and large-scale WMAP7 (ISW) data. All data are
    combined with a $H_0$ prior and small-scale CMB data, see text. From
    \citet{CFHTLenS-mod-grav}. 
    }
   \label{fig:modgrav}

  \end{minipage}

\end{figure}

%%%%%%%%%%%%%%%%%%%%%%%%%%%%%%%%%%%%%%%%%%%%%%%%%%%%%%%%%%%%
\subsection{Higher-order correlations}
\label{sec:higher-order-results}
\label{sec:third_order}
%%%%%%%%%%%%%%%%%%%%%%%%%%%%%%%%%%%%%%%%%%%%%%%%%%%%%%%%%%%%

The motivation behind higher-order shear statistics has been argued for in
Sect.~\ref{sec:higher-order}. Even though the measurement is challenging and the
overall signal-to-noise ratio is low, several significant detections of
third-order shear correlations have been made.

Several higher-order measurements resulted from CFHTLenS. First,
\citet{CFHTLenS-kappa-maps} measured the skewness of reconstructed convergence
maps and found good agreement with WMAP7 predictions. After validating the data
for shear residual third-order correlations, \citet{CFHTLenS-3pt} performed a
cosmological analysis of the aperture-mass skewness exploring a non-Gaussian
likelihood. \citet{CFHTLenS-2+3pt} combined the second- and third-order
aperture-mass combined with \survey{WMAP9} and \survey{Planck} to obtain
cosmological results (see Fig.~\ref{fig:Om_s8_Planck}), including models of
intrinsic alignment and source-lens clustering as astrophysical systematics.

%%%%%%%%%%%%%%%%%%%%%%%%%%%%%%%%%%%%%%%%%%%%%%%%%%%%%%%%%%%%
%\subsection{Peak counts}
\label{sec:peak_counts_results}
%%%%%%%%%%%%%%%%%%%%%%%%%%%%%%%%%%%%%%%%%%%%%%%%%%%%%%%%%%%%

Quite a few measurements of weak-lensing peak counts were obtained using data
from the \survey{CFHT/MegaCam Stripe-82} survey (\survey{CS82}),
\survey{CFHTLenS}, the \survey{Dark Energy Survey (DES)}, and the
\survey{Kilo-Degree Survey (KiDS)}
\cite{2013arXiv1311.1319S,2014arXiv1412.0757L,2014arXiv1412.3683L,2016arXiv160305040K,2018MNRAS.474..712M}.
Preliminary measurements on the combination of \survey{DES} (SVD; science
verification data), \survey{KiDS} data release DR1/2, and \survey{CFHTLenS},
and comparison to our fast stochastic models are presented and discussed in
Chieh-An Lin's PhD thesis \cite{2016arXiv161204041L}.

%%%%%%%%%%%%%%%%%%%%%%%%%%%%%%%%%%%%%%%%%%%%%%%%%%%%%%%%%%%%
\subsection{Intrinsic alignment}
\label{sec:ia_results}
%%%%%%%%%%%%%%%%%%%%%%%%%%%%%%%%%%%%%%%%%%%%%%%%%%%%%%%%%%%%

In \citet{FSHK08} we tried to obtain constraints on intrinsic alignment. Since
we had no redshift information and therefore one large redshift bin, the
different contributions to the shape correlation function of cosmic shear
($GG$) and shear-shape intrinsic alignment ($GI$) are not easily separated. The
shape-shape alignment ($II$) cannot be resolved with a wide redshift
distribution. We therefore obtained a non-detection of $GI$, with $A =
2.2^{+3.8}_{-4.6} \times 10^{-7} h/{\rm Mpc}$, using the simple model of
\citet{Heymans06}. No evidence pointed to the presence of $GI$ above the
statistical and systematic error level.

A $2\sigma$ detection of intrinsic alignment from early-type galaxies was
obtained by jointly fitting cosmology and the \citet{2004MNRAS.353..529H}
linear IA model to \survey{CFHTLenS} cosmic shear tomographic data
\cite{CFHTLenS-IA}. Fig.~\ref{fig:II-GI} shows a systematic lower amplitude of
shear correlation for cross-redshift correlations compared to the cosmic-shear
prediction, as expected from a negative $GI$ contribution. No detection was
found for the late-type sample.

\doifonecol{%
\begin{figure}

  \centerline{\resizebox{0.7\hsize}{!}{
    \includegraphics{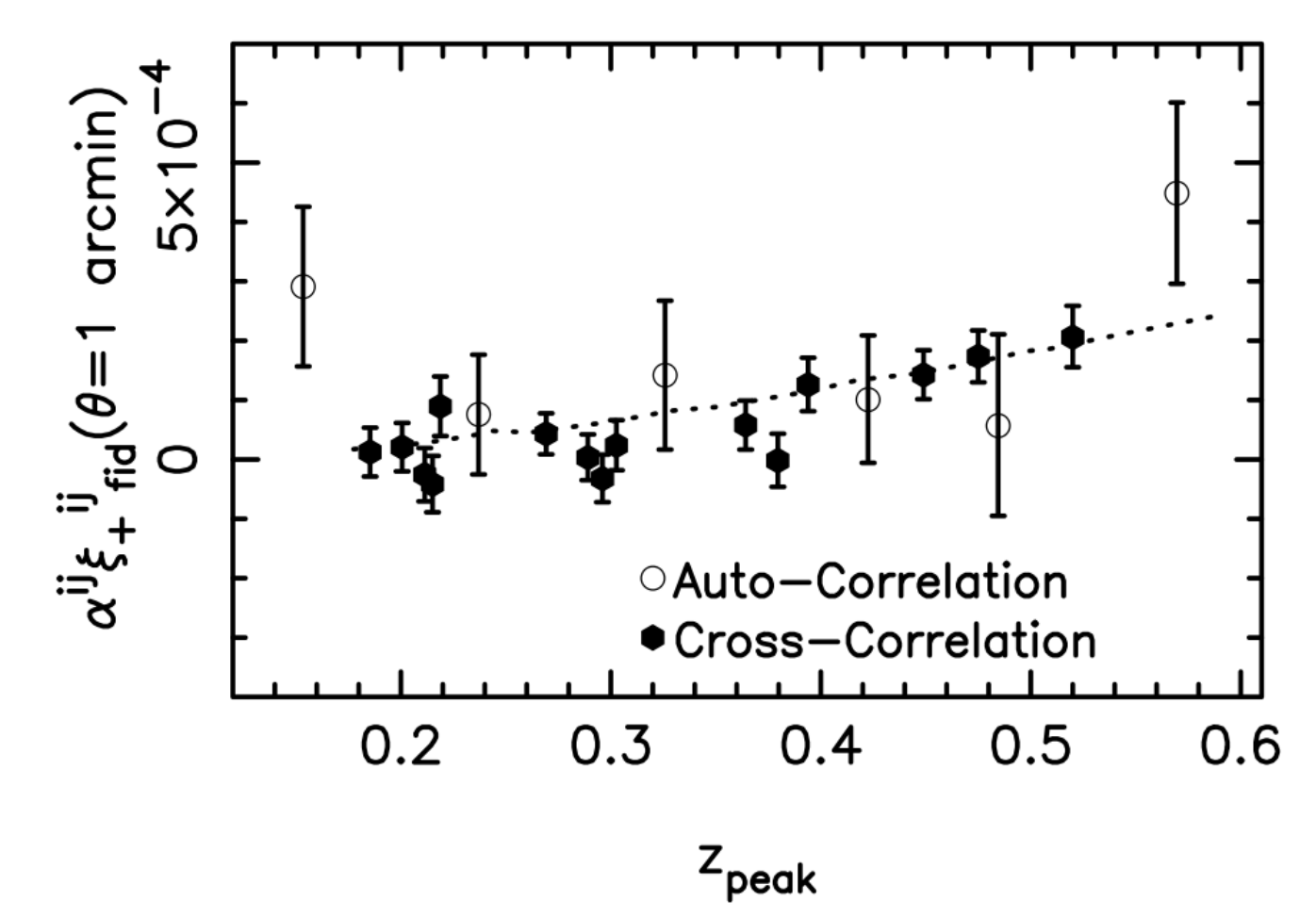}
  }}
}
\doiftwocol{%
\begin{figure}[H]

  \centerline{\resizebox{\hsize}{!}{
    \includegraphics[bb=20 0 385 280, height=10em, width=10em]{figures/fig12.pdf}
  }}
}

  \caption{Amplitudes of tomographic measures of $\xi_+$ at $\theta = 1$ arcmin
for redshift bins $(ij)$, against peak lensing efficiency redshift $z_{\rm
peak}$ for early-type galaxies, from \survey{CFHTLenS}. The free parameters
$\alpha^{ij}$ multiplied with a WMAP7 fiducial $GG$ model $\xi_{+ \rm fid}$
were fitted to $\xi_+$ and $\xi_-$, simultaneous for all redshift bins and
angular scales. At low $z$, the auto-correlations ($i = j$, open circles) lie
above the fiducial dashed line ($a^{ij} = 1$), as expected for a $II>0$
contribution. The cross-correlations ($i \ne j$, filled circles) lie
systematically below the prediction, indicating a $GI<0$ contamination. From
\citet{CFHTLenS-IA}.
}

  \label{fig:II-GI}

\end{figure}

% Chapter 8
%% future.tex

%%%%%%%%%%%%%%%%%%%%%%%%%%%%%%%%%%%%%%%%%%%%%%%%%%%%%%%%%%%%
\section{Future cosmic shear expectations and forecasts}
\label{sec:future_cs}
%%%%%%%%%%%%%%%%%%%%%%%%%%%%%%%%%%%%%%%%%%%%%%%%%%%%%%%%%%%%

%%%%%%%%%%%%%%%%%%%%%%%%%%%%%%%%%%%%%%%%%%%%%%%%%%%%%%%%%%%%
\subsection{Upcoming and future surveys}
\label{sec:upcoming_surveys}
%%%%%%%%%%%%%%%%%%%%%%%%%%%%%%%%%%%%%%%%%%%%%%%%%%%%%%%%%%%%

New instruments, either cameras, telescopes, or both, are being designed and
built specifically for the purpose of weak-lensing observations. Their design
is driven by the goal to provide superb image quality with very small, uniform,
and well-understood image distortions. The pixel scale is chosen to
sufficiently sample the PSF. In view of the enormous costs of new experiments,
in particular space missions, the instruments have to be thoroughly and
carefully designed to guarantee the desired scientific outcome, for example,
the measurement of dark-energy properties with a given accuracy.

%The KiDS survey \citeaffixed{2013ExA....35...25D}{\survey{KiDS; }} will map
%$1,500$ square degrees. Four optical bands are complemented by five deep IR
%bands observed with the \instrument{VISTA} telescope. The \survey{Dark Energy
%Survey} \citeaffixed{2005astro.ph.10346T}{\survey{DES}; }, will observe $5,000$
%square degrees in the South. The survey area overlaps with many observations in
%other wavelengths, e.g.~with the \instrument{South Pole Telescope} (SPT), and
%the (shallow) infrared \survey{Vista Hemisphere Survey} (VHS). A smaller area
%but significantly deeper limiting magnitude is provided by the
%\instrument{HyperSuprimeCam} (\instrument{HSC}) survey. Around $1,500$ square
%degrees with excellent image quality in multiple optical bands will be used for
%weak lensing, with a very high planned depth down to $i = 26$
%\cite{2012SPIE.8446E..0ZM}.

In early 2017, \survey{CFIS}, the \survey{Canada-France Imaging Survey} will
start observations. CFIS is attributed around 200 nights over three years, to
map the Northern sky in $u$ and $r$. The $u$-band part (CFIS-LUAU) will
complement the on-going LUAU program, and cover at the end $10,000$ square
degree, with a limiting magnitude of $u = 24.4$. The $r$-band survey, CFIS-WIQD
(for Wide + Image Quality + Deep)  will obverse $4,800$ deg$^2$ north of
$\delta > 30\degree$ and galactic latitude $b>25\degree$, to a depth of
$r=24.85$.

Despite only observing in two bands (with a rather shallow $u$ component), CFIS
will be very interesting for weak lensing, in particular in combination with
deep spectroscopic surveys such as eBOSS and DESI. Such deep data does not
exist in the Southern hemisphere. CFIS will also contribute to DESI target selection
and provide photometric bands for Euclid photo-$z$'s.

%These current and near-future surveys will be followed by the next generation
%of experiments that will cover most of the extragalactic sky of $15,000$ square
%degrees and more. From the ground, the $8.4 \, \mbox{m}$ Large Synoptic Survey
%Telescope \citeaffixed{2009arXiv0912.0201L}{\instrument{LSST}; } will provide
%extremely deep images down to $r = 28$. Since individual exposures are very
%short, on the order of $15$ seconds to discover transient objects, they will
%have to be stacked or otherwise combined to do weak-lensing measurements.

Going to space offers the two major advantages: Escaping atmospheric turbulence
leads to a stable and small PSF, and infrared observations provide photo-$z$s
to significantly higher redshifts than from the ground.

The accessible extra-galactic sky (the area outside the Milky Way and the ecliptic) of
$15,000$ deg$^2$ will be observed from space with the ESA satellite
mission \instrument{Euclid} \cite{2011arXiv1110.3193L}. The two main science
drivers for Euclid are cosmic shear and galaxy clustering, which will be
observed using three instruments, an optical imager, a near-infrared imager,
and a near-infrared slitless spectrograph. The optical imager on board Euclid
is designed to have a very stable PSF, both spatially as well as in the time
domain. To collect enough galaxy light from billions of high-redshift galaxies
($30 \, \mbox{arcmin}^{-2}$), the transmission curve is very broad,
corresponding to the combined $R + I + z$ filters, with a required depth of
$R+I+z = 24.5$. This poses new challenges to overcome, in particular galaxy
colour gradients and PSF calibrations).

Further obstacles unique to space-based
observations will have to be tackled \cite{2013MNRAS.431.3103C}: For example,
the very small PSF will be undersampled by the pixels of size $0.1$ arcsec.
From these undersampled stellar images, a reliable, high-resolution PSF model
has to be reconstructed. Furthermore, the detector degrades with time due to
the bombardment with cosmic rays, and the shapes of objects get distorted by
the so-called \emph{charge transfer inefficiency} (CTI). Corrections as
function of time, position on chip, and brightness of the objects have to be
applied \cite{2010MNRAS.401..371M,2014MNRAS.439..887M}.

The shear calibration of the huge data expected from Euclid to the required
formidable accuracy will be another challenge. The necessary volume of image
simulations for calibration is estimated to be huge, as well as the processing
time of those simulations \cite{2016arXiv160903281H}. It is not clear yet
whether new, alternative methods such as meta-calibration, can be used: The
undersampled galaxy images with very large wavelength range makes devonvolution
with the correct PSF challenging, and will require the development of new
methods. We are studying sparsity-approaches \cite{2017A&A...601A..66F} that we
plan to develop further for Euclid.

%A space mission proposed by NASA is \instrument{WFIRST--AFTA}
%\citeaffixed{2013arXiv1305.5422S}{Wide-Field Infrared Survey Telescope --
%Astrophysics Focused Telescope Asset; }. WFIRST--AFTA uses a $2.4$ m mirror
%with near-infrared imaging and spectroscopy capabilities. Around $2,400$ square
%degrees will be imaged for weak lensing in the near infrared, with $50$
%galaxies per square arcmin, corresponding to $J = 25.7$. The smaller area but
%higher depth compared to \instrument{Euclid} will probably result in a similar
%expected performance of \instrument{WFIRST--AFTA} with respect to constraining
%cosmological parameters.

%%%%%%%%%%%%%%%%%%%%%%%%%%%%%%%%%%%%%%%%%%%%%%%%%%%%%%%%%%%%
\subsection{Outlook}
\label{sec:outlook}
%%%%%%%%%%%%%%%%%%%%%%%%%%%%%%%%%%%%%%%%%%%%%%%%%%%%%%%%%%%%

In 2000, cosmic shear was first measured over a few square degrees
of observed sky, from some ten thousand galaxies. Fifteen years later,
surveys have increased these numbers by a factor of $100$, imaging a few million
galaxies over ${\cal O}(100)$ square degrees. Many challenges were met to analyse these
data, taking years of work. This resulted in constraints on
cosmological parameters that are competitive compared to other
cosmological probes.

In another ten years, upcoming and future experiments will cover a substantial
fraction of the entire sky, measuring billions of galaxies. This signifies
yet another data volume increase of a factor of $100$,
not to mention the data quality improvement due to
instruments dedicated to weak lensing. The formidable challenge here is reducing 
systematic errors to an acceptable level when analysing these large data sets.
New, unprecedented difficulties
have to be overcome, for example CTI for Euclid, and
blended galaxy images for LSST. To fully exploit those surveys, large
follow-up programs are needed to obtain the
necessary large samples of photometric and spectroscopic redshifts.
In addition, to interpret the results of those surveys, the accuracy of
theoretical predictions of the non-linear power spectrum including baryonic
physics need to be significantly improved.

If all these challenges can be overcome, weak cosmological lensing has the
great potential to advance our understanding of fundamental physics. It can
explore the origin of the recent accelerated expansion of the Universe, and
distinguish between dark energy models and theories of modified gravity. Cosmic
shear can measure initial conditions of the primordial Universe, constrain the
mass of neutrinos, and measure properties of dark matter. Not only that, the
study of intrinsic galaxy alignments has provided insights into the formation
and evolution of high-redshift galaxies in their dark-matter environment,
proving that cosmic shear does not only probe cosmology, but influences and
enriches other areas of astrophysics. Thus, over the last fifteen years, weak
cosmological lensing has established itself as a major tool in understanding
our Universe, and with upcoming large surveys, it will continue to be of great
value for astrophysics and cosmology.
\bigskip

\addcontentsline{toc}{section}{Acknowlegdements}
% acknowledgements.tex

\ack

I would like to thank the members of the jury for agreeing to be in the
committee of my \emph{habilitation \`a diriger des recherches} (HDR). I am very
grateful to Alain Blanchard, Martin Kunz, and Christophe Pichon, for the
thoroughness of their reports on this thesis, and thus on my work over the last
ten or so years. I would also like to thank Jim Bartlett and Nick Kaiser for
their roles as examiners, and St\'ephane Plaszczynski for being the jury
chairman. For their useful comments on earlier versions of this manuscript I
would like to thank Alain Blanchard and St\'ephane Plaszczynski.

I would also like to thank Bernhard Riedl, whose master thesis work was the
basis for some of the unpublished material in Section 4.4.
 
Although a pre-requisite for the HDR, I find supervising students a real pleasure.
Much of the work presented here was made possible by those students. These include
Tim Eifler, Liping Fu, Jean Coupon, Christopher Bonnett, Laura Wolz,
Bernhard Riedl, and Chieh-An Lin, for whom I would like to thank.

I am indebted to my colleagues at the astrophysics department (DAp) of CEA
Saclay for their continuous help and support over the last seven years. Thank
you, Pierre-Olivier Lagage, Michel Talvard, Anne Decourchelle, Pascale
Delbourgo, J\'er\^ome Rodriguez, Marc Sauvage, J\'er\^ome Amiaux, Marguerite
Pierre, Koryo Okumura, Bertrand Morin, Pierre-Antoine Frugier, and Emeric
LeFloc'h, as well as everyone I had the pleasure to work with at the CosmoStat
laboratory, Jean-Luc Starck, J\'er\^ome Bobin, Florent Sureau, Sandrine Pires,
St\'ephane Paulin-Henriksson, Valeria Pettorino, Austin Peel, Sam Farrens,
Fran\c{c}ois Lanusse, Fred Ngol\'e Mboula, Arnau Pujol, Santiago Casas, Kostas
Themelis, Axel Guinot, and Morgan Schmitz.

Finally, I thank my parents for their support, Melissa for her love and care,
and L\'eon for introducing a new element of chaos, creativity, and neverending
inspiration to my life.

%%%%%%%%%%%%%%%%%%%%%%%%%%%%%%%%%%%%%%%%%%%%%%%%%%%%%%%%%%%%
\section*{References}
\addcontentsline{toc}{section}{References}
%%%%%%%%%%%%%%%%%%%%%%%%%%%%%%%%%%%%%%%%%%%%%%%%%%%%%%%%%%%%

%\bibliographystyle{unsrtshort}
%\bibliographystyle{apsrev}
%\bibliographystyle{mn2e}
\bibliographystyle{jphysicsB}

\bibliography{astro}

\stoptwocol

%%%%%%%%%%%%%%%%%%%%%%%%%%%%%%%%%%%%%%%%%%%%%%%%%%%%%%%%%%%%
\end{document}